\documentclass[12pt]{article}
\usepackage{latexsym}
\evensidemargin \oddsidemargin
\usepackage{graphicx}
\usepackage{epstopdf}

\usepackage{longtable}
\usepackage{amsmath}
\usepackage{amssymb}
\usepackage{cite}

\usepackage{ulem}   
\usepackage{color}
\definecolor{orange}{RGB}{255,127,0}
\definecolor{brown}{RGB}{102,51,0}
\definecolor{myred}{RGB}{192,0,0}
\definecolor{Darkgreen}{RGB}{30,120,30}
\definecolor{Darkblue}{RGB}{0,0,200}
\newcommand{\comment}[1]{}

\def\Comment#1{}
\newcommand{\bean}{\begin{eqnarray*}}
\newcommand{\eean}{\end{eqnarray*}}

\newcommand{\gapproxeq}{\lower
.7ex\hbox{$\;\stackrel{\textstyle >}{\sim}\;$}}
\newcommand{\lapproxeq}{\lower
.7ex\hbox{$\;\stackrel{\textstyle <}{\sim}\;$}}

\newcommand\lsim{\mathrel{\rlap{\lower4pt\hbox{\hskip1pt$\sim$}}
    \raise1pt\hbox{$<$}}}
\newcommand\gsim{\mathrel{\rlap{\lower4pt\hbox{\hskip1pt$\sim$}}
    \raise1pt\hbox{$>$}}}
\newcommand{\ba}{\begin{array}}
\newcommand{\ea}{\end{array}}
\newcommand{\nn}{\nonumber}

\newcommand{\be}{\begin{equation}}
\newcommand{\ee}{\end{equation}}
\newcommand{\bear}{\begin{eqnarray}}
\newcommand{\eear}{\end{eqnarray}}

\newcommand{\ket}{\,\rangle}
\newcommand{\bra}{\langle \,}
\newcommand{\eqn}[1]{(\ref{#1})}
\newcommand{\cO}{{\cal O}}
\newcommand{\bel}[1]{\be\label{#1}}
\newcommand{\chpt}{$\chi$PT}
\newcommand{\mL}{\mathcal{L}}
\newcommand{\mA}{\mathcal{A}}

\newcommand{\mF}{\mathcal{F}}
\newcommand{\mG}{\mathcal{G}}
\newcommand{\mI}{\mathcal{I}}
\newcommand{\mJ}{\mathcal{J}}
\newcommand{\mM}{\mathcal{M}}
\newcommand{\mN}{\mathcal{N}}
\newcommand{\mO}{\mathcal{O}}
\newcommand{\mP}{\mathcal{P}}

\newcommand{\mT}{\mathcal{T}}
\newcommand{\mV}{\mathcal{V}}
\newcommand{\mX}{\mathcal{X}}
\newcommand{\mY}{\mathcal{Y}}
\newcommand{\Frac}[2]{\frac{\displaystyle #1}{\displaystyle #2}}
\newcommand{\Int}{\displaystyle{\int}}

\def\bat{\begin{array}{cc}}

\begin{document}
\thispagestyle{empty}
\def\thefootnote{\fnsymbol{footnote}}
\phantom{hola}

\vspace{2cm}

\begin{center}

{\huge\sc {\bf
Fingerprints of heavy scales in \\[10pt]
electroweak effective Lagrangians
}}

\vspace{1cm}

{\sc
Antonio~Pich,$^{1}$\footnote{email: pich@ific.uv.es}  \
Ignasi~Rosell,$^{2}$\footnote{email: rosell@uchceu.es} \
Joaqu\'{\i}n~Santos$^{1}$\footnote{email: Joaquin.Santos@ific.uv.es}
\\[5pt]
and  \ Juan~Jos\'e~Sanz-Cillero$^{3}$\footnote{email: jjsanzcillero@ucm.es}
}

\vspace*{.7cm}

{\sl
$^1$ Departament de F\'\i sica Te\`orica, IFIC, Universitat de Val\`encia -- CSIC,\\
Apt. Correus 22085, E-46071 Val\`encia, Spain,

\vspace*{0.1cm}

$^2$
Departamento de Matem\'aticas, F\'\i sica y Ciencias Tecnol\'ogicas,\\
Universidad CEU Cardenal Herrera,
E-46115 Alfara del Patriarca, Val\`encia, Spain

\vspace*{0.1cm}

$^3$
Departamento de F\'\i sica Te\'orica I, Universidad Complutense de Madrid, E-28040 Madrid, Spain
}

\end{center}

\vspace*{0.1cm}

\begin{abstract}
\noindent
The couplings of the electroweak effective theory contain information on the heavy-mass scales which are no-longer present in the low-energy Lagrangian. We build a general effective Lagrangian, implementing the electroweak chiral symmetry breaking $SU(2)_L\otimes SU(2)_R\to SU(2)_{L+R}$, which couples the known particle fields to heavier states with bosonic quantum numbers $J^P=0^\pm$ and $1^\pm$. We consider colour-singlet heavy fields that are in singlet or triplet representations of the electroweak group. Integrating out these heavy scales, we analyze the pattern of low-energy couplings among the light fields which are generated by the massive states. We adopt a generic non-linear realization of the electroweak symmetry breaking with a singlet Higgs, without making any assumption about its possible doublet structure.
Special attention is given to the different possible descriptions of massive spin-1 fields and the differences arising from naive implementations of these formalisms, showing their full equivalence once a proper short-distance behaviour is required.
\end{abstract}


\def\thefootnote{\arabic{footnote}}
\setcounter{page}{0}
\setcounter{footnote}{0}

\newpage

\section{Introduction}

The first LHC run has established the Standard Model (SM) as the correct theory of the fundamental interactions at the energy scales explored so far \cite{Pich:2015tqa}. A Higgs boson with the expected properties has been found and its measured mass has determined the last free parameter of the electroweak Lagrangian. All SM ingredients are now verified and the experimental results are successfully explained with high precision, exhibiting an overwhelming success of the SM paradigm. At the same time, all LHC searches for exotic objects have given negative results, putting in trouble the most fashionable theoretical scenarios for physics beyond the SM.

While new dynamics is needed to explain the many open questions which remain unanswered within the SM, the LHC data are pushing the energy scale where this new physics could sit beyond the reached experimental sensitivity, well above the TeV. The non-observa\-tion of new particle states suggests the existence of a mass gap between the electroweak and new-physics scales. This situation can be adequately described with effective field theory (EFT) methods~\cite{Georgi:1994qn,Pich:1998xt},
writing the most general Lagrangian with the SM gauge symmetries in terms of the known light fields. The lowest-order term with dimension $D=4$ corresponds to the SM, and any low-energy signals of new phenomena are parametrized in terms of higher-dimensional operators suppressed by the corresponding powers of the new-physics scale. The couplings of the effective Lagrangian contain all the dynamical information on the underlying ultraviolet (UV) dynamics which is accessible at low energies.

When building the effective Lagrangian, one needs to specify the symmetry properties of the light degrees of freedom. In particular, whether the recently discovered Higgs field belongs to a $SU(2)_L$ doublet representation, as predicted in the SM, or it is a singlet field,
detached
from the electroweak Goldstones. The first possibility is usually assumed in most phenomenological analyses, since it provides a simpler and more predictive theoretical framework, based on a linear realization of the electroweak symmetry breaking.
However, in order to actually test the validity of this assumption, the more general (and involved) non-linear realization with a singlet Higgs field must be adopted.

The main weakness of the EFT approach is the large number of unknown low-energy couplings (LECs) that need to be taken into account to perform correct (no hidden assumptions) phenomenological analyses. With a single SM family of fermions and assuming the separate conservation of the baryon and lepton numbers, the most simple linear electroweak effective Lagrangian contains\footnote{
The only operator appearing at $D=5$ (up to Hermitian conjugation and flavour assignments) violates lepton number by two units \cite{Weinberg:1979sa}. With $D=6$, there are 5 independent operators which violate  ${\rm B}$ and ${\rm L}$
\cite{Abbott:1980zj,Wilczek:1979hc}.
}
59 independent operators with $D=6$ \cite{Grzadkowski:2010es,Buchmuller:1985jz}. This number blows up to 1350 $CP$-even plus 1149 $CP$-odd operators when 3-generation flavour quantum numbers are included \cite{Alonso:2013hga}.
A much larger number of independent structures is of course present
in the more general non-linear realization~\cite{Buchalla:2013rka,Buchalla:2012qq}.

Unless new particle states are soon discovered at the LHC, we need to face the involved structure of the electroweak EFT Lagrangian and learn how to identify the dynamics underlying any possible anomalous behaviour which could be observed in the data.
In this paper we attempt a first step in this direction, exploring the low-energy consequences of generic couplings of the known particle fields to heavier states
(resonances). To simplify the analysis, we only consider colour-singlet heavy fields with bosonic quantum numbers $J^P=0^\pm$ and $1^\pm$ that are in singlet or triplet representations of the electroweak group,
and work in the limit where $CP$ is an exact symmetry. Moreover, we ignore QCD interactions and drop all operators containing gluon fields.

We build a general effective Lagrangian, implementing the electroweak chiral symmetry breaking $SU(2)_L\otimes SU(2)_R\to SU(2)_{L+R}$, which contains the SM fields and the heavier states. We adopt a generic non-linear realization of the electroweak symmetry breaking with a singlet Higgs, without making any assumption about its possible doublet structure. Integrating out the heavy particles, we recover the low-energy electroweak EFT with definite values for its LECs; they are functions of the masses and couplings of the heavy states which are no longer in the effective Lagrangian. The resulting pattern of LECs among the light fields characterizes the underlying dynamics
at higher scales \cite{Pich:2015kwa}.

These generic predictions can be made more precise, assuming a given short-distance behaviour of the unknown fundamental theory, {\it i.e.}, what is the expected fall-off at high momenta of specific Green functions. This is a very generic UV requirement, characterizing broad classes of theories. Imposing
a proper UV behaviour on the effective Lagrangian which includes the heavy states, one gets constraints on its parameters with interesting implications for the LECs of the low-energy electroweak EFT~{\cite{Pich:2015kwa}.

Our approach follows the successful methodology
\cite{Ecker:1988te,Ecker:1989yg,Pich:2002xy,Cirigliano:2006hb,Cirigliano:2004ue,Cirigliano:2005xn,RuizFemenia:2003hm,Rosell:2004mn,Rosell:2006dt,Pich:2008jm,Pich:2010sm} developed long time ago in QCD to uncover the dynamical information hidden in the LECs of Chiral Perturbation Theory ($\chi$PT)
\cite{Weinberg:1978kz,Gasser:1983yg,Gasser:1984gg,Pich:1995bw,Ecker:1994gg,Bijnens:1999sh,Bijnens:1999hw,Bijnens:2014lea}.
We can profit now from this experience to explore the much more difficult electroweak case, where the fundamental theory is still unknown.

We will first discuss the well-tested pattern of electroweak symmetry breaking (EWSB) and its associated custodial symmetry in Sec.~\ref{sec:custodial}.
The needed chiral tools to develop our formalism are given in Sec.~\ref{sec:EWEFT}, where we describe the basic ingredients of the electroweak EFT and the power counting adopted to organize the low-energy Lagrangian.
Our counting of infrared chiral dimensions differs from previous works \cite{Buchalla:2013eza} in the treatment of custodial symmetry-breaking operators. We introduce a more efficient power-counting assignment which reduces the number of relevant operators, taking into account the phenomenological
suppression of these effects.
The geometric CCWZ formalism \cite{Coleman:1969sm,Callan:1969sn}
is used in Sec.~\ref{sec:Heavy_Fields} to incorporate the heavy degrees of freedom and construct the high-energy resonance Lagrangian.
We provide a complete classification of allowed structures, satisfying all symmetry requirements, and build the corresponding effective Lagrangian which couples the light and heavy fields,
describing the massive spin-1 bosons
through the usual Proca formalism.

In Sec.~\ref{sec.R-integration}, the heavy states are integrated out with a compact (tree-level)
functional procedure and the resulting low-energy Lagrangian is worked out.
We collect there all contributions to the LECs from spin-0 and spin-1
massive fields, in the Proca four-vector representation.
In some situations, spin-1 heavy particles allow for a more economical treatment
in terms of rank-2 antisymmetric tensor fields $R^{\mu\nu}$~\cite{Ecker:1989yg}.
The alternative description of the electroweak spin-1 resonances with the antisymmetric formalism is presented in Sec.~\ref{sec.antisym}, where the corresponding predictions for the LECs are worked out. The pattern of LECs obtained through a tree-level exchange of heavy spin-1 fields turns out to be completely different with the antisymmetric and Proca descriptions. Both formalisms are of course equivalent versions of the same EFT~\cite{Ecker:1989yg,Bijnens:1995,Kampf:2006}. We give an explicit proof of this equivalence and demonstrate that the differences arising through a naive exchange of massive spin-1 fields are compensated by local operators without heavy states. In Sec.~\ref{sec.SD}, we show how the couplings of these local terms can be determined through short-distance conditions. Once a proper UV behaviour is imposed, the antisymmetric and Proca formalisms yield identical predictions for the wanted LECs. The more fashionable description of spin-1 massive bosons in terms of gauge fields is analyzed in Sec.~\ref{sec.HLS}, showing that it corresponds to a particular case of the Proca formalism (a model), where the gauge symmetry generates directly the needed local terms
to guarantee good UV properties.

Our predictions for the low-energy EWET couplings are finally compiled in
Sec.~\ref{sec.summary}. We discuss there the pattern implied by the different quantum numbers of the massive states which have been integrated out, and
conclude with a few summarizing comments.
Many technical details are given in several appendices.

\section{Custodial Symmetry}
\label{sec:custodial}

In order to generate the masses of the $W^\pm$ and $Z$ bosons, it is necessary
to enlarge the massless $SU(2)_L\otimes U(1)_Y$ gauge theory with three additional degrees
of freedom to account for the missing longitudinal polarizations of the three gauge bosons.
The SM incorporates instead a complex scalar doublet $\Phi(x)$ containing four real fields and,
therefore, one massive neutral scalar, the Higgs boson, remains in the spectrum after the EWSB.
It is convenient to collect the four scalar fields in the $2\times 2$ matrix~\cite{Appelquist:1980vg}
\be
\Sigma \,\equiv\, \left( \Phi^c, \Phi\right) \, =\,
\left( \bat \Phi^{0*} & \Phi^+  \\ -\Phi^- &  \Phi^0 \ea\right)
\label{eq:sigma_matrix}
\ee
with $\Phi^c = i \sigma_2\Phi^*$ the charge-conjugate of the scalar doublet $\Phi$.
The SM scalar Lagrangian can then be written in the form \cite{Pich:1998xt,Pich:1995bw}
\bel{eq:l_sm}
\mL(\Phi)\, =\, \frac{1}{2}\, \langle\, \left(D^\mu\Sigma\right)^\dagger D_\mu\Sigma\,\rangle
- \frac{\lambda}{16} \left(\langle\,\Sigma^\dagger\Sigma\,\rangle
- v^2\right)^2 ,
\ee
where
$D_\mu\Sigma \equiv \partial_\mu\Sigma
+ i g \,\frac{\vec{\sigma}}{2}\vec{W}_\mu \,\Sigma - i g' \,\Sigma \,\frac{\sigma_3}{2} B_\mu $
is the usual gauge-covariant derivative and $\langle A\rangle$ denotes the trace of the $2\times 2$ matrix $A$.

The Lagrangian $\mL(\Phi)$ is invariant under global
$G\equiv SU(2)_L\otimes SU(2)_R$ transformations,
\be
\Sigma \quad \longrightarrow \quad g_L \,\Sigma\, g_R^\dagger\, ,
\qquad\qquad\qquad\qquad
g_{L,R}  \in SU(2)_{L,R} \, ,
\label{eq:sigma_transf}
\ee
while the vacuum choice $\langle 0|\Phi^0|0\rangle = v$ is only preserved when $g_L=g_R$,
{\it i.e.}, by the custodial symmetry group $SU(2)_{L+R}$ \cite{Sikivie:1980hm}.
In the SM, $SU(2)_L$ is promoted to a local gauge symmetry, but only the
$U(1)_Y$ subgroup of $SU(2)_R$ is gauged. Therefore, the $U(1)_Y$ interaction in the covariant derivative breaks the $SU(2)_R$ symmetry.

Let us use the polar decomposition
\be\label{eq:polar}
\Sigma(x) \, = \, \frac{1}{\sqrt{2}}
\left[ v + h(x) \right] \; U(\varphi(x))
\ee
to parametrize the four degrees of freedom as excitations over the chosen vacuum.
This separates in a clear way the Higgs field $h(x)$, which is a singlet
under $G$ transformations, from the three Goldstones $\varphi(x)$ appearing in the $2\times 2$ matrix
$U(\varphi(x))$ which transforms as
\bel{eq:Goldstones}
U(\varphi) \, =\,  \exp{\left\{ i \vec{\sigma} \, \vec{\varphi} / v \right\} }
\quad \longrightarrow \quad g_L \,U(\varphi)\, g_R^\dagger\, .
\ee
One can rewrite $\mL(\Phi)$ in the form~\cite{Appelquist:1980vg,Longhitano:1980iz,Longhitano:1980tm}:
\be
\mL(\Phi)\, =\, \frac{v^2}{4}\,
\langle\, D_\mu U^\dagger D^\mu U \,\rangle \, +\,
\cO\left( h/v \right) ,
\label{eq:sm_goldstones}
\ee
with
$D_\mu U \equiv \partial_\mu U
+ i g \,\frac{\vec{\sigma}}{2}\vec{W}_\mu \, U - i g'\, U \,\frac{\sigma_3}{2} B_\mu$.
Dropping the terms containing the Higgs field, Eq.~\eqn{eq:sm_goldstones} is the universal Goldstone Lagrangian associated with the symmetry breaking
\bel{eq:ChiralSymmetry}
G\equiv SU(2)_L\otimes SU(2)_R\quad \longrightarrow \quad H\equiv SU(2)_{L+R}\, .
\ee
The same Lagrangian describes the low-energy dynamics of pions in two-flavour QCD, with $v\to f_\pi$ and $\vec{\varphi}\to\vec{\pi}$ \cite{Pich:1998xt}.
The electroweak precision data~\cite{Pich:2012sx}
have confirmed that \eqn{eq:ChiralSymmetry} is also the right pattern of symmetry breaking associated with the electroweak Goldstone bosons,
with $v = \left(\sqrt{2} G_F\right)^{-1/2} = 246\:\mathrm{GeV}$.

The unitary gauge, where the Goldstones are rotated away through an appropriate gauge transformation, corresponds to $U=1$. The Goldstone Lagrangian in Eq.~\eqn{eq:sm_goldstones} reduces then to a quadratic mass term for the gauge bosons, giving the SM prediction for the $W^\pm$ and $Z$ masses:
$m_W = m_Z \,\cos{\theta_W} = v g/2$, with
$Z^\mu \equiv \cos{\theta_W} W_3^\mu - \sin{\theta_W} B^\mu$ and
$\tan{\theta_W} = g'/ g$. These masses are generated by the electroweak Goldstones, not by the Higgs field (the QCD pions generate a tiny correction
$\delta m_W = \delta m_Z \,\cos{\theta_W} = f_\pi g/2$).

Before the Higgs discovery, the success of the SM mechanism of EWSB was only due to its pattern of symmetry breaking in Eq.~\eqn{eq:ChiralSymmetry}, which is well established phenomenologically. The particular dynamical structure of the SM scalar Lagrangian can only be tested through the Higgs properties. The measured Higgs mass determines the quartic coupling,
$\lambda = m_h^2/(2 v^2) = 0.13$,
while its gauge couplings are consistent with the SM prediction within the present experimental uncertainties.

The SM scalar doublet $\Phi$ gives rise to a renormalizable Lagrangian with good short-distance properties. However, one would like to test phenomenologically
whether this doublet structure is indeed the mechanism chosen by Nature to generate the EWSB or there is a different implementation of the pattern of symmetry breaking in Eq.~\eqn{eq:ChiralSymmetry}.
Therefore, we will build the electroweak effective theory (EWET)
in terms of the Goldstone matrix $U(\varphi(x))$ and a singlet scalar field $h(x)$,
without assuming any relation among them. The Goldstone dynamics can be analyzed through
an effective Lagrangian with the SM gauge symmetry realized non-linearly,\footnote{The usual linear realization is just a particular case of the more general non-linear one. Making a polar decomposition of the scalar doublet $\Phi$, the linearly-realized electroweak effective Lagrangian can be rewritten in terms of $h(x)$ and the matrix $U(\varphi)$, in the same way that has been done for the SM scalar sector in Eq.~(\ref{eq:sm_goldstones}). Since the doublet structure of $\Phi$ combines together the Goldstones and the Higgs field, it implies specific relations among the couplings of the non-linear EWET which could be tested once precise data become available.} applying momentum
expansion techniques analogous to those used in $\chi$PT to study low-energy QCD.

\section{Electroweak Effective Theory}
\label{sec:EWEFT}

The EWET is defined by the most general low-energy Lagrangian, containing the SM gauge bosons and fermions, the electroweak Goldstones and the Higgs field $h$, which satisfies the SM gauge symmetries. Our only assumption is the pattern of EWSB in Eq.~\eqn{eq:ChiralSymmetry}. The Lagrangian will be organized as an expansion in powers of derivatives (momenta) over the EWSB (and/or any new physics) scale:
\bel{eq:L_EWET}
\mL_{\mathrm{EWET}}\, =\, \mL_{\mathrm{SM}}^{(0)} +  \Delta \mL_2   +\cdots
\ee
The first piece $\mL_{\mathrm{SM}}^{(0)}$ denotes the renormalizable massless (unbroken) SM Lagrangian,
which only contains fermions and gauge bosons:
\bear
\mL_{\rm SM}^{(0)} &=& \sum_{f} i \bar{f}\gamma^\mu D_\mu f\, +\, \mL_{\rm YM}\, ,
\eear
with the sum running over all fermions $f$ in the SM, $D_\mu$ being the covariant derivative
of the $SU(3)_C\times SU(2)_L \times U(1)_Y $ SM gauge group and $\mL_{\rm YM}$
the corresponding Yang-Mills Lagrangian. When we later study the chiral low-energy counting
we will see that $\mL_{\rm SM}^{(0)}$ is part of the lowest-order (LO) Lagrangian $\mL_2$.
The remaining LO terms related with the EWSB are contained in $\Delta\mL_2$
and the dots stand for the infinite tower of higher-order operators in the chiral expansion.

A very detailed description of the EWET has already been given in Refs.~\cite{Buchalla:2013rka,Buchalla:2012qq}. We will introduce a slightly modified formalism for the Goldstone fields, which is more appropriate to study their couplings to massive states \cite{Ecker:1988te}.

\subsection{Bosonic fields}
\label{subsec:bosons}

The electroweak Goldstone bosons are parametrized by the
$SU(2)_L\times SU(2)_R/SU(2)_{L+R}$
coset coordinates $(u_L^{\phantom{\dagger}}(\varphi),u_R^{\phantom{\dagger}}(\varphi))$, which transform under $g\equiv (g_L^{\phantom{\dagger}},g_R^{\phantom{\dagger}})\in SU(2)_L\times SU(2)_R$ as
\be
u_L^{\phantom{\dagger}}(\varphi)\quad\longrightarrow\quad g_L^{\phantom{\dagger}}\, u_L^{\phantom{\dagger}}(\varphi)\, g_h^\dagger(\varphi,g)\, ,
\qquad\qquad\qquad
u_R^{\phantom{\dagger}}(\varphi)\quad\longrightarrow\quad g_R^{\phantom{\dagger}}\, u_R^{\phantom{\dagger}}(\varphi)\, g_h^\dagger(\varphi,g)\, ,
\ee
with $g_h^{\phantom{\dagger}}(\varphi,g)\equiv g_h^{\phantom{\dagger}}$ a compensating transformation to preserve the chosen coset representative, which depends both on the Goldstone coordinates and the group element $g$ \cite{Coleman:1969sm,Callan:1969sn}. Since parity interchanges left and right, leaving $SU(2)_{L+R}$ invariant, the compensating transformation $g_h^{\phantom{\dagger}}(\varphi,g)$ is the same in the two chiral sectors. We will adopt the canonical choice of coset representative
$u_L^{\phantom{\dagger}}(\varphi)=u_R^\dagger(\varphi)=u(\varphi)$~\cite{Pich:2012dv,Pich:2013fea},\footnote{
The opposite convention $u_R^{\phantom{\dagger}}(\varphi)=u_L^\dagger(\varphi)=u(\varphi)$ is usually adopted
in $\chi$PT~\cite{Ecker:1988te}.}
which transforms like
\bel{eq:u_transf}
u(\varphi)\quad \longrightarrow\quad g_L^{\phantom{\dagger}}\,  u(\varphi)\,  g_h^\dagger(\varphi,g) \; =\; g_h^{\phantom{\dagger}}(\varphi,g) \, u(\varphi) \, g_R^\dagger\, ,
\ee
with the exponential representation
$u(\varphi)=\exp\{ i\vec{\sigma}\,\vec{\varphi}/(2v)\}$.
Its relation with the matrix $U(\varphi)$ in Eq.~\eqn{eq:Goldstones} is given by
\bel{eq:U_u_rel}
U(\varphi)\,\equiv\, u_L^{\phantom{\dagger}}(\varphi)\, u_R^\dagger(\varphi)\, =\, u(\varphi)^2
\quad\longrightarrow\quad g_L^{\phantom{\dagger}}\, U(\varphi)\, g_R^\dagger \, .
\ee

We formally introduce the $SU(2)_L$ and $SU(2)_R$ matrix fields, $\hat{W}_\mu$ and $\hat{B}_\mu$ respectively, transforming as
\bel{eq:FakeTransform}
\hat{W}^\mu\quad\longrightarrow\quad g_L^{\phantom{\dagger}}\, \hat{W}^\mu g_L^\dagger + i\, g_L^{\phantom{\dagger}}\, \partial^\mu g_L^\dagger\, ,
\qquad\qquad\quad
\hat{B}^\mu\quad\longrightarrow\quad g_R^{\phantom{\dagger}}\, \hat{B}^\mu g_R^\dagger + i\, g_R^{\phantom{\dagger}}\, \partial^\mu g_R^\dagger\, ,
\ee
the covariant derivative
\bel{eq:DU}
D_\mu U \, =\, \partial_\mu U -  i\, \hat{W}_\mu  U + i\, U \hat{B}_\mu
\quad\longrightarrow\quad g_L^{\phantom{\dagger}}\, D_\mu U \, g_R^\dagger
\, ,
\ee
and the corresponding field-strength tensors
\bear
\hat{W}_{\mu\nu}  &=& \partial_\mu \hat{W}_\nu - \partial_\nu \hat{W}_\mu
- i\, [\hat{W}_\mu,\hat{W}_\nu]
\quad \longrightarrow\quad g_L^{\phantom{\dagger}}\, \hat{W}_{\mu\nu} \, g_L^\dagger
\, ,
\nn\\
\hat{B}_{\mu\nu}  &=& \partial_\mu \hat{B}_\nu - \partial_\nu \hat{B}_\mu
- i\, [\hat{B}_\mu,\hat{B}_\nu]
\quad\longrightarrow\quad
g_R^{\phantom{\dagger}}\, \hat{B}_{\mu\nu} \, g_R^\dagger
\, .
\eear
We can then build effective operators invariant under local $SU(2)_L\otimes SU(2)_R$
transformations.
The identification \cite{Pich:2012jv}
\bel{eq:SMgauge}
\hat{W}^\mu \, =\, -g\;\frac{\vec{\sigma}}{2}\, \vec{W}^\mu \, ,
\qquad\qquad\qquad\quad
\hat{B}^\mu\, =\, -g'\;\frac{\sigma_3}{2}\, B^\mu
\ee
allows us to recover the SM gauge fields, breaking explicitly the $SU(2)_R$ symmetry group while preserving the $SU(2)_L\otimes U(1)_Y$ gauge symmetry.

For the construction of the effective Lagrangian, it is convenient to define tensors transforming as $SU(2)_{L+R}$ triplets,
$\mathcal{X}\longrightarrow g_h^{\phantom{\dagger}} \mathcal{X} g_h^\dagger$, and their covariant derivatives
\bel{eq:CovDev}
\nabla_\mu \mX \, =\, \partial_\mu \mX \, +\, [\Gamma_\mu , \mX ]
\quad\longrightarrow\quad g_h^{\phantom{\dagger}} \, \nabla_\mu \mX\, g_h^\dagger
\, .
\ee
The needed connection can be easily constructed with the left and right parts of the Goldstone coset representative \cite{Pich:2012jv}:
\bel{eq:connection}
\Gamma_\mu =
\Frac{1}{2} \left(\Gamma_\mu^{L} +\Gamma_\mu^{R}\right)\, ,
\quad\;
\Gamma_\mu^{L} = u_L^\dagger(\varphi) \left(\partial_\mu - i\,\hat{W}_\mu\right) u_L^{\phantom{\dagger}}(\varphi)
\, , \quad\;
\Gamma_\mu^{R} = u_R^\dagger(\varphi) \left(\partial_\mu - i\,\hat{B}_\mu\right) u_R^{\phantom{\dagger}}(\varphi)
\, ,
\ee
which transform as
\be
\Gamma_\mu^{L,R} \quad \longrightarrow \quad g_h^{\phantom{\dagger}}\, \Gamma_\mu^{L,R} \, g_h^\dagger\, +\, g_h^{\phantom{\dagger}}\, \partial_\mu g_h^\dagger \, ,
\qquad\qquad\quad
\Gamma_\mu \quad\longrightarrow\quad g_h^{\phantom{\dagger}}\, \Gamma_\mu\, g_h^\dagger
\, +\, g_h^{\phantom{\dagger}}\, \partial_\mu g_h^\dagger \, .
\ee
The quantities
\bear\label{eq.cov-bosonic-tensors}
u_\mu  \, =\, i\,
\left( \Gamma_\mu^R - \Gamma_\mu^L\right)
&\! =&\!
i\, u\, (D_\mu U)^\dagger u \, =\, -i\, u^\dagger D_\mu U\, u^\dagger
\, =\, u_\mu^\dagger\, ,
\nn\\[5pt]
f_\pm^{\mu\nu} &\! =&\!
u^\dagger \hat{W}^{\mu\nu}  u \pm u\, \hat{B}^{\mu\nu} u^\dagger
\eear
turn out to be very useful building blocks, satisfying the required triplet transformation property:
\bel{eq:umu_transform}
u_\mu\quad\longrightarrow\quad g_h^{\phantom{\dagger}}\, u_\mu\, g_h^\dagger \, ,
\qquad\qquad\qquad
f_\pm^{\mu\nu}\quad\longrightarrow\quad g_h^{\phantom{\dagger}}\, f_\pm^{\mu\nu}\, g_h^\dagger \, .
\ee
In App.~\ref{app:discrete-transf}, we summarize how these bosonic chiral structures transform
under discrete symmetries and Hermitian conjugation.

The LO Goldstone Lagrangian in Eq.~\eqn{eq:sm_goldstones} can be written in terms
of the invariant operator $\langle u_\mu u^\mu\rangle$.
Since the Higgs field is a singlet under $SU(2)_L\times SU(2)_R$, we can multiply this structure with an arbitrary polynomial
of $h$ \cite{Grinstein:2007iv}. The powers of the Higgs field are compensated by corresponding powers of
the electroweak scale $v$, as happens for the Goldstone fields in the non-linear representation given
by $u(\varphi)$. We will show later that they do not increase the chiral dimension, leading to a consistent
power counting to organize the EWET \cite{Buchalla:2013eza}.
The bosonic part of $\Delta\mL_2$ is then given by
\bel{eq:L2}
\Delta\mL_2^{\mathrm{Bosonic}}\, =\, \frac{1}{2}\,
\partial_\mu h\,\partial^\mu h
\, -\,\frac{1}{2}\, m_h^2\, h^2 \, -\, V(h/v)
\, +\,
\frac{v^2}{4}\,\mF_u(h/v)\;\langle u_\mu u^\mu\rangle\, ,
\ee
with
\be\label{eq:Fhu_V}
  V(h/v)\, = \, v^4\;\sum_{n=3} c^{(V)}_n \left(\frac{h}{v}\right)^n\, ,
\qquad\qquad
\mF_u(h/v)\, = \, 1\, +\, \sum_{n=1} c^{(u)}_n \left(\frac{h}{v}\right)^n\, .
\ee
The SM scalar Lagrangian is recovered for
 $c^{(V)}_3 = \frac{1}{2}\, m_h^2/v^2$, $c^{(V)}_4 = \frac{1}{8}\, m_h^2/v^2$, $c^{(V)}_{n>4} = 0$,
$c^{(u)}_1 = 2$, $c^{(u)}_2 = 1$ and $c^{(u)}_{n>2} = 0$.
Since we expect the Higgs $h$ and the electroweak Goldstones to have a similar underlying origin, we assume that the coefficients $c_n^{(u)}$ are $\cO(1)$, as those governing the expansion of $u(\varphi)$ in terms of the $\vec\varphi$  fields.
This is consistent with the present experimental situation,
where the only coupling measured so far, $c_1^{(u)}$, is found to be close to its SM value.

The symmetry requirements allow one to multiply the quadratic derivative term of the Higgs with an arbitrary
function $\mF_h(h/v)$. However, this function can be always reduced to $\mF_h=1$ through an appropriate Higgs field
redefinition~\cite{SILH}. An explicit derivation is provided in App.~\ref{app:simplifications}.

\subsection{Fermionic fields}

In order to embed the SM fermion multiplets in $SU(2)_L\otimes SU(2)_R$, the symmetry group is extended
to $\mG=SU(2)_L\otimes SU(2)_R\otimes U(1)_{X}$
with $X=(\mathrm{B}-\mathrm{L})/2$, being $\mathrm{B}$ and $\mathrm{L}$ the baryon and lepton
quantum numbers, respectively~\cite{Hirn:2006}.
The left and right chiralities of the SM fermions are arranged into $SU(2)_L$ and $SU(2)_R$ doublets:
\bear
\psi_L = \left( \begin{array}{c} t_L \\ b_L \end{array}\right)\, ,
\qquad\qquad
\psi_R  =\left( \begin{array}{c} t_R \\ b_R \end{array}\right) \, ,
\eear
with $\psi_{L,R}=P_{L,R}\,\psi$ and $P_{L,R} =\frac{1}{2}\, (1\mp \gamma_5)$.
The other quark and lepton doublets are organized similarly.
The fermions transform under $\mG$ like
\bear
\psi_L  \quad\longrightarrow\quad g_X\, g_L\;  \psi_L   \, ,
\qquad\qquad\qquad
\psi_R  \quad\longrightarrow\quad g_X\, g_R\;  \psi_R   \, ,
\label{eq.psi-transformation}
\eear
with $g_X \in U(1)_X$.
The corresponding covariant derivatives of these fermion doublets are given by
\bel{eq.dpsi}
D_\mu^L\psi_L\, =\,\left(\partial_\mu - i\,\hat{W}_\mu - i\,
\hat{X}_\mu \,
\Frac{( {\rm B}-{\rm L})}{2}
\right) \psi_L\, ,
\qquad
D_\mu^R\psi_R\, =\,\left(\partial_\mu - i\,\hat{B}_\mu - i\,
\hat{X}_\mu
  \Frac{({\rm B- L})}{2}
\right) \psi_R\, ,
\ee
where the $SU(2)_{L,R}$ auxiliary matrix fields
$\hat{W}_\mu$ and $\hat{B}_\mu$ were introduced in the previous section,
$({\rm B- L})/2$ must be understood as an operator that acts on the fermions
and the $U(1)_X$ field $\hat{X}_\mu$ transforms like
\bel{eq:Xtransform}
\hat{X}^\mu\quad\longrightarrow\quad \hat{X}^\mu
+ i\, g_X^{\phantom{\dagger}}\, \partial^\mu g_X^\dagger\, .
\ee
The $U(1)_X$ field strength tensor
\be
\hat{X}_{\mu\nu}\, =\,\partial_\mu\hat{X}_\nu -\partial_\nu \hat{X}_\mu
\ee
is a singlet under $\mG$.
The SM gauge interactions are recovered when these auxiliary fields are forced to take the values given in Eq.~(\ref{eq:SMgauge}) and
\bel{eq:Xmu-fixing}
\hat{X}_\mu  \, =\, \,-\, g'\, B_\mu
 \, .
\ee
This introduces an explicit breaking of the symmetry group $\mG$
to the SM subgroup $SU(2)_L\times U(1)_Y$ with
$Y = T_{3R} +\frac{1}{2}\, (\mathrm{B}-\mathrm{L})$~\cite{MS:75}, {\it i.e.},
\bel{eq:Q_LR}
Q\, =\, T_{3L} + T_{3R} + \frac{\mathrm{B}-\mathrm{L}}{2}\, .
\ee
The bosonic formalism discussed in the previous subsection does
not get modified by this enlargement of the symmetry group as
for bosons one has
$\mathrm{B}=\mathrm{L}=0$.

In order to construct the EWET operators, it is convenient to introduce the covariant fermion doublet fields
\be
\xi_{L} \,\equiv\, u_{L}^\dagger \, \psi_{L}\, =\, u^\dagger \, \psi_L \, ,
\qquad\qquad\qquad
\xi_{R} \,\equiv\,  u_{R}^\dagger \, \psi_{R}  \, =\, u \,\psi_R \, ,
\ee
that transform with $g_h$ instead of $g_{L,R}$:
\bel{eq:xi_transf}
\xi_{L,R}\quad\longrightarrow\quad   g_X \, g_h\; \xi_{L,R}\, .
\ee
The same transformation applies obviously to the combined fermion field $\xi \equiv \xi_L + \xi_R$.
The corresponding covariant derivatives are easily found to be
\bear
d_\mu^{L} \xi_{L}   &\! =&\!
\left(  \partial_\mu + \Gamma_\mu^{L}  - i
\,\hat{X}_\mu
\Frac{({\rm B}-{\rm L})}{2}
\right) \xi_{L}
\,\, =\,\,
u_L^\dagger \left(  \partial_\mu - i\, \hat{W}_\mu - i
\,\hat{X}_\mu
\Frac{( {\rm B}  -{\rm L})}{2}
     \right) \psi_{L}
\, ,
\nn\\
d_\mu^{R} \xi_{R}   &\! =&\!
\left(  \partial_\mu + \Gamma_\mu^{R} - i
\,\hat{X}_\mu
\Frac{({\rm B}-{\rm L})}{2}
 \right) \xi_{R}
\,\, =\,\,
u_R^\dagger \left(  \partial_\mu - i\,\hat{B}_\mu - i
\, \hat{X}_\mu
\Frac{({\rm B}-{\rm L})}{2}
 \right) \psi_{R}
\, ,
\eear
and
$d_\mu \xi  =  d_\mu^R\xi_R + d_\mu^L\xi_L$.
They transform covariantly under $\mG$ in the form
\bear
d_\mu^{L,R} \xi_{L,R}
\quad\longrightarrow\quad   g_X \, g_h\; d_\mu^{L,R} \xi_{L,R}\, .
\eear

Notice that the Goldstones disappear from the covariant form of the kinetic fermion Lagrangian:
\be
\mL_{\rm Fermionic}^{(0)}   \, =\,
i\,\bar\xi \gamma^\mu d_\mu \xi \, =\,
i\,\overline\psi_L \gamma^\mu D_\mu^L\psi_L \, +\,
i\,\overline\psi_R \gamma^\mu D_\mu^R\psi_R \, .
\ee
In general, Goldstone fields are only required by the electroweak symmetry in fermionic terms that mix left and right chiralities, {\it e.g.}, scalar, pseudoscalar and tensor fermion bilinears, contrary to vector and axial-vector ones:
\bear
\bar{\xi}\,\Gamma\, \xi' &\! =&\! \left\{ \bat
\overline{\psi}_L\Gamma\psi_L' \,+\, \overline{\psi}_R\Gamma\psi_R'
& \qquad\quad\mbox{\small  ($\Gamma=\gamma^\mu, \, \gamma^\mu \gamma_5$)}\, ,
\\[10pt]
\overline{\psi}_L \Gamma\, U(\varphi)\,\psi_R'
\,+\, \overline{\psi}_R \Gamma \, U(\varphi)^\dagger \psi_L'
& \qquad\quad\mbox{\small ($\Gamma=1,\, i\gamma_5,\, \sigma^{\mu\nu}$)} \, .
\ea\right.\quad
\eear

The fermion masses are generated through Yukawa interactions that break explicitly the symmetry group $\mG$. To account for this type of symmetry breaking one introduces right-handed spurion fields
transforming as
\bel{eq:Yspurion}
\mY_R\quad\longrightarrow\quad g_R^{\phantom{\dagger}}\,\mY_R\, g_R^\dagger\, ,
\qquad\qquad\qquad
\mY\, =\, u\,\mY_R\, u^\dagger
\quad\longrightarrow\quad g_h^{\phantom{\dagger}}\,\mY\, g_h^\dagger
\, .
\ee
The Yukawa interaction takes then the form
\bel{eq:Yukawas}
\Delta\mL_2^{\mathrm{Fermionic}}\, =\, -v\; \bar\xi_L\, \mY\, \xi_R \, +\, \mathrm{h.c.}
\, =\, -v\;\bar\psi_L\, U(\varphi)\,\mY_R\,\psi_R\, +\, \mathrm{h.c.}
\ee
which is formally invariant under $\mG$ transformations. The explicit symmetry breaking incorporated into the SM Lagrangian is recovered when the spurion field adopts the value \cite{ABCH:85,BEM:99}
\bel{eq:SM_Yspurion}
\mY\, =\, \hat{Y}_t(h/v)\,\mP_+ + \hat Y_b(h/v)\,\mP_-\, ,
\qquad\qquad
\mP_\pm \,\equiv\,\frac{1}{2}\,\left( I_2\pm\sigma_3\right)\, ,
\ee
where \cite{Buchalla:2013rka}
\bel{eq:SM_Yspurion_h}
\hat{Y}_{t,b}(h/v)\, =\, \sum_{n=0}\, \hat{Y}_{t,b}^{(n)}\, \left(\frac{h}{v}\right)^n\, .
\ee

In order to incorporate the flavour structure, the fermion doublets $\xi$ must be promoted to vectors $\xi^A$ in the generation space
with family index $A$. The spurion field $\mY$ becomes then a $3\times 3$ flavour matrix \cite{Espriu:2000fq} with up-type and down-type components $\hat{Y}_u(h/v)$ and $\hat{Y}_d(h/v)$, which parametrize the custodial and flavour symmetry breaking. Moreover, different $\hat{Y}_{u,d}^{(n)}$ flavour structures could appear at every order in the expansion in powers of the Higgs field $h$, unless additional dynamical inputs are introduced (chiral symmetry alone does not fix these structures).\footnote{A minimal flavour violation scenario \cite{D'Ambrosio:2002ex} would imply a common $\hat Y_t$ or $\hat Y_b$ flavour structure for all $(h/v)^n$ terms.} For simplicity, in this article we will only consider a single fermion family and assume universality, {\it i.e.}, that all families couple in exactly the same way. We postpone the study of the EWET flavour dynamics to future works.

The fermionic fields are combined into generic bilinears $J_\Gamma$
with well-defined Lorentz transformation properties, which can be further used to build Lagrangian
operators with an even number of fermion fields. Making explicit the spinorial ($\alpha,\beta$) and $SU(2)$ ($m,n$) indices, the covariant bilinears have the general form
\be
J^\Gamma_{mn}\, = \, \bar{\eta}^\alpha_n \Gamma^{\alpha\beta} \zeta^\beta_m  \, = \, - \zeta^\beta_m \bar{\eta}^\alpha_n \Gamma^{\alpha\beta}
\, =\, - \mathrm{Tr}_D \{ \zeta_m \bar{\eta}_n \Gamma \} \, ,
\ee
where $\eta, \zeta$ are covariant spinor structures,
$\Gamma = \left\{ I, i\gamma_5, \gamma^\mu, \gamma^\mu\gamma_5, \sigma^{\mu\nu}\right\}$ the usual basis of Dirac matrices, and
$\mathrm{Tr}_D$ refers to the Dirac trace. The minus sign on the right-hand side is generated by the permutation of the two fermion fields. These bilinears transform covariantly,
\be
J_\Gamma\, = \, \bar{\eta}\, \Gamma\, \zeta
\quad \longrightarrow \quad  g_h^{\phantom{\dagger}}\, J_\Gamma\, g_h^\dagger\, ,
\ee
and can be easily combined with other tensors $\mO$ transforming like
$\mO\to g_h^{\phantom{\dagger}} \mO g_h^\dagger$ to build invariant operators under $\mG$:
\be
\bra J_\Gamma \, \mO \ket  \, =\, -  \zeta_m^\beta \bar{\eta}_n^\alpha \Gamma^{\alpha\beta}  \; \mO_{nm}
\,  =\,    \bar{\eta}  \, \Gamma \,   \mO \,  \zeta\, .
\ee

The bilinears relevant for the present work are:
\bear
(J_S)_{mn} &\equiv & \, -\, Tr_D\{ \xi_m \bar{\xi}_n \}
\,\,\, =\,\,\, \bar{\xi}_n \xi_m    \, ,
\nn\\
(J_P)_{mn}  &\equiv & \, -\, i\, Tr_D\{ \xi_m \bar{\xi}_n \gamma_5 \}
\,\,\,=\,\,\, i\,   \bar{\xi}_n \gamma_5 \xi_m  \, ,
\nn\\
(J_V^\mu)_{mn} &\equiv & \, -\, Tr_D\{ \xi_m \bar{\xi}_n \gamma^\mu \}
\,\,\, =\,\,\,  \bar{\xi}_n\gamma^\mu \xi_m \, ,
\nn\\
(J_A^\mu)_{mn} &\equiv & \, -\, Tr_D\{ \xi_m \bar{\xi}_n\gamma^\mu\gamma_5\}
\,\,\, =\,\,\, \bar{\xi}_n\gamma^\mu\gamma_5 \xi_m \, ,
\nn\\
(J_T^{\mu\nu})_{mn} &\equiv & \, -\, Tr_D\{ \xi_m \bar{\xi}_n\sigma^{\mu\nu} \}
\,\,\, =\,\,\,  \bar{\xi}_n\sigma^{\mu\nu} \xi_m \, .
\eear
Some useful transformation properties of the covariant bilinears under discrete
symmetries are compiled in App.~\ref{app:discrete-transf}.

\subsection{Chiral power counting}

The LO bosonic Lagrangian $\Delta\mL_2^{\mathrm{Bosonic}}$
involves terms with arbitrary powers of the Goldstone and Higgs fields,
which are generated through the Taylor expansions of the non-linear coset representative
$u(\varphi)$ and the polinomic functions $\mF_u(h/v)$ and $V(h/v)$ in Eq.~(\ref{eq:L2}).
Therefore, the EWET operators cannot be simply ordered according to their canonical dimensions. One must use instead the so-called chiral dimension $\hat d$ which reflects their infrared behaviour at low momenta~\cite{Weinberg:1978kz}. The effective Lagrangian is expressed as an infinite sum of terms, scaling with increasing powers of momenta in the limit $p\to 0$:
\bel{eq:EchL}
\mL_{\mathrm{EWET}}\, =\, \sum_{\hat d\ge 2}\, \mL_{\hat d}\, ,
\qquad\qquad\qquad
\mL_{\hat d} = \mO (p^{\hat d})\, .
\ee
Quantum loops are renormalized order by order in this low-energy expansion.

Owing to their non-linear transformation \eqn{eq:Goldstones}, Goldstones do not have infrared dimension and their canonical field dimension is compensated by the intrinsic electroweak scale $v$ characterizing the EWSB. Therefore $u(\varphi)\sim \varphi/v\sim\cO(p^0)$.
We assume that the same chiral counting applies to the light Higgs field.\footnote{
This assumption can be easily relaxed in weakly-coupled scenarios where the perturbative expansions in powers of $h/v$ of $\mF_u(h/v)$, $V(h/v)$ and analogous functions are suppressed by corresponding powers of some weak coupling.}

Derivatives bring one power of momenta. A consistent counting requires then that the external gauge sources $\hat W_\mu$, $\hat B_\mu$ and $\hat X_\mu$, present in the covariant derivatives, carry the same infrared dimension $\hat d = 1$. Moreover, since $p_{W,Z,h}^2=m_{W,Z,h}^2$, in the low-energy effective theory involving light $W^\pm$, $Z$ and $h$ fields, their masses must also be counted as $\cO(p)$. Since $m_W = g v/2$ and $m_Z = \sqrt{g^2+g'^2}\, v/2$, this implies that $g, g'\sim \mO(p)$, while $\vec{W}_\mu$ and $B_\mu$ are $\mO(p^0)$.\footnote{This infrared power counting is needed for a consistent loop expansion. Of course, in particular kinematical regimes such as $p\gg m_W$ one can always introduce a refined hierarchy of scales and couplings.}
With these chiral counting rules, all terms in $\mL_{\rm YM}$ and $\Delta\mL_2^{\mathrm{Bosonic}}$ are of $\mO(p^2)$, provided one assigns also this chiral dimension to the Higgs potential.\footnote{
The SM Higgs self-interactions are proportional to $m_h^2\sim \cO(p^2)$.
This counting is also consistent with strongly-coupled scenarios with a pseudo-Goldstone Higgs and models where the potential is assumed to be radiatively generated and, therefore, implicitly includes two powers of some weak coupling.}
In particular, the kinetic, cubic and quartic gauge terms have all $\hat d = 2$. Therefore,
the chiral low-energy expansion preserves gauge invariance order by order \cite{Hirn:2006}.

The infrared dimension of chiral fermion fields is also one unit less that their canonical dimension, $\xi_{L,R}\sim\mO(p^{1/2})$, so that the fermionic component of $\mL_{\rm SM}^{(0)}$ is of $\cO(p^2)$. The Yukawa couplings $y_\xi$,
and thus the SM fermion masses, are assigned chiral dimension $\hat d=1$; the fermion mass terms are then also of $\cO(p^2)$.

The EWET power-counting rules can be summarized as:
\bear\label{eq:power_counting}
v\, ,\, \Frac{\varphi}{v}\,  , \, u(\varphi)\, ,\, U(\varphi)\, , \, \Frac{h}{v}\,  , \, \Frac{\vec{W}_\mu}{v}\,  ,\, \Frac{B_\mu}{v}
& \sim & \cO\left(p^0\right)\, ,
\nn\\
\Frac{\xi}{v}\, ,\, \Frac{\bar\xi}{v}\, ,\,\Frac{\psi}{v}\, ,\,\Frac{\bar\psi}{v} & \sim & \cO\left( p^{1/2} \right)\, ,
\nn\\
D_\mu U\,  ,\, u_\mu\, , \,\partial_\mu\, , \, \hat{W}_\mu\, , \, \hat{B}_\mu\, ,\, \hat{X}_\mu\,  ,\, m_h\, , \, m_W\, , \, m_Z\, , \, m_\psi\, ,\, g\, ,\, g'\, ,\, \mY  & \sim & \cO\left( p\right)\, ,
\nn\\
\hat{W}_{\mu\nu}\, ,\, \hat{B}_{\mu\nu}\, ,\, \hat{X}_{\mu\nu}\, ,\, f_{\pm\, \mu\nu} \, ,\, c_n^{(V)} & \sim & \cO\left( p^2\right)\, ,
\nn\\
\partial_{\mu_1}\partial_{\mu_2} ... \partial_{\mu_n}\,\mF(h/v)  & \sim & \cO\left( p^n\right)\, .
\eear

The infrared power counting leads to a well-defined loop expansion,
because loops increase the chiral dimension and their divergences are then renormalized by higher-order operators.
A standard dimensional analysis \cite{Weinberg:1978kz,Georgi-Manohar}, explained in detail in App.~\ref{app:power-counting}, shows that an arbitrary Feynman diagram $\Gamma$ scales like \cite{Hirn:2006,Buchalla:2012qq,Buchalla:2013rka,Buchalla:2013eza}
\bel{eq:Gamma_scaling}
\Gamma \,\sim\, p^{\hat{d}_\Gamma}\, ,
\qquad\qquad\qquad
\hat{d}_\Gamma\, =\, 2 + 2L + \sum_{\hat{d}} (\hat{d} -2)\, N_{\hat{d}} \, ,
\ee
where $L$ is the number of loops and $N_{\hat{d}}$ indicates the number of vertices with a given value of $\hat{d}$. Loops increase the chiral dimension by two units and are suppressed by the usual geometrical factor $1/(4\pi)^2$, giving rise to a series expansion in powers of
momenta over the electroweak chiral scale $\Lambda_{\mathrm{EWET}} = 4\pi v\sim 3~\mathrm{TeV}$. There will be in addition, operators generated by short-distance contributions from new physics, suppressed by the corresponding new-physics scale $\Lambda_{\mathrm{NP}}$. When momenta are low compared with these two scales, only a finite number of operators need to be taken into account, at a given order in $p/\Lambda_{\mathrm{EWET}}$ and $p/\Lambda_{\mathrm{NP}}$. The precision can always be improved by going to the next order in the expansion, at the price of having more operators with their corresponding LECs.

The LO contribution is generated by tree-level diagrams with the $\hat d=2$ Lagrangian $\mL_2$. Next-to-leading-order (NLO) corrections have $\hat d=4$ and originate from two different sources: 1) one-loop diagrams with the LO Lagrangian $\mL_2$, and 2) tree-level diagrams with one $\hat{d}=4$ operator and an arbitrary number of insertions of $\mL_2$, which do not increase the chiral dimension.

According to the power-counting rules in Eq.~\eqn{eq:power_counting}, a four-fermion operator brings a chiral dimension 2. This is consistent with the light-boson-exchange amplitudes ($\phi=W^\pm$, $Z$, $\gamma$, $h$) from $\mL_2$, which carry a factor $g_\phi^2/(p^2-m_{\phi}^2)\sim \mO(p^0)$ with $g_\phi$ the appropriate coupling. However, those are non-local contributions. Local four-fermion operators in the EWET originate in short-distance exchanges of heavier states and will be suppressed by a factor $g_{\mathrm{NP}}^2/\Lambda_{\mathrm{NP}}^2$ \cite{Buchalla:2013eza}. The same argument applies to operators with a higher number of fermion pairs. Therefore, one must assign an additional $\mO(p)$ suppression to fermion bilinears,\footnote{
Obviously, this additional chiral power does not apply to the kinetic term. In the Yukawas it has already been assigned through the spurion $\mY$.}
originating from some new-physics coupling, in the same way we did before for the Yukawas. Therefore,
\bel{eq:4-fermion-counting}
(\bar\eta\,\Gamma\,\zeta)^n \quad\sim\quad\mO\left( p^{2n}\right)\, .
\ee
This assignment assumes that the SM fermions couple weakly to the strong sector \cite{Buchalla:2013eza}.

The specific values assigned to the gauge sources in Eqs.~\eqn{eq:SMgauge} and \eqn{eq:Xmu-fixing} introduce an explicit breaking of custodial symmetry that is transferred to higher orders through quantum loops.
This is analogous to the explicit breaking of chiral symmetry through electromagnetic interactions in \chpt\ \cite{Ecker:1988te,chpt+photons,EckerIMNP:2000}. This breaking can be easily incorporated into the effective theory through the right-handed spurion
\bel{eq:T_R}
\mT_R\quad\longrightarrow\quad g_R^{\phantom{\dagger}}\, \mT_R\, g_R^\dagger\, ,
\ee
or its covariant counterpart
\bel{eq:T_covariant}
\mT\, =\, u\, \mT_R\, u^\dagger \quad\longrightarrow\quad g_h^{\phantom{\dagger}} \mT g_h^\dagger\, .
\ee
Building invariant operators with an even number of spurion fields and making the identification
\bel{eq:T_R-value}
   \mT_R    \, =\, -g'\;\frac{\sigma_3}{2}\, ,
\ee
one formally obtains the custodial symmetry-breaking structures induced through quantum loops with internal $B_\mu$ lines.
Since each $B_\mu$ field carries a coupling $g'$,
this spurion has chiral dimension 1,\footnote{
The pioneering papers discussing the Higgsless EWET \cite{Longhitano:1980iz,Longhitano:1980tm} adopted a naive power counting in terms of derivatives where $\mT_R\sim\mO(p^0)$. This implied the presence in $\mL_2$ of a custodial symmetry-breaking operator $\langle \mT u_\mu\rangle^2$ which is very suppressed phenomenologically. Our power-counting assignment in Eq.~\eqn{eq:T_R-counting} avoids this pitfall and leads to a phenomenologically consistent expansion, even in the presence of additional (small) sources of custodial symmetry breaking.}
%
\bel{eq:T_R-counting}
\mT_R\quad\sim\quad \mT\quad\sim\quad \cO(p)\, .
\ee

\subsection{NLO Lagrangian}

\begin{table}[t!]
\begin{center}
\renewcommand{\arraystretch}{1.8}
\begin{tabular}{|c||c|c|}
\hline
\multicolumn{1}{|c||}{$i$} &
\multicolumn{1}{|c|}{${\cal O}_i$} &
\multicolumn{1}{|c|}{$\widetilde{\cal O}_i$}  \\
\hline
\hline
$1$  &
$\Frac{1}{4}\bra {f}_+^{\mu\nu} {f}_{+\, \mu\nu}
- {f}_-^{\mu\nu} {f}_{-\, \mu\nu}\ket$
&  $\Frac{i}{2} \bra {f}_-^{\mu\nu} [u_\mu, u_\nu] \ket$
\\ [1ex]
\hline
$2$  &
$ \Frac{1}{2} \bra {f}_+^{\mu\nu} {f}_{+\, \mu\nu}
+ {f}_-^{\mu\nu} {f}_{-\, \mu\nu}\ket$
& $\bra {f}_+^{\mu\nu} {f}_{-\, \mu\nu} \ket $
\\ [1ex]
\hline
$3$  &
$\Frac{i}{2}  \bra {f}_+^{\mu\nu} [u_\mu, u_\nu] \ket$
&  $\Frac{(\partial_\mu h)}{v}\,\bra f_+^{\mu\nu}u_\nu \ket$
\\ [1ex]
\hline
$4$  &
$\bra u_\mu u_\nu\ket \, \bra u^\mu u^\nu\ket $
& ---
\\ [1ex]
\hline
$5$  &
$  \bra u_\mu u^\mu\ket^2$
& ---
\\ [1ex]
\hline
$6$ &
$\Frac{(\partial_\mu h)(\partial^\mu h)}{v^2}\,\bra u_\nu u^\nu \ket$
& ---
\\ [1ex]
\hline
$7$  &
$\Frac{(\partial_\mu h)(\partial_\nu h)}{v^2} \,\bra u^\mu u^\nu \ket$
& ---
\\ [1ex]
\hline
$8$ &
$\Frac{(\partial_\mu h)(\partial^\mu h)(\partial_\nu h)(\partial^\nu h)}{v^4}$
& ---
\\ [1ex]
\hline
$9$ &
$\Frac{(\partial_\mu h)}{v}\,\bra f_-^{\mu\nu}u_\nu \ket$
&  ---
\\ [1ex]
\hline
$10$ & $\langle \mT u_\mu\rangle^2$  & ---
\\ [1ex]
\hline
$11$ & $ \hat{X}_{\mu\nu} \hat{X}^{\mu\nu}$ & ---
\\ [1ex]
\hline
\multicolumn{3}{c}{}
\end{tabular}
\end{center}
\vspace*{-1.cm}
\caption{\small
$CP$-invariant bosonic operators of the $\cO(p^4)$ EWET Lagrangian.
$P$-even ($P$-odd) operators are shown in the left (right) column.}
\label{tab:bosonic-Op4}
\end{table}

At NLO, one must consider one-loop contributions \cite{Guo:2015isa,Alonso:2015fsp,Alonso:2016oah,Espriu1,Espriu2,Espriu3,Dobado,Dobado2,Herrero,Filipuzzi:2012bv,Gavela}
with the LO Lagrangian plus $\mO(p^4)$ local structures. The $\hat d=4$ Lagrangian
for the Goldstone fields was analyzed long time ago in the case of a Higgsless effective theory \cite{Longhitano:1980iz,Longhitano:1980tm}.
Including the additional operators with the singlet Higgs field,\footnote{
A much larger number of operators appears in previous EWET studies, assuming
a slightly different chiral counting~\cite{Alonso:2012px,Buchalla:2013rka,Buchalla:2012qq}.}
the most general $CP$-invariant NLO bosonic Lagrangian has the form \cite{Pich:2015kwa}

\bel{eq:L4}
\mL_4^{\mathrm{Bosonic}}\, =\, \sum_{i=1}^{11} \mF_i(h/v)\; \mO_i
\, +\, \sum_{i=1}^{3}\widetilde\mF_i(h/v)\; \widetilde \mO_i \, .
\ee
The coefficients $\mF_i(h/v)$ and $\widetilde\mF_i(h/v)$ must be understood as polynomials
of $h/v$, {\it i.e.},
\bel{eq:mFi}
\mF_i \, =\, \sum_{n=0} \mF_{i,n} \left(\frac{h}{v}\right)^n\, ,
\qquad\qquad\qquad\quad
\widetilde\mF_i\, =\,\sum_{n=0} \widetilde\mF_{i,n} \left(\frac{h}{v}\right)^n\, .
\ee
We have distinguished two types of $CP$-invariant operators, according to their even ($\mO_i$)
or odd ($\widetilde \mO_i$) transformation property under parity. Our operator basis
is given in Table~\ref{tab:bosonic-Op4}~\cite{Pich:2015kwa}.\footnote{$\mO_{1,4,5}$ have the same structure as the corresponding
Longhitano operators $O^L_{1,4,5}$ \cite{Longhitano:1980iz,Longhitano:1980tm,Morales:94},
while $O^L_{2,3}$ correspond to $\pm \mO_3 - \widetilde\mO_1$.
The custodial-breaking structure $\mO_{10}\sim \mO_0^{L}$ was
considered to be of $\mO(p^2)$ in the Longhitano basis;
this basis included additional operators with $\mT$ spurions and more derivatives which, in our counting, are higher-order terms.
The operators $\mO_{6,7,8}$ with explicit derivatives of the Higgs field correspond to $O_{D7,D8,D11}$
in Ref.~\cite{Buchalla:2013rka}, while $\mO_{2}$ and $\widetilde\mO_2$ are equal
to $O_{Xh2}\pm O_{Xh1}/2$.}

Once the auxiliary fields are forced to take the values in Eqs.~\eqn{eq:SMgauge} and \eqn{eq:Xmu-fixing}, the Higgsless term $\mF_2[0]\,\mO_2
 + \mF_{11}[0] \, \mO_{11} +\widetilde{\mF}_2[0]\, \widetilde{\mO}_2$
is a linear combination of the $W_\mu$ and $B_\mu$
Yang-Mills Lagrangians. Its effects could then be accounted for through a modification of the corresponding gauge couplings.

The fermionic part of $\mL_4$ involves operators with one or two fermion bilinears:
\bel{eq:L4-fermionic}
\mL_4^{\mathrm{Fermionic}}\, =\, \sum_{i=1}^{  7  }
\mF_i^{\psi^2}(h/v)\; \mO_i^{\psi^2}
\, +\, \sum_{i=1}^{  3   }
\widetilde\mF_i^{\psi^2}(h/v)\; \widetilde \mO_i^{\psi^2}
\, +\, \sum_{i=1}^{10}\mF_i^{\psi^4}(h/v)\; \mO_i^{\psi^4}
\, +\, \sum_{i=1}^{2}\widetilde\mF_i^{\psi^4}(h/v)\; \widetilde \mO_i^{\psi^4}
\, .
\ee
The relevant $CP$-conserving operator structures for a single fermion doublet $\psi$, {\it i.e.}, neglecting any kind of flavour structure, are shown in Table~\ref{tab:fermion-ops}.\footnote{Using Fierz identities and $SU(2)$ relations, one could eliminate six four-fermion operators in Table~\ref{tab:fermion-ops}. However, we prefer to keep the full basis
with twelve operators because it is no-longer redundant when colour and/or flavour are included.}

\begin{table}[!t] 
\begin{center}
\renewcommand{\arraystretch}{1.8}
\begin{tabular}{|c||c|c||c|c|}
\hline
$i$ & ${\cal O}^{\psi^2}_i$ & $\widetilde{\cal O}^{\psi^2}_i$
& ${\cal O}^{\psi^4}_i$ & $\widetilde{\cal O}^{\psi^4}_i$
\\ \hline\hline
1  & $\bra J_S \ket \bra u_\mu u^\mu \ket$  &
$\bra J_T^{\mu \nu} f_{- \,\mu\nu} \ket $
& $\bra J_{S} J_{S} \ket $ &  $\bra J_V^\mu J_{A,\mu}^{\phantom{\mu}}\ket $
\\  [1ex] \hline
2 &  $ i \,  \bra J_T^{\mu\nu} \left[ u_\mu, u_\nu \right] \ket$
& $\displaystyle\frac{\partial_\mu h}{v} \, \bra u_\nu J^{\mu\nu}_T \ket $
& $\bra J_{P} J_{P} \ket $ &  $\bra J_V^\mu\ket \bra J_{A,\mu}^{\phantom{\mu}}\ket $
\\  [1ex] \hline
3 & $\bra J_T^{\mu \nu} f_{+ \,\mu\nu} \ket $ & $\bra J_V^\mu \ket \bra u_\mu \mathcal{T} \ket $ & $\bra J_{S} \ket \bra  J_{S} \ket $ & ---
\\  [1ex] \hline
4 & $\hat{X}_{\mu\nu} \bra J_T^{\mu \nu} \ket $ & ---
& $\bra J_{P} \ket \bra  J_{P} \ket $ & ---
\\ [1ex] \hline
5 & $\displaystyle\frac{\partial_\mu h}{v} \, \bra u^\mu J_P \ket $ & ---
&  $\bra J_V^\mu J_{V,\mu}^{\phantom{\mu}}\ket $ & ---
\\ [1ex] \hline
6 & $\bra J_A^\mu \ket \bra u_\mu \mathcal{T} \ket $ & --- &
 $\bra J_A^\mu J_{A,\mu}^{\phantom{\mu}}\ket $  & ---
\\ [1ex] \hline
7 &  $\Frac{(\partial_\mu h) (\partial^\mu h)}{v^2} \bra J_S\ket $
& --- &
 $\bra J_V^\mu\ket \bra J_{V,\mu}^{\phantom{\mu}}\ket $  & ---
\\ [1ex] \hline
8 & --- & --- &
 $\bra J_A^\mu\ket \bra J_{A,\mu}^{\phantom{\mu}}\ket $  & ---
\\ [1ex] \hline
9 & --- & --- &
$\bra J^{\mu\nu}_{T} J_{T\,\mu\nu}^{\phantom{\mu}} \ket $ & ---
\\ [1ex] \hline
10 & --- & --- &
$\bra J^{\mu\nu}_{T} \ket \bra J_{T\,\mu\nu}^{\phantom{\mu}} \ket $ & ---
\\ [1ex] \hline
\end{tabular}
\caption{\small
$CP$-conserving fermion operators with $\hat{d}=4$. $\mO_i^{\psi^2,\psi^4}$
($\widetilde\mO_i^{\psi^2,\psi^4}$) denote $P$-even (odd) structures.}
\label{tab:fermion-ops}
\end{center}
\end{table}

The NLO fermionic Lagrangian could also include the operators
$\bra J_S \mathcal{T}\ket$, $\bra u_\mu J_V^\mu\ket$ and $\bra u_\mu J_A^\mu\ket$,
which are of $\cO(p^3)$ and, therefore, have a smaller chiral suppression than the ones
in Eq.~\eqn{eq:L4-fermionic}.
Operators of this chiral order have in fact been previously considered
in the literature~\cite{ABCH:85,BEM:99}. As demonstrated in App.~\ref{app:simplifications},
with a single fermion doublet these operators can be removed from the effective Lagrangian through appropriate redefinitions of the auxiliary fields $\mT$, $\hat W_\mu$ and $\hat B_\mu$.
With several fermion families, the scalar-current operator could still be removed, with all its flavour dependence being reabsorbed into the spurion $\mY$.
However, a non-trivial flavour structure in the vector and axial-vector $\cO(p^3)$ operators could not be reabsorbed into the gauge sources, and would introduce interesting dynamical implications that we plan to study in future works.

\section{Effective Lagrangian with heavy states}
\label{sec:Heavy_Fields}

Our main goal is to estimate the contributions to the LECs of the NLO EWET coming from tree-level exchanges of heavy fields, not included in the low-energy effective theory.
With this purpose, we build a more general EFT incorporating, in addition to the SM particles, heavier bosonic states (the lightest new-physics resonances). While the low-energy EWET is only valid for energies smaller than the resonance masses, the high-energy resonance theory extends its validness to higher scales below the next heavier states not yet incorporated in its Lagrangian.

We will consider generic massive states, transforming under $\mG$ as $SU(2)_{L+R}$ triplets ($R= \sigma^a R^a/\sqrt{2}$) or singlets ($R_{1}$):
\bear
R \quad\longrightarrow\quad g_h^{\phantom{\dagger}}\, R\, g_h^\dagger\, ,
\qquad\qquad\qquad
R_{1}\quad\longrightarrow\quad R_{1} \, .
\label{eq.R-transform}
\eear
We will assume that the underlying strongly-coupled theory preserves charge conjugation ($C$) and parity ($P$), so that we can work with massive
eigenstates with definite $C$ and $P$ properties. For simplicity, we will restrict our present analysis to colour-singlet massive states with bosonic $J^{PC}$ quantum numbers $0^{++}$ (S), $0^{-+}$ (P), $1^{--}$ (V) and $1^{++}$ (A).
Their transformation properties~\cite{Ecker:1988te,Ecker:1989yg} under $P$, $C$ and Hermitian conjugation can be found in App.~\ref{app:discrete-transf}.
The masses of these heavy states are expected to be of the order of (or above) the electroweak chiral scale  $\Lambda_{\mathrm{EWET}} = 4\pi v\approx 3$~TeV.

We will first construct an invariant chiral Lagrangian coupling these heavy states to the SM fields, at the lowest possible order in the chiral expansion. Since LHC searches have essentially excluded the presence of new particles below 1~TeV, we will later integrate out the heavy states $R$ and $R_1$, and extract their corresponding contributions to the low-energy EWET. For our purposes, we only need to consider in the effective Lagrangian operators with a single massive state, because terms with a higher number of heavy fields do not contribute at $\mO(p^4)$.

The high-energy action is given by a Lagrangian with the structure
\bear
\mL &=& \mL_{\rm Heavy\, Fields}[R,\phi,\psi]\,\,\,+\,\,\, \mL_{\text{non-R}}[\phi,\psi]\, ,
\eear
where the first piece on the right-hand side (rhs) contains resonance fields and light degrees of freedom $\phi$ and $\psi$ whereas the second one only depends on the light fields. The term $\mL_{\text{non-R}}[\phi,\psi]$ is formally identical to the EWET Lagrangian but with different couplings, because it describes the interactions of a different EFT valid at the resonance mass scale. Resonance exchanges among
$\mL_{\rm Heavy\, Fields}[R,\phi,\psi]$ vertices will generate additional contributions to the LECs of the EWET which we want to identify.

The chiral-invariant Lagrangian for the heavy fields takes the generic form
\bear
\mL_{\mathrm{Heavy\, Fields}} &=&
\sum_{R=S,S_1,P,P_1} \mL_R\,\,\, +\,\,\, \sum_{R=V,V_1,A,A_1} \mL_R \,  ,
\eear
where the corresponding kinetic and mass terms are included in $\mL_R$.

\subsection{Spin-0 resonance Lagrangian ($S,S_1,P,P_1$)}

The relevant spin-0 resonance interactions take the form
\bear
\mL_R &\! =&\! \Frac{1}{2}\bra \nabla^\mu R\,  \nabla_\mu R \, -\, M_R^2\, R^2\ket \; +\; \bra R\, \chi_R^{\phantom{\mu}}\ket
\hskip 3.5cm (R=S,\, P)\, ,
\nn\\
\mL_{R_1} &\! =&\!  \Frac{1}{2}  \left(  \partial^\mu R_1\,  \partial_\mu R_1 \, -\, M_{R_1}^2\, R_1^2 \right)  \; +\;
R_1\, \chi_{R_1}^{\phantom{\mu}}
\hskip 3.5cm (R_1=S_1,\, P_1)\, .
\label{eq.resonance-LS}
\eear
In addition to the quadratic kinetic and mass pieces, there are terms linear in the heavy resonances with chiral structures containing light fields. At LO they
are given by
\bear
\chi_S^{\phantom{\mu}} &\! =&\!
  c_1^S\;  J_S \, ,
\nn\\
\chi_P^{\phantom{\mu}} &\! =&\!
    c_1^P\; J_P \; +\;
    d_P\; \Frac{(\partial_\mu h)}{v}\, u^\mu \, ,
\label{eq:triplet-S}\;
\eear
for the triplets $S$ and $P$, while the singlet operators are provided by
\bear
\chi_{S_1}^{\phantom{\mu}} &\! = &\!
\lambda_{hS_1} \, v \; h^2  \; +\;
\Frac{c_{d}}{\sqrt{2}}\; \bra u_\mu u^\mu \ket \; +\;
\Frac{c_1^{S_1}}{\sqrt{2}}\; \bra J_S \ket  \, ,
\nn\\
\chi_{P_1}^{\phantom{\mu}} &\! = &\!
\Frac{c_1^{P_1}}{\sqrt{2}}\;  \bra  J_P\ket
\, .
\label{eq:singlet-S}
\eear
All these structures are of $\mO(p^2)$ in the chiral power counting, except the $\lambda_{hS_1}$ term which naively appears to be of $\mO(p^0)$. We will see later that the coupling $\lambda_{hS_1}$ must be assigned a chiral dimension two, so that all terms in \eqn{eq:singlet-S} are of the same order in the momentum expansion.

Here and in what follows we reduce the number of chiral structures
through the use of field redefinitions, partial integration, equations of motion (EoM) and
algebraic Cayley-Hamilton relations. More details are given in App.~\ref{app:simplifications}.
Since one may introduce an arbitrary number of light Higgs fields without increasing the chiral dimension, all couplings must be understood as functions of $h/v$, so for instance
\be
c_i^{R} \, =\,  \sum_{n=0}\, c_i^{R\, (n)}\; \left(\Frac{h}{v}\right)^n\, .
\ee

\subsection{Proca Lagrangian for spin-1 resonances ($V,V_1,A,A_1$)}

There is some freedom in choosing an explicit representation for the spin-1 fields. Although physics is independent of the adopted formalism (Proca, antisymmetric tensor or gauge-like field), a clever choice can provide a simpler interaction Lagrangian and be more convenient for phenomenological studies~\cite{Ecker:1988te,Ecker:1989yg}. For simplicity, we start using here the more common Proca representation and will later analyze the equivalence of the three formalisms and the interesting subtleties arising with the different options.

Let us then describe the triplet and singlet spin-1 heavy particles through the Proca fields $\hat{R}^\mu$ and $\hat{R}_1^\mu$, transforming under $\mG$ as in Eq.~(\ref{eq.R-transform}), with $R=V, A$ for the vector and axial-vector states.
Including only interactions linear in the four-vector fields $\hat{R}^\mu$, the relevant chiral Lagrangians take the form
\bear
\mL_{\hat R}^{(P)} &\! =&\! - \Frac{1}{4}\,\bra \hat R_{\mu \nu}\, \hat R^{\mu \nu} \,
- \, 2\,M_{R}^2\, \hat R_\mu \hat R^\mu \ket
\; +\; \bra \hat R_{\mu}\, \hat \chi^{\mu}_{\hat R} \,
+ \hat R_{\mu \nu}\, \hat \chi_{\hat R}^{\mu \nu} \ket
\qquad \qquad\qquad (\hat R=\hat V,\, \hat A)\, ,
\nn\\
\mL_{\hat R_1}^{(P)} &\! =&\! - \Frac{1}{4} \left( \hat R_{1\, \mu \nu}\, \hat R_1^{\mu \nu} \,
- \, 2\,M_{R_1}^2\, \hat R_{1\,\mu} \hat R_1^\mu \right)
\; +\;  \hat R_{1\,\mu}\, \hat \chi^{\mu}_{\hat R_1} \,
+ \hat R_{1\,\mu \nu}\, \hat \chi_{\hat R_1}^{\mu \nu}
\qquad\qquad (\hat R_1 = \hat V_1,\, \hat A_1)\, ,
\nn\\
\label{eq.resonanceProca-L}
\eear
where
\be
\hat R_{\mu \nu} \, =\, \nabla_\mu \hat R_\nu - \nabla_\nu \hat R_\mu\, ,
\qquad\qquad\qquad\quad
\hat R_{1\,\mu \nu} \, =\, \partial_\mu \hat R_{1\,\nu} - \partial_\nu  \hat R_{1\,\mu} \, .
\label{eq:Rhat_mn}
\ee
The tensors $\hat \chi^{\mu}_{\hat R}$, $\hat \chi_{\hat R}^{\mu \nu}$
($\hat \chi^{\mu}_{\hat R_1}$, $\hat \chi_{\hat R_1}^{\mu \nu}$) denote the most general triplet (singlet) chiral structures constructed with the SM fields, with the appropriate quantum numbers $R=V,A$ ($R_1=V_1,A_1$). Assuming invariance under the $CP$ symmetry, their LO expressions
involve the $\mO(p^2)$ terms:
\bear
\hat \chi_{\hat V}^{\mu \nu}  &\! =&\!
\Frac{f_{\hat V}}{2 \sqrt 2} \, f_{+}^{\mu\nu}
\; + \; \Frac{i\, g_{\hat V}}{2 \sqrt 2} \, [u^\mu,u^\nu]
\; + \; \Frac{\widetilde f_{\hat V}}{2\sqrt 2} \, f_{-}^{\mu\nu}
\; +\;
\Frac{  \widetilde{\lambda}_1^{h\hat{V}} }{\sqrt{2}}\;\left[
(\partial^\mu h)\, u^\nu-(\partial^\nu h)\, u^\mu \right]
\;+\;
c_{0}^{\hat{V}}
J_T^{\mu\nu}
\, ,
\nn\\[10pt]
\hat \chi_{\hat A}^{\mu \nu}  &\! =&\!
 \Frac{f_{\hat A}}{2 \sqrt 2} \, f_{-}^{\mu\nu}
\; +\;
\Frac{ \lambda_1^{h\hat{A}} }{\sqrt{2}}\;\left[
(\partial^\mu h)\, u^\nu-(\partial^\nu h)\, u^\mu \right]
\; + \; \Frac{\widetilde f_{\hat A}}{2\sqrt 2} \, f_{+}^{\mu\nu}
\; + \; \Frac{i\, \widetilde g_{\hat A}}{2 \sqrt 2} \, [u^\mu,u^\nu]
\;+\;
\widetilde{c}_0^{\hat{A}} J_T^{\mu\nu}
\, ,
\nn\\[10pt]
\hat \chi_{\hat V_1}^{\mu\nu}  &\! =&\!
 f_{\hat{V}_1} X^{\mu\nu}
\;+\;
 \Frac{c_0^{{\hat{V}}_1} }{\sqrt{2}} \bra J_T^{\mu\nu}\ket
\, ,
\qquad \qquad \qquad \qquad
\hat \chi_{\hat A_1}^{\mu\nu}  =
  \widetilde{f}_{\hat{A}_1} X^{\mu\nu}
\;+\;
  \Frac{\widetilde c_0^{{\hat{A}}_1}
}{\sqrt{2}} \bra J_T^{\mu\nu}\ket\, ,
\label{eq.Proca-chimunu}
\eear
and
\bear
 \hat \chi_{\hat V}^\mu  &\! =&\!
c_1^{\hat{V}}\, J^\mu_V \; +\; \widetilde c_1^{\hat{V}}\, J^\mu_A\, ,
\qquad\qquad\qquad\qquad\qquad \;\;\;
\hat \chi_{\hat A}^\mu =
c_1^{\hat{A}}\, J^\mu_A \; +\; \widetilde c_1^{\hat{A}}\, J^\mu_V\, ,
\nn\\[10pt]
 \hat \chi_{\hat V_1}^\mu  &\! =&\!
 \widetilde{c}_{\mathcal{T}}^{\hat{V}_1} \bra u^\mu \mathcal{T} \ket \;+\;
 \Frac{c_1^{{\hat{V}}_1}}{\sqrt{2}}
\, \bra J^\mu_V\ket \; +\;
 \Frac{   \widetilde c_1^{{\hat{V}}_1}  }{\sqrt{2}}
\, \bra J^\mu_A\ket\, ,
\nn\\[10pt]
 \hat \chi_{\hat A_1}^\mu  &\!=&\!
 c_{\mathcal{T}}^{\hat{A}_1} \bra u^\mu \mathcal{T} \ket \;+\;
 \Frac{  c_1^{{\hat{A}}_1}  }{\sqrt{2}}
\, \bra J^\mu_A\ket \; +\;
 \Frac{  \widetilde c_1^{{\hat{A}}_1}  }{\sqrt{2}}
\, \bra J^\mu_V\ket \, .
\label{eq.Proca-chi}
\eear
In principle one could also write down the $\cO (p^1)$ operators $\bra \hat V^\mu u_\mu \ket$
and $\bra \hat A^\mu u_\mu \ket$ ($P$-odd and $P$-even, respectively),
but they can be removed from the action by means of the field redefinitions
described in App.~\ref{app:simplifications}. The structure of the $P$-even part of the Lagrangian agrees with that found in resonance
models of QCD in the Proca formalism~\cite{Ecker:1989yg,Op3-res-Lagrangian}.

\section{Integrating out the heavy states}
\label{sec.R-integration}

At energies much smaller than the resonance masses, the presence of the heavy states can be only inferred from their contributions to the LECs
 of the EWET Lagrangian.
These effects can be formally computed integrating out the heavy fields from the generating functional and expanding the resulting non-local action in powers of momenta over the heavy scales. For sake of clarity we are going to separate the analysis of spin-0 and spin-1 resonance contributions.
Furthermore, in what follows we will implicitly assume that the relevant chiral structures $\chi_{R}$ do not contain couplings growing with the resonance mass.
This is the decoupling behaviour expected in strongly-coupled scenarios. Therefore, our generic expressions for the LECs do not apply to renormalizable Higgsed models which require a more specific treatment.\footnote{An enlightening discussion within a simple model with one doublet and one singlet scalar multiplets has been given in ref.~\cite{Buchalla:2016bse}.}

\subsection{Spin-0 resonance contributions to the EWET}

The LO contributions to the LECs correspond to tree-level exchanges of heavy fields. They can be easily obtained through the EoM of the massive resonances, which in the spin-0 case take the form:
\bear
(\nabla^2  + M_R^2)\, R  &\! =&\!
\chi_R - \frac{1}{2} \,\bra \chi_R^{\phantom{\mu}}\ket
\hskip 2.5cm (R=S,\, P)\, ,
\nn\\
(\partial^2  + M_{R_1}^2)\, R_1
&\! =&\! \,   \chi_{R_1}^{\phantom{\mu}}
\hskip 4.0cm (R_1=S_1,\, P_1)\,  .\quad
\label{eq.resonance-EoM1S}
\eear
We have employed the generic Lagrangians in Eq.~\eqref{eq.resonance-LS}
which only take into account interactions with a single heavy state.
Moreover, we will only consider contributions to the tensors
$\chi_R^{\phantom{\mu}}$ and $\chi_{R_1}^{\phantom{\mu}}$ which are at most of $\cO(p^2)$.
The trace term ensures that the rhs of the first equation is traceless, as it happens with the left-hand side (lhs).

In the low-energy limit, the solutions for the heavy field EoM can be expanded in terms of local operators which only contain light
fields~\cite{Ecker:1988te}:
\bear
R &\! =&\!   \Frac{1}{M_R^2 }   \,  \left( \chi_R^{\phantom{\mu}}-  \frac{1}{2}  \bra \chi_R^{\phantom{\mu}}\ket \right)
\; +\; \mO\left(\frac{p^4}{M_R^4}\right) \hskip 2.84cm (R=S,\, P)\, ,
\nn\\
R_1 &\! =&\!  \Frac{1}{M_{R_1}^2 }   \,    \chi_{R_1^{\phantom{\mu}}}
\; +\; \mO\left(\frac{p^4}{M_R^4}\right) \hskip 4.7cm (R_1=S_1,\, P_1)\,   .
\label{eq.resonance-EoM2S}
\eear
Substituting these solutions back into the resonance Lagrangian
$\mL_R$ and $\mL_{R_1}$    in Eq.~\eqref{eq.resonance-LS},
one obtains the corresponding contributions to the  low-energy effective Lagrangian of the EWET:
\bear
\Delta \mL_{R}^{\cO(p^4)} &\! =&\! \Frac{1}{2M_R^2} \,
 \left(  \bra  \chi_R^{\phantom{\mu}}\, \chi_R^{\phantom{\mu}}\ket - \frac{1}{2}
 \bra \chi_R^{\phantom{\mu}}\ket^2\right)
\hskip 2.4cm (R=S,\, P)\, ,
\nn\\
\Delta \mL_{R_1}^{\cO(p^4)} &\! =&\! \Frac{1}{2M_{R_1}^2} \,
 ( \chi_{R_1}^{\phantom{\mu}})^2
\hskip 5.1cm (R_1=S_1,\, P_1)\,   .
\label{integration0}
\eear
These results must be finally simplified and written in our basis of $\cO(p^4)$ operators.

The singlet scalar $S_1$ couples directly to the Higgs field through the term $\chi_{S_1}^{h} S_1 = \lambda_{hS_1} v \, h^2 S_1$, which does not contain any explicit chiral suppression. The tree-level exchange of the massive $S_1$ state generates then the following correction to the $\mO(p^2)$ EWET Lagrangian,
\bel{eq:S1-p2}
\Delta\mL_{S_1}^{\mO(p^2)}\, =\, \frac{1}{2 M_{S_1}^2}\,\left\{
(\lambda_{hS_1})^2 v^2 h^4\, +\,
\sqrt{2}\, \lambda_{hS_1} v\, h^2 \left[c_d\, \bra u_\mu u^\mu \ket
+ c_1^{S_1}\,\bra J_S \ket \right]\right\}\, ,
\ee
which is suppressed by two powers of the heavy mass scale $M_{S_1}$.
A consistent power counting requires to assign a chiral dimension 2 to the function $\lambda_{hS_1}(h/v)$, so that the three terms in Eq.~\eqn{eq:S1-p2} have the same chiral order $\mO(p^4)$, as all other resonance-exchange contributions in Eq.~\eqn{integration0}. Eq.~\eqn{eq:S1-p2} should then be considered as an $\mO(p^4)$ correction to the lowest-order operators in $\mL_2$.

The $(\lambda_{hS_1})^2$ term represents a correction to the Higgs potential $V(h/v)$ in Eq.~\eqn{eq:L2}, while the term proportional to $\lambda_{hS_1} c_d$ contributes to $\mF^{(u)}(h/v)$. In terms of the corresponding series-expansion coefficients in powers of $h/v$, one gets
\bel{eq:p2corr-h}
\Delta c^{(V)}_{n\ge 4}\, =\, -\frac{v^2}{2 M_{S_1}^2}\;\sum_{k=0}^{n-4}
\lambda_{hS_1}^{(k)} \,\lambda_{hS_1}^{(n-k-4)}\, ,
\qquad\qquad
\Delta c^{(u)}_{n\ge 2}\, =\, \frac{2\sqrt{2} v}{M_{S_1}^2}\;
\sum_{k=0}^{n-2} \lambda_{hS_1}^{(k)}\, c_d^{(n-k-2)}\, .
\ee

The third term proportional to $\lambda_{hS_1} c_1^{S_1}$ contributes to the LO fermionic Lagrangian, {\it i.e.}, to the Yukawa coupling in Eq.~\eqn{eq:Yukawas}. However, it only starts to contribute at $\mO(h^2)$:
\bel{eq:DYukawa-p2}
\Delta\mY\, =\, -\frac{1}{\sqrt{2} M_{S_1}^2}\; h^2\, \lambda_{hS_1}(h/v)\; c_1^{S_1}(h/v)\, .\ee
%

\begin{table}[!t]  
\begin{center}
\renewcommand{\arraystretch}{2.1}
\begin{tabular}{|c||c||c||c|} 
\hline
$i$ & $\Delta\mF_i$ &  $\Delta\mF^{\psi^2 }_i$ &
$\Delta\mF^{\psi^4 }_i$
\\  \hline\hline
1
& 0
& $\displaystyle\frac{c_d c_1^{S_1}}{2M_{S_1}^2}$
& $\displaystyle\frac{(c_1^S)^2}{2M_S^2}$
\\[1ex] \hline
2
& 0
& 0
& $\displaystyle\frac{(c_1^P)^2}{2M_P^2}$
\\[1ex] \hline
3
& 0
& 0
& $-\displaystyle\frac{(c_1^S)^2}{4M_S^2}+\displaystyle\frac{(c_1^{S_1})^2}{4M_{S_1}^2}$
\\[1ex] \hline
4
& 0
& 0
& $-\displaystyle\frac{(c_1^P)^2}{4M_P^2}+\displaystyle\frac{(c_1^{P_1})^2}{4M_{P_1}^2}$
\\[1ex] \hline
5
&
$\displaystyle\frac{c_{d}^2}{4M_{S_1}^2}$
& $\displaystyle\frac{d_P c_1^{P}}{M_P^2}$
& 0
\\[1ex] \hline
7
&
$\displaystyle      \Frac{ d_P^2}{2 M_P^2} $
& 0 & 0
\\[1ex] \hline
\end{tabular}
\caption{{\small
$\Delta \mL_R^{\cO(p^4)}$ contributions to the
$\cO(p^4)$ LECs from heavy $S$, $S_1$, $P$, $P_1$
exchanges. The remaining $\cO(p^4)$ LECs, not listed here, do not receive contributions from these spin-0 resonances.}}
\label{tab:spin-0-Op4-LEC}
\end{center}
\end{table}

The contributions to the $\cO(p^4)$ operators in the EFT coming from spin-0 resonance exchanges
are given in Table~\ref{tab:spin-0-Op4-LEC}.
The LECs not listed in the table are not sensitive to the exchange of scalar or pseudoscalar heavy bosons, which only generates $P$-even structures.
The bosonic LECs in the first column were already presented in Ref.~\cite{Pich:2015kwa}.
The triplet scalar field only contributes to the four-fermion operators $\mO_1^{\psi^4}$ and $\mO_3^{\psi^4}$, while $S_1$-exchange generates $\mO_5$, $\mO_1^{\psi^2}$ and $\mO_3^{\psi^4}$. The operators $\mO_7$, $\mO_5^{\psi^2}$, $\mO_2^{\psi^4}$ and $\mO_4^{\psi^4}$ receive pseudoscalar-triplet contributions, and the only manifestation of the singlet pseudoscalar appears in $\mO_4^{\psi^4}$.

\subsection{Spin-1 resonance contributions to the EWET in the Proca representation $\mathbf{(P)}$}

The classical EoM for the Proca resonance fields are
\bear
\nabla_\mu \hat R^{\mu \nu} + M^2_{R}\, \hat R^\nu & = &
- \left(
\hat\chi^\nu_{\hat R} - 2\, \nabla_\mu\hat \chi_{\hat R}^{\mu \nu}
-  \frac{1}{2}
\,\bra \hat\chi^\nu_{\hat R} - 2\, \nabla_\mu\hat \chi_{\hat R}^{\mu \nu}\ket \right)
\qquad  (\hat R=\hat V,\, \hat A)\, ,
\nn\\
\partial_\mu \hat R_1^{\mu \nu} + M^2_{R_1}\, \hat R_1^\nu & = &
- \left(
\hat\chi^\nu_{\hat R_1} - 2\, \partial_\mu\hat \chi_{\hat R_1}^{\mu \nu}\right)
\qquad\qquad\qquad\qquad\qquad\quad (\hat R_1 = \hat V_1,\, \hat A_1)\, .
\eear
For $p\ll M_R$, the solutions of the EoM for the heavy fields are given at LO by
\bear
\hat R^{\nu} =
- \frac{1}{M^2_{R}}\;\left(
\hat\chi^\nu_{\hat R} - \frac{1}{2}\,\bra \hat\chi^\nu_{\hat R}\ket\right)
\, , \qquad\qquad\qquad\qquad
\hat{R}_1^{\nu}       = - \frac{1}{M^2_{R_1}}\; \hat\chi^\nu_{\hat R_1}\, .
\eear
Substituting them back into the Lagrangians $\mL^{(P)}_{\hat R}$ and $\mL^{(P)}_{\hat R_1}$
in Eq.~(\ref{eq.resonanceProca-L}), one obtains the contributions to the EWET coming from one-resonance spin-1 exchanges at low energies,
\bear\label{eq:Proca-EL}
\Delta\mL^{\mO(p^4)}_{\hat R} & = & - \frac{1}{2 M^2_{R}}\;\left\{
\bra  \hat\chi^\mu_{\hat R}  \,\hat\chi_{\hat R\,\mu}^{\phantom{\mu}}  \ket
-  \frac{1}{2}\,\bra \hat\chi^\mu_{\hat R} \ket
\bra \hat\chi_{\hat R\,\mu}^{\phantom{\mu}} \ket\right\}\, ,
\nn\\
\Delta\mL^{\mO(p^4)}_{\hat R_1} & = & - \frac{1}{2 M^2_{R_1}}\;
\hat\chi^\mu_{\hat R_1}  \hat\chi_{\hat R_1\,\mu}^{\phantom{\mu}}  \, .
\eear
Expanding these results on our basis of EWET operators, one obtains the resonance-exchange predictions for their LECs shown in Tables~\ref{tab:spin-1-Op4-LEC-newterms} and \ref{tab:Op4-fromR-Proca}. The LECs not listed in the tables do not receive any contribution from the exchange of heavy spin-1 Proca fields.

\begin{table}[!t]  
\begin{center}
\renewcommand{\arraystretch}{2.4}
\begin{tabular}{|c||c||c|c|} 
\hline
$i$ & $\Delta\mF_i^{( P ) }$ &  $\Delta\mF^{\psi^2 \, (P)}_i$ &
$\Delta\widetilde{\mF}^{\psi^2\, (P) }_i$
\\  \hline\hline
3
& 0
& 0
& $-\displaystyle\frac{\widetilde{c}_{\mathcal{T}}^{\hat{V}_1} c_{1}^{\hat{V}_1}    }{\sqrt{2} M_{V_1}^2} - \displaystyle\frac{c_{\mathcal{T}}^{\hat{A}_1} \widetilde{c}_{1}^{\hat{A}_1}    }{\sqrt{2} M_{A_1}^2}$
\\[1ex] \hline
6
& 0
& $-\displaystyle\frac{\widetilde{c}_{\mathcal{T}}^{\hat{V}_1} \widetilde{c}_{1}^{\hat{V}_1}    }{\sqrt{2} M_{V_1}^2} - \displaystyle\frac{c_{\mathcal{T}}^{\hat{A}_1} c_{1}^{\hat{A}_1}    }{\sqrt{2} M_{A_1}^2}$
& ---
\\[1ex] \hline
10
& $-\displaystyle\frac{(\widetilde{c}_{\mathcal{T}}^{\hat{V}_1})^2}{2M_{V_1}^2}-\displaystyle\frac{(c_{\mathcal{T}}^{\hat{A}_1})^2}{2M_{A_1}^2}$
& ---
& ---
\\[1ex] \hline
\end{tabular}
\caption{{\small
$\Delta \mL_R^{\cO(p^4)}$ contributions to the purely bosonic and two-fermion
$\cO(p^4)$ LECs from heavy $V$, $V_1$, $A$, $A_1$
exchanges in the Proca formalism. The remaining $\cO(p^4)$ LECs, not listed here, do not receive contributions from these spin-1 resonances.}}
\label{tab:spin-1-Op4-LEC-newterms}
\end{center}
\end{table}

\begin{table}[!t]  
\begin{center}
\renewcommand{\arraystretch}{2.4}
\begin{tabular}{|c||c|c|}
\hline
$i$
& $\Delta\mF_i^{\psi^4\, (P)}$
&  $\Delta\widetilde{\mF}_i^{\psi^4\, (P)}$
\\ \hline\hline
1
& 0
&
$ -
\displaystyle\frac{c_1^{\hat{V}} \widetilde{c}_1^{\hat{V}} }{M_V^2}
-\frac{ c_1^{\hat{A}}\widetilde{c}_1^{\hat{A}}}{M_A^2}\;$
\\  \hline
2
& 0
&
$ \displaystyle\frac{c_1^{\hat{V}}\widetilde{c}_1^{\hat{V}}}{2M_V^2}
+\frac{c_1^{\hat{A}}\widetilde{c}_1^{\hat{A}}}{2M_A^2}
   -\frac{c_1^{\hat{V}_1}\widetilde{c}_1^{\hat{V}_1}}{2M_{V_1}^2} -\frac{c_1^{\hat{A}_1}\widetilde{c}_1^{\hat{A}_1}}{2M_{A_1}^2}
$
\\  \hline
5
& $ -\displaystyle\frac{(c_1^{\hat{V}})^2}{2M_V^2}
-\frac{(\widetilde{c}_1^{\hat{A}})^2}{2M_A^2}$
& ---
\\ \hline
6
& $- \displaystyle\frac{(\widetilde{c}_1^{\hat{V}})^2}{2M_V^2}
-\frac{({c}_1^{\hat{A}})^2}{2M_A^2} $
& ---
\\  \hline
7
& $\displaystyle\frac{({c}_1^{\hat{V}})^2}{4M_V^2}
+\frac{(\widetilde{c}_1^{\hat{A}})^2}{4M_A^2}
   -\frac{({c}_1^{\hat{V}_1})^2}{4M_{V_1}^2}  -\frac{(\widetilde{c}_1^{\hat{A}_1})^2}{4M_{A_1}^2}   $
& ---
\\  \hline
8
& $ \displaystyle\frac{(\widetilde{c}_1^{\hat{V}})^2}{4M_V^2}
+\frac{({c}_1^{\hat{A}})^2}{4M_A^2}
   -\frac{(\widetilde{c}_1^{\hat{V}_1})^2}{4M_{V_1}^2}  -\frac{({c}_1^{\hat{A}_1})^2}{4M_{A_1}^2} $
& ---
\\  \hline
\end{tabular}
\caption{\small $\Delta \mL_R^{\cO(p^4)}$
contributions to the four-fermion $\cO(p^4)$ LECs from $V$, $V_1$, $A$ and $A_1$ heavy-boson exchanges in the Proca formalism.}
\label{tab:Op4-fromR-Proca}
\end{center}
\end{table}

Notice that the tree-level exchange of heavy Proca fields can only generate $\mO(p^4)$ EWET operators through the chiral structures $\hat\chi_{\hat R}^\mu$ and $\hat\chi_{\hat R_1}^\mu$ in Eq.~\eqn{eq.Proca-chi}. Owing to the additional derivative present in $\hat R_{\mu\nu}$, the contributions from the rank-two tensors $\hat\chi_{\hat R}^{\mu\nu}$ and $\hat\chi_{\hat R_1}^{\mu\nu}$ in Eq.~\eqn{eq.Proca-chimunu} are at least of $\mO(p^6)$. Therefore, the tree-level exchange of $\hat R^\mu$ and $\hat R_1^\mu$ fields has a quite reduced impact on the low-energy EWET Lagrangian $\mL_4$. The custodial-breaking interactions of the singlet vector and axial-vector fields, $\widetilde c_\mT^{\hat V_1}$ and
$c_\mT^{\hat A_1}$, leave their imprints on $\mO_{10}$, $\mO_{6}^{\psi^2}$ and
 $\widetilde\mO_{3}^{\psi^2}$,
the last two operators requiring also the presence of
$\widetilde c_1^{\hat V_1}$ ($c_1^{\hat V_1}$) and $c_1^{\hat A_1}$
($\widetilde c_1^{\hat A_1}$), for $\mO_{6}^{\psi^2}$
 ($\widetilde\mO_{3}^{\psi^2}$).
The singlet vertices $c_1^{\hat V_1}$, $\widetilde c_1^{\hat V_1}$, $c_1^{\hat A_1}$ and $\widetilde c_1^{\hat A_1}$ also manifest in
$\mO_{7}^{\psi^4}$, $\mO_{8}^{\psi^4}$ and $\widetilde \mO_{2}^{\psi^4}$, while the $c_1^{\hat V}$,
$\widetilde c_1^{\hat V}$, $c_1^{\hat A}$ and $\widetilde c_1^{\hat A}$ interactions of the triplet vector and axial-vector states contribute to  $\mO_{5,6,7,8}^{\psi^4}$ and $\widetilde\mO_{1,2}^{\psi^4}$.

\section{Antisymmetric spin-1 resonance fields $(A)$}
\label{sec.antisym}

Until this point we have described all the spin-1 resonances through 4-vector Proca fields $\hat{R}^\mu$. However, it is sometimes convenient to express the massive spin-1 fields in terms of rank-2 antisymmetric tensors $R^{\mu\nu}$, a formalism widely used in \chpt~\cite{Ecker:1988te,Ecker:1989yg} which is reviewed in App.~\ref{app:antisymmetric}. A comparative analysis of the two descriptions turns out to be very enlightening.

In terms of tensor $R_{\mu\nu}$ fields, the spin-1 resonance Lagrangian takes the form
\bear
\mL_R^{(A)} &\! =&\! - \Frac{1}{2}\bra \nabla^\lambda R_{\lambda\mu} \,  \nabla_\sigma R^{\sigma \mu}
\, -\, \frac{1}{2} M_R^2\, R_{\mu\nu} R^{\mu\nu} \ket \; +\; \bra R_{\mu\nu} \chi^{\mu\nu}_R\ket
\hskip 1.5cm (R=V,\, A)\, ,
\nn\\
\mL_{R_1}^{(A)} &\! =&\!   -   \Frac{1}{2} \left( \partial^\lambda R_{1\, \lambda\mu}^{\phantom{\mu}} \,  \partial_\sigma R_1^{\sigma \mu}
\, -\, \frac{1}{2} M_{R_1}^2\, R_{1\, \mu\nu}^{\phantom{\mu}} R_1^{\mu\nu} \right)
\; +\; R_{1\, \mu\nu}^{\phantom{\mu}}\, \chi^{\mu\nu}_{R_1}
\hskip 1.2cm (R_1=V_1,\, A_1)\, .\qquad
\label{eq.resonance-LVA}
\eear
At $\mO(p^2)$, the most general expressions of the chiral tensors $\chi^{\mu\nu}_R$ and $\chi^{\mu\nu}_{R_1}$ ($R=V,A$) are:
\bear
\chi_{V}^{\mu\nu\, (2)}  &\!\! =&\!
\Frac{F_V}{2\sqrt{2}}\;  f_+^{\mu\nu}\; +\;
\Frac{i\, G_V}{2\sqrt{2}}\; [u^\mu, u^\nu]
\; +\; \Frac{\widetilde{F}_V }{2\sqrt{2}}\; f_-^{\mu\nu} \; +\;
\Frac{ \widetilde{\lambda}_1^{hV} }{\sqrt{2}}\;\left[
(\partial^\mu h)\, u^\nu-(\partial^\nu h)\, u^\mu \right]
\; +\; C_{0}^V\; J_T^{\mu\nu} \, ,
\nn\\
\chi_{A}^{\mu\nu\, (2)}  &\!\! =&\!
\Frac{F_A}{2\sqrt{2}}\;  f_-^{\mu\nu} \; +\;
\Frac{ \lambda_1^{hA} }{\sqrt{2}}\;\left[
(\partial^\mu h)\, u^\nu-(\partial^\nu h)\, u^\mu \right]
+\; \Frac{\widetilde{F}_A}{2\sqrt{2}}\; f_+^{\mu\nu}\; +\;
\Frac{i\, \widetilde{G}_A}{2\sqrt{2}}\; [u^{\mu}, u^{\nu}]
\; +\;  \widetilde{C}_{0}^A\;  J_{T}^{\mu\nu}\, ,
\nn\\
\chi_{V_1}^{\mu\nu\, (2)} &\!\! = &\!
  F_{V_1}\; X^{\mu\nu}
\; +\; \Frac{C^{V_1}_{0}}{\sqrt{2}}\;  \bra J_T^{\mu\nu} \ket
\, ,
\nn \\
\chi_{A_1}^{\mu\nu\, (2)}  &\!\! = &\!
    \widetilde{F}_{A_1}\; X^{\mu\nu}
\; +\; \Frac{\widetilde C^{A_1}_{0}}{\sqrt{2}}\;  \bra J_T^{\mu\nu} \ket
\, .
\label{eq:Achi}
\eear

All these structures have an exact correspondence with the rank-two Proca tensors $\hat\chi^{\mu\nu}_{\hat R}$ in Eq.~\eqn{eq.Proca-chimunu}. However, at $\mO(p^2)$ the antisymmetric description cannot incorporate chiral interactions with a single Lorentz index, analogous to the $\hat\chi^{\mu}_{\hat R}$ terms in Eq.~\eqn{eq.Proca-chi}.

\subsection{Integrating out the heavy spin-1 antisymmetric fields}

The LO contributions to the LECs of the EWET
can be easily obtained through the EoM associated with the generic Lagrangians in Eq.~\eqref{eq.resonance-LVA}:
\bear
\nabla^\mu \nabla_\rho  R^{\rho \nu}
-\nabla^\nu \nabla_\rho   R^{\rho\mu}
+ M_R^2\, R^{\mu\nu} &\! =&\!
-\, 2\, \left( \chi^{\mu\nu}_R-  \frac{1}{2}
\bra \chi^{\mu\nu}_R\ket \right)
\hskip 1.1cm (R=V,\, A)\, ,
\nn\\
\partial^\mu \partial_\rho  R_1^{\rho \nu}
-\partial^\nu \partial_\rho   R_1^{\rho\mu}
+ M_{R_1}^2 R_1^{\mu\nu}
&\! =&\! \,   -\, 2\,  \chi^{\mu\nu}_{R_1}
\hskip 3.4cm (R_1=V_1,\, A_1)\, .
\label{eq.resonance-EoM1A}
\eear
%

\begin{table}[tb]  
{\renewcommand{\arraystretch}{2.1}
\begin{center}
\begin{tabular}{|c||c|c|}
\hline
$i$ &   $\Delta\mF^{(A)}_i$ &  $\Delta\widetilde{\mF}^{(A)}_i$  \\  \hline \hline
1 &  $- \Frac{F_V^2-\widetilde{F}_V^2}{4M_V^2}
+ \Frac{F_A^2-\widetilde{F}_A^2}{4M_A^2} $
&
 $- \Frac{\widetilde{F}_VG_V}{2M_V^2}
- \Frac{F_A\widetilde{G}_A}{2M_A^2}$
\\ [1ex] \hline
2 &
 $- \Frac{F_V^2+{\widetilde{F}_V}^2}{8M_V^2}
- \Frac{F_A^2+{\widetilde{F}_A}^2}{8M_A^2}$
&
 $- \Frac{F_V \widetilde{F}_V}{4M_V^2}
- \Frac{F_A \widetilde{F}_A}{4M_A^2}$
\\ [1ex] \hline
3 &
$-  \Frac{F_VG_V}{2M_V^2} - \Frac{\widetilde{F}_A\widetilde{G}_A}{2M_A^2}$
&
 $- \Frac{F_V \widetilde{\lambda}_1^{hV} v}{M_V^2}
 - \Frac{\widetilde{F}_A \lambda_1^{hA} v}{M_A^2}$
\\ [1ex] \hline
4 &
 $\Frac{G_V^2}{4M_V^2} + \Frac{{\widetilde{G}_A}^2}{4M_A^2} $   & ---
\\ [1ex] \hline
5 &
$
-\Frac{G_V^2}{4M_V^2} - \Frac{{\widetilde{G}_A}^2}{4M_A^2} $   & ---
\\ [1ex] \hline
6 &
 $ - \Frac{\widetilde{\lambda}_1^{hV\,\, 2}v^2}{M_V^2}
- \Frac{\lambda_1^{hA\,\, 2}v^2}{M_A^2}$  & ---
\\ [1ex] \hline
7 & $
 \Frac{\lambda_1^{hA\,\, 2}v^2}{M_A^2}
+  \Frac{\widetilde{\lambda}_1^{hV\,\, 2}v^2}{M_V^2}$ & ---
\\ [1ex] \hline
9 &  $  - \Frac{F_A \lambda_1^{hA} v}{M_A^2}
- \Frac{\widetilde{F}_V \widetilde{\lambda}_1^{hV} v}{M_V^2}$  & ---
\\ [1ex] \hline
11 &  $- \Frac{F_{V_1}^2}{M_{V_1}^2} - \Frac{\widetilde{F}_{A_1}^2}{M_{A_1}^2} $  & ---
\\ [1ex] \hline
\end{tabular}
\end{center}
}
\caption{\small $\Delta \mL_R^{\cO(p^4)}$ contributions to the $\cO(p^4)$ LECs of bosonic operators from $V$, $V_1$, $A$ and $A_1$  heavy-boson exchanges in the antisymmetric formalism.
\label{tab:bLECs-A}}
\end{table}

\begin{table}[!t]  
\begin{center}
\renewcommand{\arraystretch}{2.1}
\begin{tabular}{|c||c|c||c|}
\hline
$i$ &  $\Delta\mF^{\psi^2\,\,(A)}_i$ &  $\Delta\widetilde{\mF}^{\psi^2\,\,(A)}_i$ & $\Delta\mF_i^{\psi^4\,\, (A)}$
\\  \hline\hline
1
& 0
& $-\displaystyle\frac{\widetilde{F}_V C_0^V}{\sqrt{2}M_V^2}\! -\! \displaystyle\frac{F_A \widetilde{C}_0^A}{\sqrt{2}M_A^2} $ & 0
\\[1ex] \hline
2
&  $-\displaystyle\frac{G_V C_0^V}{\sqrt{2}M_V^2} \!-\! \displaystyle\frac{\widetilde{G}_A \widetilde{C}_0^A}{\sqrt{2}M_A^2} $
&\! \!\!$-\displaystyle\frac{2\sqrt{2}v\widetilde{\lambda}_1^{hV}C_0^V}{M_V^2} \!-\! \frac{2\sqrt{2}v\lambda_1^{hA}\widetilde{C}_0^A}{M_A^2}$ & 0
\\[1ex] \hline
3
& $-\displaystyle\frac{F_V C_0^V}{\sqrt{2}M_V^2} \!-\! \displaystyle\frac{\widetilde{F}_A \widetilde{C}_0^A}{\sqrt{2}M_A^2} $
& 0 & 0
\\[1ex] \hline
4 &\!\!\!$-\displaystyle\frac{\sqrt{2}F_{V_1} C_0^{V_1}}{M_{V_1}^2} \!- \!\displaystyle\frac{\sqrt{2}\widetilde{F}_{A_1} \widetilde{C}_0^{A_1}}{M_{A_1}^2} $
& --- & 0
\\[1ex] \hline
9 & --- & --- &
$-\displaystyle\frac{(C_0^V)^2}{M_V^2}-\displaystyle\frac{(\widetilde{C}_0^A)^2}{M_A^2}$
\\[1ex] \hline
10 & --- & --- &
\!\! \!$\displaystyle\frac{(C_0^V)^2}{2M_V^2}\!-\!\displaystyle\frac{(C_0^{V_1})^2}{2M_{V_1}^2}
\!+\!\displaystyle\frac{(\widetilde{C}_0^A)^2}{2M_A^2}\!-\!\displaystyle\frac{(\widetilde{C}_0^{A_1})^2}{2M_{A_1}^2}$\!\!\!
\\[1ex] \hline
\end{tabular}
\end{center}
\caption{\small $\Delta \mL_R^{\cO(p^4)}$ contributions to the $\cO(p^4)$ LECs of fermionic operators from $V$, $V_1$, $A$ and $A_1$  heavy-boson exchanges in the antisymmetric formalism.}
 \label{tab:fLECs-A}
\end{table}

\noindent
Expanding them in powers of momenta,
\bear
R^{\mu\nu}  &\! =&\! -\, \Frac{2}{M_R^2} \,
 \left( \chi^{\mu\nu}_R-  \frac{1}{2}
 \bra \chi^{\mu\nu}_R\ket \right)
\; +\; \mO\left(\frac{p^4}{M_R^4}\right)
\hskip 2.2cm (R=V,\, A)\, ,
\nn\\
R_1^{\mu\nu}  &\! =&\! -\, \Frac{2}{M_{R_1}^2} \, \chi^{\mu\nu}_{R_1}
\; +\; \mO\left(\frac{p^4}{M_{R_1}^4}\right)
\hskip 4.25cm (R_1=V_1,\, A_1)\, ,
\label{eq.resonance-EoM2A}
\eear
and substituting these expressions back into the resonance Lagrangian~\eqref{eq.resonance-LVA}, one obtains the corresponding contributions to the low-energy Lagrangian of the EWET:
\bear
\Delta \mL_{R}^{\cO(p^4)} &\! =&\! -\, \Frac{1}{M_R^2} \,
\left(  \bra  \chi_R^{\mu\nu} \, \chi_{R \, \mu\nu}^{\phantom{\mu}} \ket
- \frac{1}{2} \,
\bra  \chi_R^{\mu\nu}\ket \bra  \chi_{R\,\mu\nu}^{\phantom{\mu}}\ket \right)
\hskip 1.6cm (R=V,\, A)\, ,
\nn\\
\Delta \mL_{R_1}^{\cO(p^4)} &\! =&\! -\, \Frac{1}{M_{R_1}^2} \;
    \chi_{R_1}^{\mu\nu} \, \chi_{R_1 \, \mu\nu}^{\phantom{\mu}}
\hskip 5.5cm (R_1=V_1,\, A_1)  \, .
\label{integration}
\eear
Expressing these results in our basis of $\cO(p^4)$ operators, one obtains the predictions for their LECs listed in Tables~\ref{tab:bLECs-A} and \ref{tab:fLECs-A}, for the bosonic and fermion operators, respectively.
Only those LECs receiving non-zero contributions are shown in the tables.
The $P$-even contributions to the first column of Table~\ref{tab:bLECs-A} agree with the results obtained previously in Ref.~\cite{Pich:2015kwa}.
The low-energy contributions from exotic $J^{PC}=1^{+-}$ heavy states
were analyzed in a similar way in Ref.~\cite{pseudovector-Cata}.

The predicted pattern of LECs is very rich with the antisymmetric description of heavy spin-1 bosons. The exchange of vector and axial-vector triplet states gives rise to the operators $\mO_{1,2,3,4,5,6,7,9}$, $\widetilde\mO_{1,2,3}$,
$\mO_{2,3}^{\psi^2}$, $\widetilde\mO_{1,2}^{\psi^2}$ and $\mO_{9,10}^{\psi^4}$,
while the singlet states only leave their fingerprints in $\mO_{11}$,
$\mO_{4}^{\psi^2}$ and $\mO_{10}^{\psi^4}$. In all cases the $1^{--}$ and $1^{++}$ massive states contribute simultaneously to the LECs.

The $\mO(p^4)$ LECs which receive contributions from the tree-level exchange of antisymmetric spin-1 fields are different from the ones generated through Proca-exchange. This is not surprising, since the two mechanisms refer to completely different dynamical structures. In the antisymmetric formalism the LECs originate in $\chi_R^{\mu\nu}$ chiral structures, while in the Proca description only the  $\hat\chi_{\hat R}^{\mu}$ terms contribute.

\subsection{Equivalence of the antisymmetric and Proca descriptions}
\label{subsec:equivalence}

The results shown in Tables~\ref{tab:spin-1-Op4-LEC-newterms}, \ref{tab:Op4-fromR-Proca},  \ref{tab:bLECs-A} and \ref{tab:fLECs-A} look quite different. A naive resonance-exchange calculation leads to a pattern of EWET LECs which depends on the adopted representation to describe the heavy spin-1 fields, either Proca or antisymmetric. Clearly, we are still missing some important ingredient, because physically meaningful results must be independent of the particular mathematical formalism used in their description.

As explicitly shown in App.~\ref{app:PA-correspondence}, the Proca and antisymmetric formalisms can be related through a simple change of variables in the corresponding path integral~\cite{Bijnens:1995,Kampf:2006}, transforming the Proca Lagrangian $\mL_R^{(P)} +\mL_{\text{non-R}}^{(P)}$
into an equivalent antisymmetric Lagrangian
$\mL_R^{(A)} +\mL_{\text{non-R}}^{(A)}$, with (linear) resonance interactions determined by the chiral tensors
\bear
\chi_R^{\mu\nu}  &\! =&\!
\Frac{1}{2 M_R}\, (\nabla^\mu \hat{\chi}_{\hat{R}}^\nu
-\nabla^\nu \hat{\chi}_{\hat{R}}^\mu)
\, +\, M_R\, \hat{\chi}_{\hat{R}}^{\mu\nu}
\hskip 2.1cm (R=V,\, A)\, ,
\nn\\
\chi_{R_1}^{\mu\nu}  &\! =&\!
\Frac{1}{2  M_{R_1}}\, (\partial^\mu \hat{\chi}_{\hat{R}_1}^\nu
-\partial^\nu \hat{\chi}_{\hat{R}_1}^\mu)
\, +\, M_{R_1}\, \hat{\chi}_{\hat{R}_1}^{\mu\nu}
\hskip 1.5cm (R_1=V_1,\, A_1)\, .
\label{eq.equivP-R}
\eear
The operators with only light fields in the antisymmetric representation $(A)$ are related to those in the Proca Lagrangian
$\mL^{(P)}_{\text{non-R}}$ through~\cite{Bijnens:1995,Kampf:2006}
\bear
\mL^{\rm (A)}_{\text{non-R}} &\! =&\!
\sum_{R=V,A} \left[ \bra \hat{\chi}_{\hat{R}\, \mu\nu}\,
\hat{\chi}_{\hat{R}}^{\mu\nu}\ket
\, -\, \Frac{1}{2}\, \bra  \hat{\chi}_{\hat{R}}^{\mu\nu}\ket
\bra  \hat{\chi}_{\hat{R}\,\mu\nu}^{\phantom{\mu}}\ket
\, -\,
\Frac{1}{2M_R^2}\, \left(  \bra  \hat{\chi}_{\hat{R}}^{\mu}\,
\hat{\chi}_{\hat{R}\,\mu}^{\phantom{\mu}} \ket
-\Frac{1}{2}\, \bra \hat{\chi}_{\hat{R}}^{\mu}\ket
\bra \hat{\chi}_{\hat{R}\, \mu}^{\phantom{\mu}}\ket\right)
\right]
\nn\\
&\! +&\!
\sum_{R_1=V_1,A_1} \left[  ( \hat{\chi}_{\hat{R}_1}^{\mu\nu}\,\hat{\chi}_{\hat{R}_1\,\mu\nu}^{\phantom{\mu}} )  \, -\,
\Frac{1}{2M_{R_1}^2}\, (\hat{\chi}_{\hat{R}_1}^{\mu}\, \hat{\chi}_{\hat{R}_1\,\mu}^{\phantom{\mu}})\right]
\,\, +\,\, \mL^{\rm (P)}_{\text{non-R}}
\, .
\label{eq.equivP-nonR}
\eear
Expressions~(\ref{eq.equivP-R}) and (\ref{eq.equivP-nonR}) provide an exact general relation between the Proca and antisymmetric representations, without any approximation or truncation. Therefore, the two descriptions are mathematically equivalent.

More precisely, inserting the $\mO(p^2)$ Proca chiral tensors of
Eqs.~\eqn{eq.Proca-chimunu} and (\ref{eq.Proca-chi}) into \eqn{eq.equivP-R} yields the following resonance interactions in the antisymmetric formalism:
\bear
\chi_V^{\mu\nu}  &\! =&\! \chi_V^{\mu\nu\, (2)} \, +\,
\Frac{C_{1}^V}{2}\, \left(  \nabla^\mu J_V^\nu   - \nabla^\nu J_V^\mu\right)
\, +\,  \Frac{\widetilde{C}_{1}^V}{2}\, \left(  \nabla^\mu J_A^\nu - \nabla^\nu J_A^\mu \right)\, ,
\nn\\
\chi_A^{\mu\nu} &\! =&\!  \chi_A^{\mu\nu\, (2)} \, +\,
\Frac{C_{1}^A }{2}\, \left( \nabla^\mu J_A^\nu - \nabla^\nu J_A^\mu \right)
\, +\, \Frac{\widetilde{C}_{1}^A}{2}\, \left( \nabla^\mu J_V^\nu - \nabla^\nu J_V^\mu \right) \, ,
\nn\\
\chi_{V_1}^{\mu\nu} &\! = &\! \chi_{V_1}^{\mu\nu\, (2)}\, +\,
\Frac{C_{1}^{V_1 }}{2\sqrt{2}} \,
\bra \partial^\mu J_V^\nu -  \partial^\nu J_V^\mu \ket \, +\,
\Frac{\widetilde{C}_{ 1}^{V_1}}{2\sqrt{2}}\, \bra\partial^\mu J_A^\nu - \partial^\nu J_A^\mu \ket
\nn\\ &+&
\frac{\widetilde{C}_{\mathcal{T}}^{V_1}}{2}\, \left( \partial^\mu \bra u^\nu \mathcal{T} \ket - \partial^\nu \bra u^\mu \mathcal{T} \ket \right)
\, ,
\nn \\
\chi_{A_1}^{\mu\nu} &\! = &\! \chi_{A_1}^{\mu\nu\, (2)}\, +\,
\Frac{C_{1}^{A_1 }}{2\sqrt{2}} \,\bra\partial^\mu  J_A^\nu - \partial^\nu J_A^\mu \ket \,+\,
\Frac{\widetilde{C}_{ 1}^{A_1}}{2\sqrt{2}}\, \bra\partial^\mu J_V^\nu - \partial^\nu J_V^\mu \ket
\nn\\ &+&
\frac{C_{\mathcal{T}}^{A_1}}{2}\, \left( \partial^\mu \bra u^\nu \mathcal{T} \ket - \partial^\nu \bra u^\mu \mathcal{T} \ket \right)
\, ,
\label{eq:BosAequiv}
\eear
where $\chi_{R}^{\mu\nu\, (2)}$ are the $\mO(p^2)$ structures in Eq.~\eqn{eq:Achi}, with the relations
\begin{align}
F_R\,=&\; f_{\hat{R}}\, M_R\, , \qquad\; &
G_R\,=&\; g_{\hat{R}}\, M_R\, , \qquad\; &
\lambda_1^{hR} \, =&\; \lambda_1^{h\hat{R}}\, M_R\, , \qquad\; &
C_0^R \,=&\; c_0^{\hat{R}}\,  M_R\, ,
\nn\\
\widetilde{F}_R \,=&\; \widetilde{f}_{\hat{R}}\, M_R\, , &
 \widetilde{G}_R \,=&\; \widetilde{g}_{\hat{R}}\, M_R\, , &
\widetilde{\lambda}_1^{hR} \, =&\; \widetilde{\lambda}_1^{h\hat{R}}\, M_R \, ,&
\widetilde{C}_0^R \,=&\; \widetilde{c}_0^{\hat{R}}\, M_R\, ,
\nn\\
C_{\mathcal{T}}^R \,=&\; c_{\mathcal{T}}^{\hat{R}} /M_R\, , &
\widetilde{C}_{\mathcal{T}}^R \,=&\; \widetilde{c}_{\mathcal{T}}^{\hat{R}} /M_R\, , &
C_1^R \,=&\; c_1^{\hat{R}} /M_R\, ,  &
\widetilde{C}_1^R \,=&\; \widetilde{c}_1^{\hat{R}} /M_R\, ,\;
\label{eq.P-A-relations}
\end{align}
for $R=V,A,V_1,A_1$.

The rank-two Proca tensors $\hat \chi_{\hat R}^{\mu\nu}$ transform into the antisymmetric structures
$\chi_{R}^{\mu\nu\, (2)}$. The additional derivative present in the $\hat R_{\mu\nu}$ fields gets traded by the factor $M_R$ in the couplings of the corresponding antisymmetric operators, reducing the overall chiral dimension. Therefore, the tree-level exchange of a spin-1 heavy boson between this type of chiral structures carries two powers of momenta less
in the antisymmetric formalism, allowing it to generate contributions to the $\mO(p^4)$ LECS which are absent in the Proca description. This behaviour gets reversed for the $\hat \chi_{\hat R}^{\mu}$ Proca structures, which transform into the $\mO(p^3)$ terms in Eq.~\eqn{eq:BosAequiv}. The antisymmetric formalism requires an additional derivative to carry the missing Lorentz index, compensating its dimension with a $1/M_R$ factor in the corresponding couplings $C_{\mathcal{T}}^R$, $\widetilde{C}_{\mathcal{T}}^R$, $C_1^R$ and $\widetilde{C}_1^R$. For these vertices, the spin-1 boson exchange carries two powers of momenta more in the antisymmetric description which, therefore, can only induce  LECs with chiral dimension $\hat d\ge 6$, while the Proca formalism generates
$\mO(p^4)$ LECs. All differences among the two scenarios are of course compensated by the local structure in Eq.~\eqn{eq.equivP-nonR}.

Thus, both formalisms give obviously the same predictions for the LECs. However, the splitting between `resonance-exchange' and `local' contributions depends on the adopted prescription and, therefore, is unphysical \cite{Ecker:1989yg}.
Quantum fields are just integration variables in the corresponding path-integral formulation of the generating functional, and the effective Lagrangian takes different explicit forms with different (equivalent) choices of functional field representations.

Tables~\ref{tab:spin-1-Op4-LEC-newterms}, \ref{tab:Op4-fromR-Proca},  \ref{tab:bLECs-A} and \ref{tab:fLECs-A} only contain the contributions to the EWET LECs
generated through resonance exchange in the two spin-1 formalisms. To those predictions one should add local contributions from operators without explicit resonance fields. Unfortunately, the relation \eqn{eq.equivP-nonR} only determines the difference
$\mL^{\rm (A)}_{\text{non-R}} -\mL^{\rm (P)}_{\text{non-R}}$.
This is not enough to decide which ones of the values quoted in the tables (if any) are the correct predictions for the LECs. We need additional dynamical information in order to pin down those pieces of the short-distance Proca and antisymmetric Lagrangians which only contain light fields.

We have already noticed in Eq.~\eqn{eq:BosAequiv} that, starting from $\mO(p^2)$ chiral tensors in the Proca representation, one gets $\mO(p^2)$ and $\mO(p^3)$ contributions to the $\chi^{\mu\nu}_R$ tensors in the antisymmetric formalism. This just reflect the different momentum dependence of these two spin-1 field representations. The UV behaviour of the adopted resonance EFT turns out to be crucial to correctly determine the predicted LECs \cite{Ecker:1989yg}. We are going to analyze it in the next section.

\section{Short-distance constraints}
\label{sec.SD}

Let us denote the antisymmetric and Proca short-distance effective theories as SDET-A and SDET-P, respectively. They contain the SM fields plus the heavy spin-1 vector and axial-vector states in their corresponding formulations (antisymmetric or Proca),
and the spin-0 resonances which are the same in both effective theories.
In addition to operators including the heavy fields, the two effective theories
contain terms with just light degrees of freedom, which are formally identical to those present
in the low-energy EWET. However, their couplings are obviously different,
since they belong to different effective theories. For every generic coupling $\mF_i$ of the EWET, we will denote as $\mF_i^{\,\mathrm{SDA}}$ and $\mF_i^{\,\mathrm{SDP}}$ the corresponding couplings in SDET-A and SDET-P:
\be
\mL_{\text{non-R}}^{(A)} \; =\; \sum_i \, \mF_i^{\,\mathrm{SDA}}\, \cO_i[\phi,\psi] \, ,
\qquad\qquad\qquad
\mL_{\text{non-R}}^{(P)} \; =\; \sum_i \, \mF_i^{\,\mathrm{SDP}}\, \cO_i[\phi,\psi] \, ,
\label{eq.SDA-P}
\ee
{where we have implicitly summed over all bosonic and fermionic operators.
SDET-A and SDET-P are equivalent formulations of the same dynamical theory,
{\it i.e.}, they must contain the same physics. In order to relate the two descriptions, one must analyze their predictions for specific Green functions.

\subsection{Purely bosonic sector}
\label{sec.SD-bosonic}

Let us consider the vector and axial-vector currents, defined through functional derivatives of the action with respect to the corresponding external sources:
\bear
\label{eq:Vcurrent}
\mV^\mu_a &\equiv & \Frac{\partial S}{\partial  v_\mu^a}\, ,
\qquad\qquad\qquad\qquad
\hat v_\mu\; =\; \frac{1}{2}\,\left( \hat{B}^\mu + \hat{W}^\mu\right)
\; =\; \frac{1}{2}\,\vec\sigma\, \vec v_\mu\, ,
\\
\mA^\mu_a &\equiv & \Frac{\partial S}{\partial  a_\mu^a}\, ,
\qquad\qquad\qquad\qquad
\hat a_\mu\; =\; \frac{1}{2}\,\left( \hat{B}^\mu - \hat{W}^\mu\right)
\; =\; \frac{1}{2}\,\vec\sigma\, \vec a_\mu\, .
\label{eq:Acurrent}
\eear
Their 2-Goldstone matrix elements are characterized by the vector and axial-vector form functions,
\be
\bra \varphi^+ (p_1)\, \varphi^- (p_2) \,|\, \mJ^\mu_3 \,|\, 0 \ket\; =\;
(p_1 - p_2)^\mu\; \mathbb{F}^\mJ_{\varphi\varphi} (s) \qquad\qquad\quad (\mJ = \mV,\,\mA )\, ,
\ee
with $s = (p_1+p_2)^2$.
A simple tree-level calculation gives the results:
\bear
\mathbb{F}^\mV_{\varphi\varphi}(s) &\! =&\! \left\{ \bat
1\, +\,\Frac{F_V\,G_V}{v^2}\,\Frac{s}{M_V^2-s}\,
  +\,\Frac{\widetilde{F}_A\,\widetilde{G}_A}{v^2}\,\Frac{s}{M_A^2-s}
  \,  -\, 2\, \mF_3^{\,\mathrm{SDA}}\,\Frac{s}{v^2}
& \qquad\quad\mbox{\small  (SDET-A)}\, ,
\nonumber \\[10pt]
1\, +\,\Frac{f_{\hat{V}}\, g_{\hat{V}}}{  v^2  }\,\Frac{ s^2 }{M_V^2-s}\,
  + \,\Frac{\widetilde{f}_{\hat{A}}\,\widetilde{g}_{\hat{A}}   }{  v^2 }\,\Frac{  s^2   }{M_A^2-s}\,
  -\, 2\, \mF_3^{\,\mathrm{SDP}}\,\Frac{s}{v^2}
& \qquad\quad\mbox{\small (SDET-P)} \, ,
\ea\right.\quad
\\[15pt]
\mathbb{F}^\mA_{\varphi\varphi}(s) &\! =&\! \left\{ \bat
\Frac{\widetilde{F}_V\, G_V}{v^2}\,\Frac{s}{M_V^2-s} \, +\,
\Frac{F_A\,\widetilde{G}_A}{v^2}\,\Frac{s}{M_A^2-s}
\, -\, 2\, \widetilde\mF_1^{\,\mathrm{SDA}}\,\Frac{s}{v^2}
& \qquad\qquad\mbox{\small  (SDET-A)}\, ,
\\[10pt]
\Frac{\widetilde{f}_{\hat{V}}\, g_{\hat{V}} }{v^2}\,\Frac{s^2}{M_V^2-s}
\, +\,
 \Frac{f_{\hat{A}}  \,\widetilde g_{\hat{A}}}{ v^2 }\,\Frac{s^2}{M_A^2-s}
 \, -\, 2\, \widetilde\mF_1^{\,\mathrm{SDP}}\,\Frac{s}{v^2}
& \qquad\qquad\mbox{\small (SDET-P)} \, .
\ea\right.
\eear

The form functions exhibit an unacceptable UV behaviour, growing linearly with the
squared momentum transfer. In SDET-A the unphysical linear dependence with $s$ is only generated
by the local operators $\mO_3$ and $\widetilde O_1$, while in SDET-P the non-local exchange
of Proca fields also contributes.
Requiring that $\mathbb{F}^\mV_{\varphi\varphi}(s)$ and $\mathbb{F}^\mA_{\varphi\varphi}(s)$ should not grow at large energies, we get the conditions:
\bear\label{eq:F3SDA}
\mF_3^{\,\mathrm{SDA}} & = & \widetilde\mF_1^{\,\mathrm{SDA}} \; =\; 0\, ,
\\
\mF_3^{\,\mathrm{SDP}} & = & -\,\Frac{f_{\hat V}\,g_{\hat V}}{2}\,
- \, \Frac{\widetilde f_{\hat A}\,\widetilde g_{\hat A}}{2}\, ,
\qquad \qquad\quad
\widetilde\mF_1^{\,\mathrm{SDP}} =
-\,\Frac{\widetilde{f}_{\hat V}\, g_{\hat V}}{2}\,
- \, \Frac{f_{\hat A}\, \widetilde g_{\hat A}}{2}\, .
\label{eq:F3SDP}
\eear
The two formalisms give then identical form functions with the identifications:
\begin{align}
& F_V\, G_V \; = \; f_{\hat V}\,g_{\hat V}\, M_V^2\, ,
\qquad\qquad
& \widetilde{F}_A\, \widetilde{G}_A \; = \; \widetilde{f}_{\hat A}\,\widetilde{g}_{\hat A}\, M_A^2\, ,
\label{eq:FGeven}
\\
& \widetilde{F}_V\, G_V \; = \; \widetilde{f}_{\hat V}\,g_{\hat V}\, M_V^2\, ,
\qquad\qquad
& {F}_A\, \widetilde{G}_A \; = \; {f}_{\hat A}\,\widetilde{g}_{\hat A}\, M_A^2\, .
\label{eq:FGodd}
\end{align}
These equalities are fully consistent with the relations between the Proca and antisymmetric couplings obtained in Eq.~(\ref{eq.P-A-relations}).
Moreover, the differences $\mF_3^{\,\mathrm{SDA}} - \mF_3^{\,\mathrm{SDP}}$ and $\widetilde\mF_1^{\,\mathrm{SDA}}-\widetilde\mF_1^{\,\mathrm{SDP}}$ are in agreement with Eq.~\eqn{eq.equivP-nonR}.

Thus, the requirement of a good UV behaviour carries a very interesting implication.
The $\cO(p^4)$ Goldstone couplings $\mF_3^{\,\mathrm{SDA}}$ and $\widetilde\mF_1^{\,\mathrm{SDA}}$ of SDET-A
must be zero, and the exchange of the heavy antisymmetric fields saturates the values of the corresponding LECs in the low-energy EWET. However, in SDET-P things work the opposite way:
the exchange of heavy spin-1 Proca particles does not give any contribution to the $\cO(p^4)$ LECs of the EWET, but a proper UV behaviour forces the presence of direct $\mF_3^{\,\mathrm{SDP}}$ and $\widetilde\mF_1^{\,\mathrm{SDP}}$ contributions. The final predictions for the LECs of the EWET, $\mF_3$ and $\widetilde\mF_1$, are exactly the same in both formalisms.

Studying other Green functions, it is easy to prove the
equivalence of the two formalisms in the bosonic sector.
For instance, the high-energy behaviour of the two-Goldstone scattering amplitudes determines
 the LECs $\mF_4$ and $\mF_5$,
and a similar thing occurs with the $hh\to\varphi\varphi, hh$ scattering and $\mF_{6,7,8}$.
On the other hand,
$\mF_1$, $\mF_2$ and $\widetilde\mF_2$ can be fixed with the two-point correlators of vector and axial-vector currents. A detailed analysis of
Higgsless bosonic operators is presented in App.~\ref{app:more-high-energy}, following the same procedure used before in QCD to exhibit the resonance saturation of the $\chi$PT LECs~\cite{Ecker:1989yg}.

Imposing a proper UV behaviour, one finds that the $\cO(p^4)$ LECs of SDET-A corresponding to bosonic operators must vanish,
\bear
\mF_i^{\,\mathrm{SDA}}\, =\,  \widetilde{\mF}_i^{\,\mathrm{SDA}} &=& 0
\qquad\qquad\qquad  (i\not= 10) \, ,
\eear
and the exchange of massive spin-1 antisymmetric fields saturates the values
of the corresponding EWET LECs $\mF_i$ and $\widetilde{\mF}_i$.
On the other hand, in SDET-P the same predictions are obtained through direct local couplings in the Lagrangian, {\it i.e.},
\be
\mF_i = \mF_i^{\,\mathrm{SDP}}\, ,\qquad\qquad \widetilde{\mF}_i = \widetilde{\mF}_i^{\,\mathrm{SDP}}
\qquad\qquad\qquad (i\not= 10)\, ,
\ee
are in general non-zero, as the first spin-1 resonance-exchange contributions start
at $\cO(p^6)$ at low energies.
The coupling $\mF_{10}$ is studied in a later section.

One can easily understand the physics behind this equivalence because the bosonic Proca couplings can be written in the form $\mL_{\hat R}^{(P)} \,\dot=\, \bra \hat{R}_{\mu\nu}\hat{\chi}_{\hat R}^{\mu\nu}\ket$, with $\hat{R}_{\mu\nu}$ defined in Eq.~\eqn{eq:Rhat_mn}, which is formally analogous to the interaction Lagrangian of the antisymmetric spin-1 fields, $\mL_R^{(A)} \,\dot=\, \bra R_{\mu\nu}\chi_R^{\mu\nu}\ket$. The effective action
$S^{(X)}$ ($X=A,P$) for the exchange of a single heavy spin-1 particle can then be written
in a compact way as~\cite{Ecker:1989yg}
\bel{eq:S_X}
S^{(X)}\; =\;-\frac{1}{2}\,\int d^4x\, d^4y\;\,
\bra \chi^{\mu\nu}_{(X)}(x)\,\Delta^{(X)}_{\mu\nu,\rho\sigma}(x-y)\,\chi^{\rho\sigma}_{(X)}(y)\ket\, ,
\ee
with $\chi^{\mu\nu}_{(A)} = \chi_R^{\mu\nu}$ and $\chi^{\mu\nu}_{(P)} = \hat{\chi}_{\hat R}^{\mu\nu}$. Taking into account the derivatives included in the definition \eqn{eq:Rhat_mn}, the Proca propagator adopts the form
\bel{eq:ProcaProg}
\Delta^{(P)}_{\mu\nu,\rho\sigma}(x)\; = \; \int\frac{d^4 k}{(2\pi)^4}\;\frac{\mathrm{e}^{-ikx}}{M_R^2-k^2}\; \left[g_{\mu\rho}\, k_\nu k_\sigma -g_{\mu\sigma}\, k_\nu k_\rho - (\mu\leftrightarrow\nu)\right]\, ,
\ee
while in the antisymmetric formulation one has (see App.~\ref{app:antisymmetric} for further details)
\bel{eq:AntisymProg}
\Delta^{(A)}_{\mu\nu,\rho\sigma}(x)\; = \; \frac{1}{M_R^2}\;\left\{
\Delta^{(P)}_{\mu\nu,\rho\sigma}(x)\, +\, \delta^{(4)}(x)\;\left(g_{\mu\rho}\, g_{\nu\sigma} -
g_{\mu\sigma}\, g_{\nu\rho} \right)\right\}\, .
\ee
The two spin-1 resonance exchanges are then equivalent up to a local contribution. For a given chiral structure (determined by the external
legs of the Green function), the identification of the pole residues at $k^2=M_R^2$ relates the corresponding chiral couplings
in the two formalisms with the appropriate power of $M_R$ to compensate the different canonical dimensions, as
indicated in Eq.~\eqn{eq.P-A-relations}. The local contributions are adjusted to satisfy a proper UV behaviour, which results in identical Green functions in both formalisms. The EWET LECs are finally obtained from the infrared limit of the Green functions.

\subsection{Two-fermion operators}
\label{sec.SD-two-fermion}

We can distinguish three different types of $\cO(p^4)$ two-fermion operators. The first group
($\mO_3^{\psi^2}$,  $\mO_4^{\psi^2}$ and $\widetilde\mO_1^{\psi^2}$) contribute to fermion form factors.
The second ($\mO_1^{\psi^2}$,  $\mO_2^{\psi^2}$,  $\mO_5^{\psi^2}$ and $\widetilde\mO_2^{\psi^2}$)
are relevant for $\psi\varphi\to \psi \varphi , \psi h$ scattering amplitudes.
 $\mO_7^{\psi^2}$ is of a similar type and is relevant for the $\psi h \to \psi h$ scattering.
There is finally
a third group formed by the custodial symmetry-breaking
operators $\mO_6^{\psi^2}$ and $\widetilde\mO_3^{\psi^2}$.

We will focus here the discussion on the first two types of operators, which get contributions from vector and axial-vector exchanges between $\chi^{\mu\nu}$ vertices, in the antisymmetric formalism. The general structure of these spin-1 exchanges in the $\hat V_\mu$ and $V_{\mu\nu}$ descriptions is then also given
by Eqs.~\eqn{eq:S_X}, \eqn{eq:ProcaProg} and \eqn{eq:AntisymProg}. A few explicit examples are enough to check that the LECs of the EWET are saturated by the resonance-exchange contributions in SDET-A, without any need for additional local terms. In
SDET-P, the same results for the LECs must necessarily originate  in local couplings, since the Proca-exchange contributions are at least of $O(p^6)$.

The third group of operators only receive contributions from the exchanges of Proca fields between
$\hat \chi_{\hat R}^\mu$ vertices, which are not present in the antisymmetric description.
They will be analyzed in the next subsection, together with the four-fermion
vector and axial-vector structures which have a similar origin.

\subsubsection{Form-factors}

The terms $\mF_3^{\psi^2}\bra J_T^{\mu\nu} f_{+\, \mu\nu}\ket$, \
$\mF_4^{\psi^2} \bra J_T^{\mu\nu} \ket \hat{X}_{\mu\nu}$, and
$\widetilde{\mF}_1^{\psi^2}\bra J_T^{\mu\nu} f_{-\, \mu\nu}\ket$ involve the fermionic tensor
bilinear.
They can be studied considering
again the vector and axial-vector currents, defined in Eqs.~\eqn{eq:Vcurrent} and \eqn{eq:Acurrent}, and
\be
\mV_{(0)}^\mu \;\equiv\; \Frac{\partial S}{\partial \hat v^{(0)}_\mu}\, ,
\qquad\qquad\qquad\qquad
\hat v^{(0)}_\mu\; =\; \hat{X}_\mu  \, .
\ee
Assuming $CP$ conservation,
the corresponding two-fermion matrix elements are characterized by the form factors
$\mathbb{F}^\mJ_{1,2}(q^2)$,
\be
\bra \psi (p_1) \,|\, \mJ^\mu \,| \, \psi (p_2)\,  \ket\; =\;
 \bar{u}(p_1) \left[ \Gamma_\mJ^\mu\, \mathbb{F}^\mJ_{1}(q^2)
+  \frac{i}{2}\, q_\nu\, \sigma^{\mu\nu}\, \mathbb{F}^\mJ_{2}(q^2)
 \right] u(p_2)
\qquad (\mJ = \mV_3,\,\mA_3,\, \mV_{(0)} )\, ,
\ee
with $q = p_1-p_2$, $s=q^2$, $\Gamma^\mu _{\mV_3,\,\mV_{(0)}}=\gamma^\mu$
and $\Gamma^\mu_{\mA_3}=\gamma^\mu \gamma_5$.
We will focus on the second form-factor in the massless fermion limit.\footnote{
The magnetic form-factor is usually shown with the normalization
$\mathbb{F}^\mJ_{2'} (q^2)= m_\psi \,\mathbb{F}^\mJ_{2} (q^2)$.}
At tree-level, {\it e.g.}, for $\mJ = \mV_3$, one has
\bear
 \mathbb{F}^{\mV_3}_{2}(s)  &\! =&\! \left\{ \bat
 -4\sqrt{2}\; T^3_\psi \,  \left( \Frac{F_V C_0^V}{M_V^2-s} +\Frac{\widetilde{F}_A \widetilde{C}_0^A}{M_A^2-s} - \sqrt{2} \,\mF_3^{\psi^2 ,\, {\rm SDA}} \right)
& \qquad\quad\mbox{\small  (SDET-A)}\, ,
\\[10pt]
-4\sqrt{2}\; T^3_\psi \,
\left( \Frac{f_{\hat{V}} c_0^{\hat{V}} s }{M_V^2-s} +\Frac{\widetilde{f}_{\hat{A}} \widetilde{c}_0^{\hat{A}} s}{M_A^2-s} - \sqrt{2}\, \mF_3^{\psi^2 ,\, {\rm SDP}} \right)
& \qquad\quad\mbox{\small (SDET-P)} \, .
\ea\right.\quad
\eear
Demanding that $\mathbb{F}^{\mV_3}_{2}(s)$ vanishes at high energies, we get the conditions:
\bear
\mF_3^{\psi^2,\,\mathrm{SDA}} & = &  0\, ,
\nonumber \\
\mF_3^{\psi^2,\,\mathrm{SDP}} & = &-\frac{1}{\sqrt{2}}\, \left( f_{\hat V}\,
c_0^{{\hat{V}}} \, + \, \widetilde f_{\hat A}\,
\widetilde{c}_0^{{\hat{A}}} \right) \, .
\eear
The two formalisms give the same form-factor (and low-energy predictions) with the identifications,
\begin{align}
f_{\hat V}\; = \;  F_V/  M_V \, ,
\qquad
\widetilde{f}_{\hat A}  \; = \;   \widetilde{F}_A/  M_A\, ,
\qquad
c_0^{{\hat{V}}} \; = \; C_0^V/M_V\,,
\qquad
\widetilde{c}_0^{{\hat{A}}} \; = \; \widetilde{C}_0^A/M_A \,,
\end{align}
in agreement with the general relations in Eqs.~(\ref{eq.equivP-nonR}) and (\ref{eq.P-A-relations}).

A similar result is obtained for $\mJ=\mA_3, \mV_{(0)}$. In the three cases
one finds that the corresponding $\cO(p^4)$ LECs of SDET-A must vanish,}
\bear
\mF_3^{\psi^2,\,\mathrm{SDA}}\, =\,\mF_4^{\psi^2,\,\mathrm{SDA}}\, =\, \widetilde{\mF}_1^{\psi^2,\,\mathrm{SDA}}\, =\,0\, ,
\eear
and the exchange of massive spin-1 fields in the antisymmetric formalism saturates the values of the EWET LECs of the analogous two-fermion operators.
On the other hand, in SDET-P the same predictions are obtained through direct local couplings in the Lagrangian, {\it i.e.},
\be
\mF_i^{\psi^2} \,  =\, \mF_i^{\psi^2,\,\mathrm{SDP}}\, \qquad (i=3,4),
\qquad\qquad\qquad  \widetilde{\mF}_1^{\psi^2}\,  =\, \widetilde{\mF}_1^{\psi^2 ,\,\mathrm{SDP}}\, .
\ee
The direct exchange of spin-1 Proca fields does not give any contribution to these $\cO(p^4)$ LECs.

\subsubsection{$\boldsymbol{\psi\, \varphi/h \to \psi \, \varphi/h} $ scattering}

The scattering amplitudes for $\psi(p_1)\, \varphi(p_2) \to \psi(p_3)\, \varphi(p_4)$
and  $ \psi(p_1)\, \varphi(p_2) \to  \psi(p_3)\, h(p_4) $
receive contributions from heavy resonance exchanges and from the local 2-fermion operators
$\mF_1^{\psi^2}\bra J_S\ket \bra u_{\mu}u^{\mu}\ket$, \
$i\,\mF_2^{\psi^2} \bra J_T^{\mu\nu} [ u_{\mu},u_{\nu}]\ket $, \
$\mF_5^{\psi^2} \bra J_P  u_{\mu}\ket \,\partial^\mu h/v $, and
$\widetilde{\mF}_2^{\psi^2}\bra J_T^{\mu\nu} u_{\nu}\ket \,\partial^\mu h/v $.
Similarly,
$ \psi(p_1)\, h(p_2) \to  \psi(p_3)\, h(p_4)$ gets a local contribution from the
2-fermion operator $\mF_7^{\psi^2}\bra J_S\ket  (\partial_{\mu} h)(\partial^{\mu}h)/v^2$,
in addition to the resonance-exchange amplitudes.
The exchange of spin-1 Proca fields does not contribute to any of these chiral structures,
while only $\mF_2^{\psi^2}$ and $\widetilde{\mF}_2^{\psi^2}$ get  contributions
in the antisymmetric formalism.
The exchange of spin-0 resonances contributes to the LECs $\mF_1^{\psi^2}$ and $\mF_5^{\psi^2}$.

In general, the spin-0 resonance-exchange amplitudes behave
like $\mM_{\psi \varphi \to \psi \varphi ,\,  \psi  h}\sim E$
at high energies and do not violate the Froissart bound on the cross section, $\sigma(s)< C\, \ln^2(s/s_0)$
(further constraints can be nevertheless imposed through a more thorough analysis of, {\it e.g.}, partial-wave projections or forward scattering).
This is not generally true for the spin-1 interactions through the $J_T^{\mu\nu}$ resonance term. For instance,
in the antisymmetric (Proca) case, the exchange of a triplet vector resonance between a fermionic tensor vertex $C_0^V \bra V_{\mu\nu} J_T^{\mu\nu}\ket$ ($c_0^{{\hat{V}}} \bra \hat{V}_{\mu\nu} J_T^{\mu\nu}\ket$) and the two-Goldstone vertex
$\frac{i G_V}{2\sqrt{2}}\bra V^{\mu\nu} [u_\mu, u_\nu]\ket$
($\frac{i g_{\hat{V}}}{2\sqrt{2}}\bra \hat{V}^{\mu\nu} [u_\mu, u_\nu]\ket$)
scales at high energies
like
\be
\mM_{\psi \varphi \to \psi \varphi  } \bigg|_{\rm V\, through\, J_T} \; \sim \; \left\{ \bat
\Frac{C_0^V G_V}{v^2}\;  E  & \qquad\quad\mbox{\small  (SDET-A)}\, ,
\\[10pt]
\Frac{c_0^{{\hat{V}}} g_{\hat{V}} }{v^2}\;  E^3
& \qquad\quad\mbox{\small (SDET-P)} \, .
\ea\right.\quad
\ee
A similar behaviour can be derived for the other contributions from $J_T$ terms to this type of processes,
showing that the antisymmetric prediction does not violate
the Froissart bound (in its simplest approach),
on the contrary to what happens in the Proca realization which requires additional contributions
to regulate the UV behaviour.

Non-resonant contributions from the local $\cO(p^4)$ terms $
\mF_{1,2,5  ,7  }^{\psi^2 ,\, {\rm SDA}},\, \widetilde{\mF}_2^{\psi^2 ,\, {\rm SDA}}$
scale at high energies like
\be
\mM_{\psi \varphi \to \psi \varphi  } \bigg|_{\text{non-R}} \; \sim \;
\frac{ \mF_i^{\psi^2,\, {\rm SDA}} }{v^2}\;  E^3   \, ,
\ee
and the same happens for the analogous $
\mF_{1,2,5  ,7  }^{\psi^2 ,\, {\rm SDP}},\,
\widetilde{\mF}_2^{\psi^2 ,\, {\rm SDP}}$ contributions, in the Proca formalism.

Hence, in order to preserve the good short-distance behaviour,
the non-resonant contributions must vanish in the antisymmetric tensor realization,
{\it i.e.},
\be
\mF_{1,2,5   ,7  }^{\psi^2 ,\, {\rm SDA}}\, =\, \widetilde{\mF}_2^{\psi^2 ,\, {\rm SDA}}\, =\, 0\, ,
\ee
while in SDET-P appropriate non-zero values of
$\mF_{2}^{\psi^2 ,\, {\rm SDP}}$ and $\widetilde{\mF}_2^{\psi^2 ,\, {\rm SDP}}$
must be present to compensate the bad UV behaviour of the Proca-exchange contributions.
The other couplings must also be zero in the Proca formalism: $\mF_{1,5   ,7  }^{\psi^2 ,\, {\rm SDP}}=0$
(the exchange of vector or axial-vector bosons does not contribute to these operators)
.

\bigskip
\subsection{$\boldsymbol{\hat\chi_{\hat R}^\mu\,\hat\chi_{\hat R\,\mu}}$ chiral structures}
\label{sec.SD-chimuchimu}

The four-fermion operators $\mO^{\psi^4}_{5,6,7,8}$ and $\widetilde\mO^{\psi^4}_{1,2}$ and the custodial
symmetry-breaking structures $\mO_{10} = \bra u_\mu\mT\ket^2$,
$\mO_6^{\psi^2} = \bra u_\mu\mT\ket \bra J_A^\mu\ket$ and
$\widetilde\mO_3^{\psi^2}= \bra u_\mu\mT\ket \bra J_V^\mu\ket$ cannot be generated through
the exchange of antisymmetric spin-1 fields, but receive contributions from Proca-exchange.
They originate in the linear couplings $\bra\hat R_{\mu}\, \hat\chi_{\hat R}^\mu\ket$
and/or $\hat R_{1\,\mu}\, \hat\chi_{\hat R_1}^\mu$, which can only be present in the Proca formulation.
The short-distance behaviour generated by these structures is quite different from the one we studied before for the $\chi_{R}^{\mu\nu}$ terms.

Let us consider a generic Green function associated with these chiral structures, in the Proca formulation. At tree-level it can be formally written as
\bear\label{eq:4F-GreenP}
G^{(P)}(x-y)&\!\! =&\!\! \int
\frac{d^4 k}{(2\pi)^4}\;\mathrm{e}^{-ik(x-y)}\,\left\{
\sum_{i=5}^{8} \mF_i^{\psi^{4},\, {\rm SDP}}\,\mO^{\psi^{4}}_i(x)
 + \sum_{i=1}^{2} \widetilde\mF_i^{\psi^{4},\, {\rm SDP}}\,\widetilde\mO^{\psi^{4}}_i(x)
 + \mF_{10}^{\rm SDP}\,\mO_{10}(x)
\right.\nn\\[5pt] &&\hskip -.9cm\left.\mbox{}
+\mF_6^{\psi^{2},\, {\rm SDP}}\,\mO^{\psi^{2}}_6(x)
 + \widetilde\mF_3^{\psi^{2},\, {\rm SDP}}\,\widetilde\mO^{\psi^{2}}_3(x)
 +\frac{1}{2}\,\sum_{R_1=V_1,A_1}
\frac{g_{\mu\nu}-k_\mu k_\nu/M_{R_1}^2}{k^2-M_{R_1}^2}\,
\hat{\chi}_{\hat R_1}^{\mu}(x)\, \hat{\chi}_{\hat R_1}^\nu (y)
\right.\nn\\ &&\hskip -.9cm\left.\mbox{}
+\frac{1}{2}\,\sum_{R=V,A}
\frac{g_{\mu\nu}-k_\mu k_\nu/M_{R}^2}{k^2-M_R^2}\,
\left[ \bra \hat{\chi}_{\hat R}^{\mu}(x)\,\hat{\chi}_{\hat R}^\nu(y)\ket\!
-\!\frac{1}{2}\,\bra \hat{\chi}_{\hat R}^{\mu}(x)\ket \bra \hat{\chi}_{\hat R}^\mu (y)\ket\right]
\right\} ,
\eear
which includes the local contribution from $\cO(p^4)$ operators and the non-local exchanges of spin-1 fields.
Using partial integration, $k_\mu k_\nu \hat{\chi}_{\hat R}^{\mu}(x) \hat{\chi}_{\hat R}^{\nu}(y) = \partial_\mu\hat{\chi}_{\hat R}^{\mu}(x)\,\partial_\nu\hat{\chi}_{\hat R}^{\nu}(y)$.

In four-fermion amplitudes the momentum-dependent pieces in the numerators of the spin-1 propagators transform into fermion masses because $k_\mu J^\mu_{V,A}\sim m_f$. Therefore, the non-local contributions are well behaved at large energies.
Working for simplicity with massless fermions, the same happens in processes with only two fermions and Goldstones.
On the other side, the corresponding local operators have the same algebraic structure $\hat{\chi}_{\hat R}^{\mu}\,\hat{\chi}_{\hat R\,\mu}^{\phantom{\mu}}$ but without the propagator momentum suppression $(k^2-M_R^2)^{-1}$, giving rise to cross sections which would violate unitarity:
$\mM(\psi\bar\psi\to \psi\bar\psi,\varphi\varphi)\sim E^2$.
Therefore, a good UV behaviour requires\footnote{
This generic short-distance behaviour is not expected to be modified in the presence of Higgs fields. The Proca-exchange amplitude generating the bosonic structure $\mO_{10}$ does not introduce UV problems and does not need to be subtracted with local terms.
}
\bear
\mF_{5,6,7,8}^{\psi^4,\, {\rm SDP}}\, =\,
\widetilde\mF_{1,2}^{\psi^4,\, {\rm SDP}}\, =\,
\mF_{6}^{\psi^2,\, {\rm SDP}}\, =\,
\widetilde\mF_{3}^{\psi^2,\, {\rm SDP}}\, =\,
\mF_{10}^{\rm SDP}\, =\,
0\, .
\eear
The limit of small momenta ($k^2\ll M_{R}^2, M_{R_1}^2$) reproduces then the predictions for the corresponding EWET LECs in Tables~\ref{tab:spin-1-Op4-LEC-newterms} and \ref{tab:Op4-fromR-Proca}.

The relation \eqn{eq.equivP-nonR} determines the corresponding local terms in the antisymmetric formulation,
\be
\mL^{\rm (A)}_{\text{non-R}} \; \dot= \;
\, -\,
 \sum_{R=V,A}
\Frac{1}{2M_R^2}\, \left(  \bra  \hat{\chi}_{\hat{R}}^{\mu}\,
\hat{\chi}_{\hat{R}\,\mu}^{\phantom{\mu}} \ket
-\Frac{1}{2}\, \bra \hat{\chi}_{\hat{R}}^{\mu}\ket
\bra \hat{\chi}_{\hat{R}\, \mu}^{\phantom{\mu}}\ket\right)
\, -\, \sum_{R_1=V_1,A_1}
\Frac{1}{2M_{R_1}^2}\, \hat{\chi}_{\hat{R}_1}^{\mu}\, \hat{\chi}_{\hat{R}_1\,\mu}^{\phantom{\mu}}\, ,
\label{eq.AxiLECS}
\ee
which give identical predictions for the $\mO(p^4)$ LECs of the EWET.
The exchange of $R_{\mu\nu}$ fields involves in this case the $\mO(p^3)$ pieces of the chiral structures $\chi^{\mu\nu}$ in Eq.~\eqn{eq:BosAequiv} and, therefore, generates non-local contributions with a bad UV behaviour plus
local operators of $\mO(p^6)$. The combined effect of these local  $\mO(p^6)$ terms and the $\mO(p^4)$
operators in Eq.~\eqn{eq.AxiLECS} restores the good unitarity properties, giving finally
the same Green function than the Proca formalism.

\subsection{Short-distance summary}

The different Lorentz structure of the antisymmetric $R_{\mu\nu}$ tensors and the Proca $\hat R_\mu$ fields implies a different energy scaling of the corresponding spin-1 boson-exchange amplitudes. Although both descriptions are mathematically equivalent, once local terms are taken into account, the same physics gets splitted differently in local and non-local contributions. For any given Green function, a correct comparison of the two formalisms makes necessary to analyze the same physics at different chiral orders.

In general, the description in terms of antisymmetric tensors $R_{\mu\nu}$ and $\chi^{\mu\nu}_R$ chiral structures is more efficient, giving a proper UV behaviour, which does not need to be corrected with local terms, and directly generating the wanted $\mO(p^4)$ LECs through resonance exchange. The Proca description, on the other side, induces resonance-exchange amplitudes with a worse high-energy behaviour, which must be canceled by local operators with precisely the same values for their LECs.

The situation is slightly different for the few $\mO(p^4)$ LECs receiving direct contributions from the tree-level exchange of Proca $\hat R_\mu$ fields. In all cases, these contributions are generated by $\hat\chi_{\hat R}^\mu$ structures, which cannot be present in the antisymmetric formulation. The corresponding Proca-exchange amplitudes have a good UV behaviour, implying the absence of the associated local operators in SDET-P and directly leading to the wanted LECs in the infrared. The antisymmetric tensor formalism can only account for these contributions through $\mO(p^3)$
chiral structures of the type $\nabla^\mu
\mJ^\nu - \nabla^\nu \mJ^\mu$, with $\mJ^\mu = J_{V,A}^\mu, u^\mu\mT$,
requiring an $\mO(p^6)$ analysis to pin down the corresponding $\mO(p^4)$ LECs. The final results are obviously the same, since both formalisms are fully-equivalent effective descriptions of the same physics.

Since the naive exchange of antisymmetric and Proca fields generates different chiral structures, the final values for the $\mO(p^4)$ LECs are simply given by the sum of all spin-1 contributions collected in Tables~\ref{tab:spin-1-Op4-LEC-newterms}, \ref{tab:Op4-fromR-Proca},  \ref{tab:bLECs-A} and \ref{tab:fLECs-A}.
There could be in addition other contributions not related to these vector and axial-vector heavy states. For instance, the spin-0 contributions in Table~\ref{tab:spin-0-Op4-LEC}.

\section{Gauge-like formulation of spin-1 massive states}
\label{sec.HLS}

In many fashionable models the heavy vector states are introduced as massive Yang-Mills fields
or hidden local symmetry (HLS) gauge vectors~\cite{BKUY:85,BKY:88,HY:92,HY:03,Casalbuoni:1985kq,Feruglio:1988,Casalbuoni:93,ME:88},
{\it i.e.}, a triplet spin-1 vector is represented by a field
$\bar{V}_\mu$, transforming under $\mG$ as
\be
\bar{V}_\mu \quad\longrightarrow\quad g_h \,\bar{V}_\mu\, g_h^\dagger
\, +\, \Frac{i}{g_\rho} \, g_h\, \partial_\mu g_h^\dagger\, ,
\ee
and described by the Lagrangian~\cite{Ecker:1989yg}
\bel{eq:L_HLS}
\mL_V^{\rm (H)} \; =\;
-\Frac{1}{4}\, \bra \bar{V}_{\mu\nu} \bar{V}^{\mu\nu}\ket
\, +\, \Frac{1}{2}\, M_V^2\; \bra \left(\bar{V}_\mu - \Frac{i}{g_\rho}\, \Gamma_\mu \right) \left(\bar{V}^\mu - \Frac{i}{g_\rho}\, \Gamma^\mu \right)\ket\, ,
\ee
with the gauge field strength tensor $\bar{V}_{\mu\nu}=\partial_\mu \bar{V}_\nu
-  \partial_\nu \bar{V}_\mu - ig_\rho\, [\bar{V}_\mu,\bar{V}_\nu ]$
and the HLS gauge coupling $g_\rho$.

The first term in \eqn{eq:L_HLS} is just the renormalizable dimension-4 Yangs-Mills Lagrangian. Renormalizability guarantees very good UV properties which are only softly modified by the second term, incorporating the vector mass in a gauge-invariant way. The connection $\Gamma_\mu$, defined in \eqn{eq:connection}, introduces non-linear interactions with the Goldstone fields but, thanks to the underlying local symmetry, they generate scattering amplitudes which are well behaved at short distances.

One can easily recover the Proca representation with the field redefinition
\be
\bar{V}_\mu \; =\; \hat{V}_\mu \,+\, \Frac{i}{g_\rho}\,\Gamma_\mu \, ,
\label{eq.HLS-Proca}
\ee
where $\hat{V}_\mu$ transforms under $\mG$ as
$\hat{V}_\mu \to g_h \,\hat{V}_\mu\, g_h^\dagger$.
This implies~\cite{Ecker:1989yg}
\be
\bar{V}_{\mu\nu} \; =\; \hat{V}_{\mu\nu}
\, +\, \Frac{i}{g_\rho}\, \Gamma_{\mu\nu} \, - i g_\rho\, [\hat{V}_\mu ,\hat{V}_\nu]\, ,
\ee
with $\hat{V}_{\mu\nu} =\nabla_\mu \hat{V}_\nu -\nabla_\nu \hat{V}_\mu$ and
$\Gamma_{\mu\nu} = \frac{1}{4}\, [u_\mu,u_\nu]\, -\, \frac{i}{2}\, f_{+\, \mu\nu}$.
With this change of variables the resonance Lagrangian $\mL_V^{\rm (H)}$ takes the form~\cite{Ecker:1989yg}
\bear
\mL_V^{\rm (H)} &=&
-\Frac{1}{4}\, \bra \hat{V}_{\mu\nu} \hat{V}^{\mu\nu}\ket
\,  +\, \Frac{1}{2}\, M_V^2\, \bra \hat{V}_\mu \hat{V}^\mu \ket
\, -\, \Frac{i}{2 g_\rho}\, \bra \hat{V}^{\mu\nu}\, \Gamma_{\mu\nu}\ket
\, +\, \Frac{1}{4 g_\rho^2}\, \bra \Gamma_{\mu\nu}\, \Gamma^{\mu\nu} \ket
\nn\\
&& -\,  \Frac{1}{2}\,\bra \Gamma_{\mu\nu}\, [\hat{V}^\mu,\hat{V}^\nu] \ket
\, +\,\Frac{i g_\rho}{2}\, \bra  \hat{V}_{\mu\nu}\, [\hat{V}^\mu, \hat{V}^\nu ]\ket
\, +\,\Frac{g_\rho^2}{4}\,\bra  [\hat{V}_\mu,\hat{V}_\nu]\, [\hat{V}^\mu,\hat{V}^\nu]\ket\, .
\label{eq.redefined-HLS-Lagr}
\eear
Thus, one gets the free Proca Lagrangian for the field $\hat V_\mu$ plus specific interaction terms. Dropping the operators on the second line which involve two or more massive vector fields, we are left with the Proca Lagrangian
$\mL_{\hat V}^{\rm (P)} + \mL_{\text{non-R}}^{\rm (P)}$
with its couplings determined in terms of $g_\rho$:
\bear
f_{\hat V} &=& 2\, g_{\hat V}\, =  \, -\, \Frac{1}{\sqrt{2}\, g_\rho}\, ,
\nn\\
\mF_1^{\rm SDP} &=&
2 \,\mF_2^{\rm SDP} \, =\,\mF_3^{\rm SDP} \, =\,
 \,-\, 4\,\mF_4^{\rm SDP} \, =\, 4 \,\mF_5^{\rm SDP} \, =\, - \Frac{1}{8g_\rho^2}\, ,
\label{eq.HLS-Proca-couplings}
\eear
and all the other couplings zero. The $\hat V_\mu$ interactions in $\mL_{\hat V}^{\rm (P)}$ are a particular version of the triplet vector Lagrangian in Eqs.~\eqn{eq.resonanceProca-L}
to \eqn{eq.Proca-chi}, without the Higgs field, fermions and $P$-odd terms, and with the additional constraint $f_{\hat V} = 2\, g_{\hat V}$. This relation is a consequence of the specific HLS model \eqn{eq:L_HLS}, which is not required by the assumed chiral symmetry.

The predicted local terms $\mF_i^{\rm SDP}$ are in perfect agreement with our short-distance considerations in the previous section. The $\mF_i^{\rm SDP}$ values in Eq.~\eqn{eq.HLS-Proca-couplings} reproduce our more general results
 in Eq.~\eqn{eq:F3SDP} and Eqs.~\eqn{eq:F45SDP}, \eqn{eq:F1SDP} and \eqn{eq:F2SDP} in App.~\ref{app:more-high-energy},
when particularized to the specific HLS couplings. Thanks to the underlying gauge symmetry, the term without the vector field in Eq.~\eqn{eq.redefined-HLS-Lagr}, {\it i.e.},
$\mL_{\text{non-R}}^{\rm (P)} =(4 g_\rho^2)^{-1}\, \bra \Gamma_{\mu\nu}\, \Gamma^{\mu\nu} \ket$, has the precise structure and couplings needed to compensate the bad UV behaviour
of the Proca-exchange contributions and render the model well behaved at large momenta.
Since vector-exchange only starts to contribute to the EWET LECs  at $\mO(p^6)$, the $\mO(p^4)$ LECs are also fully
determined by $\mL_{\text{non-R}}^{\rm (P)}$, in nice agreement with the values quoted in Table~\ref{tab:bLECs-A}.

One could easily extend the HLS model, using the difference
$\hat{V}_\mu \; =\; \bar{V}_\mu \,-\, i g_\rho^{-1}\,\Gamma_\mu$ to build all additional invariants allowed by symmetry considerations, including the Higgs, fermions
and $P$-odd operators. The terms linear in $\hat V_\mu$ would be formally identical to the expressions in Eqs.~\eqn{eq.resonanceProca-L},
\eqn{eq.Proca-chimunu} and \eqn{eq.Proca-chi}, with couplings $f_{\bar V}$, $g_{\bar V}$, $\widetilde f_{\bar V}$, etc.
Therefore, one would just reproduce the more general Proca Lagrangian with
$f_{\hat V} \not= 2\, g_{\hat V}$. The additional interaction vertices are no longer soft terms and would need to be corrected with another $\Delta\mL_{\text{non-R}}^{\rm (P)}$ term in order to guarantee a proper UV behaviour of Green functions with light SM fields. The final result would be identical to the Proca formalism discussed in previous sections.

Likewise, using the left and right connections in Eq.~\eqn{eq:connection}, it is possible to assign different transformation properties to the hidden gauge field.
For instance, a $SU(2)_L$ triplet gauge field was considered in Ref.~\cite{Contino}.

\section{Summary}
\label{sec.summary}

Direct searches for physics beyond the SM at the electroweak scale have been unsuccessful, pointing out the existence of a mass gap in the energy spectrum. The LHC is rising up the experimental sensitivity, but no clear hint for exotic phenomena has emerged so far, pushing the new physics frontier above the TeV. Unless a new discovery is made soon, EFT methods constitute for the time being the most efficient way to become sensitive to mass scales above the energy reach of present experimental facilities.

In this article, the EWET has been formulated as the most general EFT containing the SM symmetries and its low-energy
degrees of freedom. It includes the SM bosons and fermions embedded in the extended symmetry group
$\mG=SU(2)_L\otimes SU(2)_R\otimes U(1)_{  X  }$,
with $L$ and $R$ the left and right
 chiralities and $X=({\rm B}-{\rm L})/2$, given by
the conserved baryon and lepton numbers, respectively.
The Higgs is incorporated as a light scalar boson $h$, singlet under this group. Our only premise is the symmetry breaking pattern $SU(2)_L\otimes SU(2)_R \rightarrow SU(2)_{L+R}$, which has been confirmed  phenomenologically as the right dynamical framework for the electroweak Goldstone bosons.

The low-energy EWET operators are organized according to their infrared behaviour,
as an expansion in powers of derivatives over some higher energy scale. We have carefully analyzed the power counting of the EWET, introducing a more efficient assignment for the chiral dimension of custodial symmetry breaking operators that takes into account the phenomenological suppression of these effects. This allows for a sizeable reduction in the number of NLO structures that need to be handled. With a single fermion family, assuming
${\rm B}$ and ${\rm L}$ conservation and ignoring any QCD effects, the $CP$-invariant,
$\mO(p^4)$ effective Lagrangian only contains 11 (3)  $P$-even ($P$-odd) operators in the bosonic sector (Table \ref{tab:bosonic-Op4}), and  17 (5)  operators containing fermions (Table \ref{tab:fermion-ops}).

All accessible informations on heavier new-physics states are encoded in the LECs of the EWET operators, which parametrize any possible deviations from the SM predictions at low energies. We have explored the low-energy consequences of
generic heavy states with different quantum numbers, coupled to the SM particles, {\it i.e.}, the fingerprints they leave on the LECs. Similar studies have been done before for specific weakly-coupled models of new physics
\cite{Brehmer:2015rna,deBlas:2014mba,delAguila:2010mx,delAguila:2008pw,delAguila:2000rc,Corbett:2015lfa,Bar-Shalom:2014taa,Pappadopulo:2014qza}, within the much simpler linear framework with a SM doublet Higgs;
in some cases, even at the one-loop level
\cite{HLM:16,FRW:15,DEQ:15,Drozd:2015rsp,Huo:15,Huo:15b,dAKS:16,Boggia:2016asg,Fuentes-Martin:2016uol,BS:94}
in the usual perturbative expansion in powers of small couplings. However, the LECs of the generic non-linear EWET have remained largely unexplored until now \cite{Pich:2015kwa,pseudovector-Cata,Buchalla:2016bse}.

To simplify the discussion, we have focused on colour-singlet heavy bosons with $J^{PC}= 0^{++}, 0^{-+}, 1^{--}, 1^{++}$, assuming a $CP$-invariant underlying dynamics. In addition, we have considered a single SM fermion family, leaving for future works the more involved study of a non-trivial flavour structure. We have first built a general short-distance effective Lagrangian, involving the resonances and the SM fields, which incorporates the assumed pattern of EWSB and has the minimum possible number of derivatives.
We have also assumed that the resonance couplings to the light fields do not increase with the resonance mass; {\it i.e.}, we have assumed a decoupling behaviour as expected in strongly-coupled scenarios. Therefore, our generic results cannot be directly applied to renormalizable Higgsed models, which require a more specific treatment of $O(M_R^2)$ terms.

At $\mO(1/M_R^2)$, which is the accuracy needed to determine the $\mO(p^4)$ LECs, one only needs to consider operators with at most one heavy field. In order to compute the LECs, one must integrate out from the action the heavy fields and expand in powers of momenta the resulting non-local expression. Using the classical EoM of the massive states, all low-energy implications of tree-level resonance exchanges among the SM fields can be easily determined, and expressed as a sum of EWET operators multiplied by LECs with a structure $\sim g_1 g_2/M_R^2$, where $g_{1,2}$ are the specific short-distance resonance couplings contributing to a given operator.

While the analysis of spin-0 boson exchanges is straightforward, the spin-1 contributions to the EWET Lagrangian need a more careful treatment, since there exist several formalisms to describe massive spin-1 fields, and a naive evaluation of tree-level exchange amplitudes gives results which depend on the adopted representation. We have presented a very detailed study of this potential ambiguity, demonstrating the equivalence of the different formalisms, once a good UV behaviour is required.

The final predictions for the $\cO(p^4)$ LECs of the EWET Lagrangian, generated through the exchanges of colourless (triplet and singlet) spin-0 and spin-1 heavy particles, are compiled in Tables~\ref{tab:bLECs-final}, \ref{tab:2fLECs-final} and \ref{tab:4fLECs-final}. They contain the resonance contributions to bosonic, two-fermion and four-fermion operators, respectively.
Note that all ``couplings'' here must be understood as functions of $h/v$.
The values of the bosonic LECs in Table~\ref{tab:bLECs-final} agree with the results found previously
in Ref.~\cite{Pich:2015kwa}, which only considered $P$-even operators and exact custodial symmetry. The couplings which do not contain Higgs fields are also in agreement with those found in QCD through the large-$N_C$ matching of Resonance Chiral Theory
and $\chi$PT~\cite{Ecker:1988te,Ecker:1989yg}
($\widetilde{F}_V=\widetilde{F}_A=\widetilde{G}_A = \mT =0$ in QCD).

\begin{table}[tb]  
{\renewcommand{\arraystretch}{2.1}
\begin{center}
\begin{tabular}{|c||c|c|}
\hline
$i$ &    $\mF_i$  &   $\widetilde{\mF}_i$  \\  \hline \hline
1 &  $- \Frac{F_V^2-\widetilde{F}_V^2}{4M_V^2}
+ \Frac{F_A^2-\widetilde{F}_A^2}{4M_A^2} $
&
 $- \Frac{\widetilde{F}_VG_V}{2M_V^2}
- \Frac{F_A\widetilde{G}_A}{2M_A^2}$
\\ \hline
2 &
 $- \Frac{F_V^2+{\widetilde{F}_V}^2}{8M_V^2}
- \Frac{F_A^2+{\widetilde{F}_A}^2}{8M_A^2}$
&
 $- \Frac{F_V \widetilde{F}_V}{4M_V^2}
- \Frac{F_A \widetilde{F}_A}{4M_A^2}$
\\ \hline
3 &
$-  \Frac{F_VG_V}{2M_V^2} - \Frac{\widetilde{F}_A\widetilde{G}_A}{2M_A^2}$
&
 $- \Frac{F_V \widetilde{\lambda}_1^{hV} v}{M_V^2}
 - \Frac{\widetilde{F}_A \lambda_1^{hA} v}{M_A^2}$
\\ \hline
4 &
 $\Frac{G_V^2}{4M_V^2} + \Frac{{\widetilde{G}_A}^2}{4M_A^2} $
& ---
\\ \hline
5 &
$        \Frac{c_{d}^2}{4M_{S_1}^2}
-\Frac{G_V^2}{4M_V^2} - \Frac{{\widetilde{G}_A}^2}{4M_A^2} $
& ---
\\ \hline
6 &
 $ - \Frac{\widetilde{\lambda}_1^{hV\,\, 2}v^2}{M_V^2}
- \Frac{\lambda_1^{hA\,\, 2}v^2}{M_A^2}$
& ---
\\ \hline
7 & $        \Frac{ d_P^2}{2 M_P^2}
+ \Frac{\lambda_1^{hA\,\, 2}v^2}{M_A^2}
+  \Frac{\widetilde{\lambda}_1^{hV\,\, 2}v^2}{M_V^2}$
& ---
\\ \hline
8 & 0 & ---
\\ \hline
9 &  $  - \Frac{F_A \lambda_1^{hA} v}{M_A^2}
- \Frac{\widetilde{F}_V \widetilde{\lambda}_1^{hV} v}{M_V^2}$
& ---
\\ \hline
10 &  $-\displaystyle\frac{(\widetilde{c}_{\mathcal{T}}^{\hat{V}_1})^2}{2M_{V_1}^2}-\displaystyle\frac{(c_{\mathcal{T}}^{\hat{A}_1})^2}{2M_{A_1}^2}$ & ---
\\ \hline
11 &  $- \Frac{F_{V_1}^2}{M_{V_1}^2} - \Frac{\widetilde{F}_{A_1}^2}{M_{A_1}^2} $  & ---
\\ \hline
\end{tabular}
\end{center}
}
\caption{\small
Final predictions for the massive resonance contributions to the bosonic $\cO(p^4)$ LECs of the EWET Lagrangian.  }
\label{tab:bLECs-final}
\end{table}

\begin{table}[!t]
\begin{center}
\renewcommand{\arraystretch}{2.1}
\begin{tabular}{|c||c|c|}
\hline
$i$ &   $\mF^{\psi^2}_i$   &   $\widetilde{\mF}^{\psi^2}_i$
\\  \hline\hline
1
& $\Frac{c_d c^{S_1}_1}{2 M_{S_1}^2}$
& $-\displaystyle\frac{\widetilde{F}_V C_0^V}{\sqrt{2}M_V^2}\! -\! \displaystyle\frac{F_A \widetilde{C}_0^A}{\sqrt{2}M_A^2} $
\\[1ex] \hline
2
&  $-\displaystyle\frac{G_V C_0^V}{\sqrt{2}M_V^2} \!-\! \displaystyle\frac{\widetilde{G}_A \widetilde{C}_0^A}{\sqrt{2}M_A^2} $
&\! \!\!$-\displaystyle\frac{2\sqrt{2}v\widetilde{\lambda}_1^{hV}C_0^V}{M_V^2} \!-\! \frac{2\sqrt{2}v\lambda_1^{hA}\widetilde{C}_0^A}{M_A^2}$
\\[1ex] \hline
3
& $-\displaystyle\frac{F_V C_0^V}{\sqrt{2}M_V^2} \!-\! \displaystyle\frac{\widetilde{F}_A \widetilde{C}_0^A}{\sqrt{2}M_A^2} $
& $-\displaystyle\frac{\widetilde{c}_{\mathcal{T}}^{\hat{V}_1} c_{1}^{\hat{V}_1}    }{\sqrt{2} M_{V_1}^2} - \displaystyle\frac{c_{\mathcal{T}}^{\hat{A}_1} \widetilde{c}_{1}^{\hat{A}_1}    }{\sqrt{2} M_{A_1}^2}$
\\[1ex] \hline
4 & $-\displaystyle\frac{\sqrt{2}F_{V_1} C_0^{V_1}}{M_{V_1}^2} \!- \!\displaystyle\frac{\sqrt{2}\widetilde{F}_{A_1} \widetilde{C}_0^{A_1}}{M_{A_1}^2} $ & ---
\\[1ex] \hline
5 & $\Frac{d_P c^{P}_1}{M_{P}^2}$ & ---
\\[1ex] \hline
6 & $-\displaystyle\frac{\widetilde{c}_{\mathcal{T}}^{\hat{V}_1} \widetilde{c}_{1}^{\hat{V}_1}    }{\sqrt{2} M_{V_1}^2} - \displaystyle\frac{c_{\mathcal{T}}^{\hat{A}_1} c_{1}^{\hat{A}_1}    }{\sqrt{2} M_{A_1}^2}$
& ---
\\[1ex] \hline
7 & $0$ & ---
\\[1ex] \hline
\end{tabular}
\end{center}
\caption{\small
Final predictions for the massive resonance contributions to the two-fermion $\cO(p^4)$ LECs of the EWET Lagrangian. }
 \label{tab:2fLECs-final}
\end{table}

\begin{table}[!t]
\begin{center}
\renewcommand{\arraystretch}{2.1}
\begin{tabular}{|c||c|c|}
\hline
$i$
&  $\mF_i^{\psi^4}$
&  $\widetilde{\mF}_i^{\psi^4}$
\\ \hline\hline
1
&
$\displaystyle\frac{(c_1^S)^2}{2M_S^2} $
&
$ -
\displaystyle\frac{c_1^{\hat{V}} \widetilde{c}_1^{\hat{V}} }{M_V^2}
-\frac{ c_1^{\hat{A}}\widetilde{c}_1^{\hat{A}}}{M_A^2}$
\\ \hline
2
&
$\displaystyle\frac{(c_1^P)^2}{2M_P^2}$
&
$ \displaystyle\frac{c_1^{\hat{V}}\widetilde{c}_1^{\hat{V}}}{2M_V^2}
+\frac{c_1^{\hat{A}}\widetilde{c}_1^{\hat{A}}}{2M_A^2}
   -\frac{c_1^{\hat{V}_1}\widetilde{c}_1^{\hat{V}_1}}{2M_{V_1}^2} -\frac{c_1^{\hat{A}_1}\widetilde{c}_1^{\hat{A}_1}}{2M_{A_1}^2}
$
\\ \hline
3 & $-\displaystyle\frac{(c_1^S)^2}{4M_S^2}+\displaystyle\frac{(c_1^{S_1})^2}{4M_{S_1}^2}$  & ---
\\ \hline
4 & $-\displaystyle\frac{(c_1^P)^2}{4M_P^2}+\displaystyle\frac{(c_1^{P_1})^2}{4M_{P_1}^2}$  & ---
\\ \hline
5
& $ -\displaystyle\frac{(c_1^{\hat{V}})^2}{2M_V^2}
-\frac{(\widetilde{c}_1^{\hat{A}})^2}{2M_A^2}$
& ---
\\ \hline
6
& $- \displaystyle\frac{(\widetilde{c}_1^{\hat{V}})^2}{2M_V^2}
-\frac{({c}_1^{\hat{A}})^2}{2M_A^2} $
& ---
\\ \hline
7
& $\displaystyle\frac{({c}_1^{\hat{V}})^2}{4M_V^2}
+\frac{(\widetilde{c}_1^{\hat{A}})^2}{4M_A^2}
   -\frac{({c}_1^{\hat{V}_1})^2}{4M_{V_1}^2}  -\frac{(\widetilde{c}_1^{\hat{A}_1})^2}{4M_{A_1}^2} $
& ---
\\ \hline
8
& $ \displaystyle\frac{(\widetilde{c}_1^{\hat{V}})^2}{4M_V^2}
+\frac{({c}_1^{\hat{A}})^2}{4M_A^2}
   -\frac{(\widetilde{c}_1^{\hat{V}_1})^2}{4M_{V_1}^2}  -\frac{({c}_1^{\hat{A}_1})^2}{4M_{A_1}^2} $
& ---
\\ \hline
9 &
$-\displaystyle\frac{(C_0^V)^2}{M_V^2}-\displaystyle\frac{(\widetilde{C}_0^A)^2}{M_A^2}$
& ---
\\ \hline
10 &
$\displaystyle\frac{(C_0^V)^2}{2M_V^2}\!-\!\displaystyle\frac{(C_0^{V_1})^2}{2M_{V_1}^2}
\!+\!\displaystyle\frac{(\widetilde{C}_0^A)^2}{2M_A^2}\!-\!\displaystyle\frac{(\widetilde{C}_0^{A_1})^2}{2M_{A_1}^2}$
& ---
\\ \hline
\end{tabular}
\caption{\small
Final predictions for the massive resonance contributions to the four-fermion $\cO(p^4)$ LECs of the EWET Lagrangian. }
\label{tab:4fLECs-final}
\end{center}
\end{table}

The tree-level resonance exchanges that we have analyzed contribute to all $\mO(p^4)$ operators,
except $\mO_8$, which only contains Higgs fields,
and $\mO^{\psi^2}_7$, which also contains a fermion bilinear.
While most of the LECs receive contributions from vector and axial-vector resonances, the exchange of heavy spin-0 particles only manifests in a few $P$-even LECs. A triplet scalar leaves its fingerprints on the four-fermion operators $\mO_1^{\psi^4}$ and $\mO_3^{\psi^4}$, a singlet scalar shows up in $\mO_5$, $\mO_1^{\psi^2}$ and $\mO_3^{\psi^4}$, a triplet pseudoscalar contributes to
$\mO_7$, $\mO_5^{\psi^2}$, $\mO_2^{\psi^4}$ and $\mO_4^{\psi^4}$, while a
singlet pseudoscalar can only be spotted through $\mO_4^{\psi^4}$. Obviously, if there exist several heavy states with the same $J^{PC}$ quantum numbers, each of them will give separate contributions to the LECs as indicated in the tables (appropriate sums over similar resonance states must then be understood, whenever needed).

If any anomalous (non SM) behaviour is observed in the data, the identification of its physical origin will require a detailed phenomenological study of the fitted LECs. The pattern of non-zero LECs should allow to infer the quantum numbers of the underlying dynamics. From our results, it is possible to extract a few interesting features:
\begin{enumerate}
\item A non-zero $P$-odd LEC indicates a spin-1 particle with both $P$-odd and $P$-even couplings.
\item A non-zero value of any of the LECs
$\mF_{1\text{-}4,6,9\text{-}11}$, $\mF^{\psi^2}_{2\text{-}4,6}$ and $\mF^{\psi^4}_{5\text{-}10}$ indicates spin 1.
\item A non-zero value for $\mF^{\psi^2}_1$ ($\mF^{\psi^4}_1$) signals a singlet (triplet) scalar.
\item A non-zero value for $\mF_5^{\psi^2}$ or $\mF^{\psi^4}_2$ is a signal of a triplet pseudoscalar.
\item $\mF^{\psi^4}_3$ ($\mF^{\psi^4}_4$) indicates a scalar (pseudoscalar) boson.
\item The custodial-breaking LEC $\mF^{\psi^2}_6$ ($\widetilde\mF^{\psi^2}_3$) manifests a singlet $P$-odd (even) vector  or $P$-even (odd) axial-vector coupling preserving custodial symmetry, combined with a custodial-breaking
$P$-odd (odd) vector  or $P$-even (even) axial-vector coupling.
\item A non-zero value of $\mF_4 + \mF_5$ ($\mF_6 + \mF_7$) indicates a singlet scalar (triplet pseudoscalar).
\item A non-zero value $\mF_{10}$ ($\mF_{11}$) indicates a singlet $P$-odd (even) vector or $P$-even (odd) axial-vector coupling.
\item $\mF^{\psi^4}_{5,9}$ ($\mF^{\psi^4}_6$) manifest a triplet $P$-even (odd) vector  or $P$-odd (even) axial-vector coupling.
\item $\widetilde\mF_{1\text{-}3}$, $\widetilde\mF^{\psi^2}_{1,2}$ and $\widetilde\mF^{\psi^4}_1$ signal a triplet spin-1 particle.
\item A non-zero value of $\mF^{\psi^4}_1 + 2\,\mF^{\psi^4}_3$  ($\mF^{\psi^4}_2 + 2\,\mF^{\psi^4}_4$) indicates a singlet scalar (pseudoscalar).
\end{enumerate}

There could be, in addition, other contributions not included
in the generic scenario that we have studied. Obvious extensions
of this analysis, to be investigated in future works, include spin-2 bosons,
coloured heavy states and massive fermions. A first necessary step is
to complete our minimal basis of EWET operators with the additional terms
involving QCD structures. The study of the flavour dynamics within
the EWET framework is a more challenging enterprise that we plan also to address.

When deriving these results, we have only required very mild UV conditions on the spin-1 fields which should be fulfilled in any sensible dynamical framework. As shown in Ref.~\cite{Pich:2015kwa}, additional constraints can be obtained, imposing stronger short-distance conditions on specific Green functions. In this way, one can get relations among different resonance couplings, which are valid in broad classes of underlying dynamical theories.
For instance, in the absence of $P$--odd couplings, requiring the two Weinberg sum rules (WSR)~\cite{Weinberg:1967kj} to be valid
for the $W^3B$ correlator
(they are fulfilled in asymptotically free theories~\cite{Bernard:1975cd}) leads
to a more predictive tree-level result for the oblique $S$ parameter
and its relevant LEC~\cite{Pich:2015kwa,Peskin:90,Peskin:92},
$\mF_1[0]= - v^2 (M_V^{-2}+M_A^{-2})/4$.
Comparing the experimental bounds on the $S$ parameter~\cite{Baak:2012kk,Baak:2013fwa}
with the one-loop resonance calculation~\cite{Pich:2012dv,Pich:2013fea}, one then obtains
the determination of $\mF_1[0]$ in terms of $M_V$ shown in Fig.~\ref{fig:F1}~\cite{Pich:2015kwa}.
One can also derive positivity constraints, based on generic properties such as unitarity, analyticity and crossing, which get translated into restrictions on the LECs~\cite{Pham:1985cr,Comellas:1995hq,Adams:2006sv}. A well-known example are the LECs involved in the Goldstone scattering amplitudes, which must obey the relations
$\mF_4>0$ and $\mF_4+\mF_5>0$~\cite{Filipuzzi:2012bv,Pham:1985cr,Pennington:1994kc,Manohar:2008tc} that are of course satisfied by our predictions in Table~\ref{tab:bLECs-final}.
The study of these additional high-energy conditions and their phenomenological implications is beyond the scope of the present analysis and will be pursued in future works.

\begin{figure}
\begin{center}
\includegraphics[width=7.5cm]{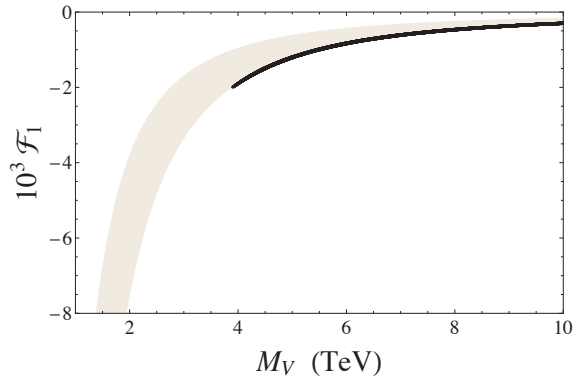}
\caption{{\small
Determination of the $\mO(p^4)$ LEC $\mF_1[0]$
in asymptotically-free theories, as function of $M_V$~\cite{Pich:2015kwa}.
The light-shaded region shows the two-WSR prediction
for $M_A>M_V$~\cite{Pich:2013fea,Pich:2015kwa}.
The experimental bounds on the $S$ parameter~\cite{Baak:2012kk,Baak:2013fwa}
restrict the allowed region to the black narrow area.
}}
\label{fig:F1}
\end{center}
\end{figure}

At present, the experimental information on the LECs is rather scarce.
$\mF_1$ is the most constrained one, since it contributes
at tree level to the oblique S parameter.
The bosonic LECs $\mF_{1\text{-}5}$ and $\widetilde \mF_{1,2}$ account for anomalous gauge couplings. The quartic gauge couplings $\mF_{4,5}$ are expected to be significantly bounded by forthcoming
run-II data at the LHC and its future high-luminosity upgrade. The Higgs-related LECs $\mF_{6\text{-}9}$ and $\widetilde \mF_3$ are still poorly constrained or unbounded. In the fermion sector, the constraints on (flavour-conserving) pure vector and axial-vector structures are probably similar to the ones derived within the more studied linear realization of the electroweak EFT, while the scalar and pseudoscalar cases require, however, a careful investigation. A global phenomenological analysis of the EWET LECs, including flavour constraints, is a necessary and highly non-trivial task to be addressed in future works.

\section*{Acknowledgements}
We thank Claudius Krause for his useful comments on the manuscript. This work has been
supported by the Spanish Government and ERDF funds from the European Commission 
(FPA2013-44773-P, FPA2014-53631-C2-1-P, 
FPA2016-75654-C2-1-P);   
by the Spanish Centro de Excelencia Severo Ochoa Programme (SEV-2012-0249, SEV-2014-0398); the Generalitat Valenciana (PrometeoII/2013/007);
by the Universidad CEU Cardenal Herrera and Banco Santander (PRCEU-UCH CON-15/03, INDI15/08);
and La Caixa (Ph.D. grant for Spanish universities).

\appendix

\section{Transformation properties of chiral structures under discrete symmetries}
\label{app:discrete-transf}

In this appendix we compile some useful transformation properties of the different chiral structures defined in the paper.
Table~\ref{tab:U-trans} shows how the basic Goldstone tensors transform under parity ($P$), charge conjugation ($C$), $CP$ and Hermitian conjugation. The analogous transformation properties of the fermion bilinears are given in Table~\ref{tab:J-trans}, while Table~\ref{tab:gamma-trans} exhibits the Dirac algebra entering into play for each of these transformations.
Finally, Table~\ref{tab:R-trans} shows the transformation properties of the different massive multiplets considered in this paper.
When building invariant operators, we have assumed that the custodial symmetry-breaking spurion $\mT$ transforms like a scalar $S$.

\begin{table}[ht]
\begin{center}
\begin{tabular}{ |c||c|c|c|c| }
\hline
\rule{0pt}{3ex}
 & $P$    &     $C$ & $CP$ & h.c.
\\[5pt] \hline\hline
\rule{0pt}{3ex}
$U$ & $U^\dagger$  & $U^t$ & $U^*$ & $U^\dagger$
\\[5pt] \hline
\rule{0pt}{3ex}
$u$ & $u^\dagger$  & $u^t$ & $u^*$ & $u^\dagger$
\\[5pt] \hline
\rule{0pt}{3ex}
$u^\mu$ & $- u_\mu$ &  $u^{\mu\, t}$ & $-u_{\mu}^{t}$ & $u^\mu$
\\[5pt] \hline
\rule{0pt}{3ex}
$f_\pm^{\mu\nu} $ & $\pm f_{\pm\, \mu\nu}$ &  $ \mp f_{\pm}^{\mu\nu\,  t} $ &   $- f_{\pm\, \mu\nu}^t $  & $f_\pm^{\mu\nu}$
\\[5pt] \hline
\end{tabular}
\caption{\small Transformation properties of the Goldstone tensors. The superindex $t$ denotes matrix transposition.
}
\label{tab:U-trans}
\end{center}
\end{table}

\begin{table}[ht]
\begin{center}
\begin{tabular}{ |c||c|c|c|c| }
\hline
\rule{0pt}{3ex}
 & $P$    &  $C$ & $CP$ & h.c.
\\[5pt] \hline\hline
\rule{0pt}{3ex}
$J_S$ & $J_S$  & $(J_S)^t$ & $(J_S)^t$ & $J_S$
\\[5pt] \hline
\rule{0pt}{3ex}
$J_P$ & $- J_P$ &  $(J_P)^t$ & $-(J_P)^t$ & $J_P$
\\[5pt] \hline
\rule{0pt}{3ex}
$J_V^{\mu}$ & $J_{V\, \mu}$ &  $-J_V^{\mu \, t}$ & $-J_{V\, \mu}^{t}$ & $J_V^{\mu}$
\\[5pt] \hline
\rule{0pt}{3ex}
$J_A^{\mu}$ & $- J_{A\, \mu }$ &  $J_A^{\mu \, t}$ & $-J_{A\, \mu}^{t}$ & $J_A^{\mu}$
\\[5pt] \hline
\rule{0pt}{3ex}
$J_T^{\mu\nu}$ & $J_{T\, \mu\nu }$ &  $-J_T^{\mu\nu \, t}$ & $-J_{T\, \mu\nu}^{t}$ & $J_T^{\mu\nu}$
\\[5pt] \hline
\end{tabular}
\caption{\small
Transformation properties of the fermionic bilinears
$(J_S)_{mn}=\bar{\xi}_n\xi_m$, $(J_P)_{mn}=i\bar{\xi}_n\gamma_5\xi_m$,
$(J_V^\mu)_{mn}=\bar{\xi}_n\gamma^\mu \xi_m$,
$(J_A^\mu)_{mn}=\bar{\xi}_n\gamma^\mu\gamma_5\xi_m$ and
$(J_T^\mu)_{mn}=\bar{\xi}_n\sigma^{\mu\nu}\xi_m$.}
\label{tab:J-trans}
\end{center}
\end{table}

\begin{table}[ht]
\begin{center}
\begin{tabular}{ |c||c|c|c|c| }
\hline \hline
\rule{0pt}{3ex}
$\Gamma$ & $P$ algebra    &  $C$ algebra & $CP$ algebra & h.c. algebra
\\[5pt]
\rule{0pt}{3ex}
 &  ( $\gamma^0\Gamma \gamma^0$ )   &     ( $-\gamma^0\gamma^2\Gamma^t \gamma^2\gamma^0$ ) &
  ( $-\gamma^2\Gamma^t \gamma^2$ )  & ( $\gamma^0\Gamma^\dagger \gamma^0$ )
\\[5pt] \hline\hline
\rule{0pt}{3ex}
1 & 1  & 1 & 1 & 1
\\[5pt] \hline
\rule{0pt}{3ex}
$i\gamma_5$  & $- i\gamma_5$ &  $i\gamma_5$ & $-i\gamma_5$ & $i\gamma_5$
\\[5pt] \hline
\rule{0pt}{3ex}
$\gamma^{\mu}$ & $\gamma_{\mu}$ &  $-\gamma^{\mu }$ & $-\gamma_\mu $ & $\gamma^{\mu}$
\\[5pt] \hline
\rule{0pt}{3ex}
$\gamma^{\mu}\gamma_5$ & $- \gamma_{\mu }\gamma_5$ &  $\gamma^{\mu}\gamma_5$ & $-\gamma_{\mu}\gamma_5$
& $\gamma^{\mu}\gamma_5$
\\[5pt] \hline
\rule{0pt}{3ex}
$\sigma^{\mu\nu}$ & $\sigma_{\mu\nu }$ &  $-\sigma^{\mu\nu}$ & $-\sigma_{\mu\nu} $ & $\sigma^{\mu\nu}$
\\[5pt] \hline
\end{tabular}
\caption{\small
Related Dirac algebra for transformation properties of the fermionic bilinears.
}
\label{tab:gamma-trans}
\end{center}
\end{table}

\begin{table}[!t]
\begin{center}
\begin{tabular}{ |c||c|c|c|c| }
\hline
\rule{0pt}{3ex}
 & $P$    & $C$ & $CP$ & h.c.
\\[5pt] \hline\hline
\rule{0pt}{3ex}
$S$ & $S$  & $S^t$ & $S^t$ & $S$
\\[5pt] \hline
\rule{0pt}{3ex}
$P$ & $- P$ &  $P^t$ & $-P^t$ & $P$
\\[5pt] \hline
\rule{0pt}{3ex}
$V^{\mu\nu}$ & $V_{\mu\nu}$ &  $-V^{\mu\nu\, t}$ & $-V_{\mu\nu}^{t}$ & $V^{\mu\nu}$
\\[5pt] \hline
\rule{0pt}{3ex}
$A^{\mu\nu}$ & $- A_{\mu\nu}$ &  $A^{\mu\nu\, t}$ & $-A_{\mu\nu}^{t}$ & $A^{\mu\nu}$
\\[5pt] \hline
\end{tabular}
\caption{\small
Transformation properties of $J^{PC}= 0^{++}$ ($S$), $0^{-+}$ ($P$), $1^{--}$ ($V$) and $1^{++}$ ($A$) multiplets~\cite{Ecker:1988te,Ecker:1989yg}.
The transposition operation $t$ is absent for singlet resonances.
}
\label{tab:R-trans}
\end{center}
\end{table}

\section{Lagrangian simplifications}
\label{app:simplifications}

Many redundant operators can be eliminated from the effective Lagrangian by using partial integration, field redefinitions, the classical
EoM or algebraic identities~\cite{Ecker:1988te,Bijnens:1999sh,odd-Op6-chpt}. We provide next a few illustrative examples.

The kinetic derivative term of the Higgs in Eq.~\eqn{eq:L2} can be multiplied with an arbitrary function $\mF_h(h/v)$; {\it i.e.}, an operator of the form

\bel{eq:L2Fh}
\widetilde\mL_2\, =\, \frac{1}{2}\,\mF_h(h/v)\;
\partial_\mu h\,\partial^\mu h\, ,
\qquad\qquad\qquad
\mF_h(h/v)\, = \, 1\, +\, \sum_{n=1} c^{(h)}_n
\left(\frac{h}{v}\right)^n\, ,
\ee
satisfies all symmetry requirements. However, the function $\mF_h(h/v)$ can be eliminated through a non-linear redefinition of the Higgs field:
\bel{eq:h-redefinition}
h'(x) \, =\, v\; \sum_{n=1} a_n\left(\frac{h}{v}\right)^n\, ,
\qquad\qquad\qquad a_1 = 1\, .
\ee
Imposing that Eq.~\eqn{eq:L2Fh} reduces to the canonic kinetic term, $\widetilde\mL_2 = \frac{1}{2}\,\partial_\mu h' \partial^\mu h'$, determines the coefficients $a_n$ through the iterative relations
\bel{eq:a_n}
c^{(h)}_n\, =\,\sum_{k=1}^{n+1}\, k\, (n+2-k)\, a_k\, a_{n+2-k}\, .
\ee

The massive singlet scalar $S_1$ could couple to the Higgs through terms of the form
\bel{eq:S-h_mixing}
\Delta\mL_{S_1h}\, =\, a\; S_1\, h \, +\, b\;\partial_\mu S_1\,\partial_\mu h \, +\, S_1\,\partial_\mu h \,\partial^\mu h \;\sum_{n=0} c_n\,\left(\frac{h}{v}\right)^n\, .
\ee
The couplings $a$ and $b$ would generate a mixing between $S_1$ and $h$; they can be eliminated through a proper redefinition of both scalar fields and their masses. The $c_n$ operators can be written through partial integration in the form:
\bel{eq:O_n}
O_n\, \equiv\, S_1\, h^n\,\partial_\mu h \,\partial^\mu h
\; =\; \frac{1}{(n+1)(n+2)}\, h^{n+2}\,\Box S_1\, -\,
\frac{1}{(n+1)}\, h^{n+1}\, S_1\,\Box h\, .
\ee
Applying the $S_1$ and $h$ EoM on the rhs, $O_n$ can be expressed in terms of other operators included in the effective Lagrangian.

In general, any coupling of the form $\bra \partial_\mu R\,\chi^\mu\ket$
can be written through partial integration as
$-\bra R\,\partial_\mu \chi^\mu\ket$. Therefore, the interaction terms in Eqs.~\eqn{eq.resonance-LS} do not include operators with derivatives of the heavy states.

When using the Proca description of vector and axial-vector fields,
the effective Lagrangians $\mL_{\hat R}$ ($R=V,A$) in Eq.~\eqn{eq.resonanceProca-L}
could also include the $\cO (p)$ operators $\bra \hat V^\mu u_\mu \ket$
and $\bra \hat A^\mu u_\mu \ket$ ($P$-odd and $P$-even, respectively). This type of operators lead to
$\hat A^\mu$-$\varphi$ mixing terms between the spin-0 components of the axial-vector Proca fields and the Goldstones. These operators can be removed from the action by means of the field redefinitions
\be
\hat R^\mu \quad\rightarrow\quad \hat R'^\mu\, =\,\hat R^\mu + \alpha_{\hat R}\; u^\mu
\qquad\qquad\qquad (R=V,A)\, ,
\ee
with $\hat{R}'_{\mu\nu} = \hat{R}_{\mu\nu}  - \alpha_{\hat R}\, f_{-\,\mu\nu}$.
Tuning $\alpha_{\hat R}$ conveniently, one can remove the undesired terms
while keeping the same formal structures in the Lagrangian~(\ref{eq.resonanceProca-L}). These redefinitions are not needed in the antisymmetric formalism
because the tensor field representation does not allow for these $\cO (p)$ operators.

The $\mO(p)$ operator $\bra S \mT\ket$, involving the custodial symmetry breaking spurion $\mT$,
could also be present in the triplet scalar Lagrangian in Eq.~\eqn{eq.resonance-LS}.
Taking the appropriate value of $\alpha_S$, the scalar field redefinition $S = S' - \alpha_S\,\mT$
allows one to trade this operator by the $\mO(p^3)$ structure $\langle \nabla_\mu S\,\nabla^\mu\mT\ket$.

\subsection{$\mO(p^3)$ fermionic operators in the EWET}
\label{app:p3Foperators}

If present, the $O(p)$ operators $\bra S \,\mT\ket$, $\bra \hat V^\mu u_\mu \ket$ and $\bra \hat A^\mu u_\mu \ket$ would generate resonance-exchange contributions to the LECs of the $O(p^3)$ fermionic Lagrangian
\bel{eq:L3F}
\mL_3^{\mathrm{Fermionic}}\; =\; \beta_S\,\bra\mT J_S\ket
+ \beta_V\,\bra u_\mu J_V^{  \mu  }  \ket + \beta_A\,\bra u_\mu J_A^{  \mu  }  \ket\, .
\ee
Since we have just seen that the three $O(p)$ operators can be eliminated from the resonance effective theory through appropriate redefinitions of the heavy $S$, $V_\mu$ and $A_\mu$ fields,
one could wonder whether there are corresponding field transformations in the low-energy EWET that remove the $O(p^3)$ fermionic operators in \eqn{eq:L3F}.

The scalar $\beta_S$ term can be easily reabsorbed into the following redefinition of the LO Yukawa coupling in Eq.~\eqn{eq:Yukawas},
\be
\mY\; =\; \mY' + \frac{\beta_S}{v}\,\mT\, ,
\ee
{\it i.e.},
\be
-v\,\left(\bar\xi_L\,\mY\,\xi_R + \mathrm{h.c.}\right) + \beta_S\,\bra\mT J_S\ket
\; =\; -v\,\left(\bar\xi_L\,\mY'\,\xi_R + \mathrm{h.c.}\right)\, .
\ee

Similarly, redefining the auxiliary gauge sources through
\bear\label{eq:WBtransfP3}
\hat W_\mu\, =\; \hat W'_\mu -
(\beta_V - \beta_A)
\: u u_\mu u^\dagger\, ,
\qquad\qquad\quad
\hat B_\mu\, =\; \hat B'_\mu -
(\beta_V + \beta_A)
\: u^\dagger u_\mu u\, ,
\eear
one can reabsorb the $\beta_V$ and $\beta_A$ terms into the kinetic fermion Lagrangian:
\be
i\, \bar\xi\gamma^\mu d_\mu\xi
+ \beta_V\,\bra u_\mu J_V^{  \mu  }  \ket
+ \beta_A\,\bra u_\mu J_A^{  \mu  }  \ket
\; =\; i\, \bar\xi\gamma^\mu d'_\mu\xi\,  .
\ee
This redefinition, when applied to the Yang-Mills Lagrangian
$\mL_{\mathrm{YM}}$,
generates contributions to some of the $\mO(p^4)$ operators $\mO_i$
in Table~\ref{tab:bosonic-Op4}, suppressed by factors of $\beta_V^n$
or $\beta_A^n$ with $1\le n\le 4$.
The axial part of the transformation \eqn{eq:WBtransfP3} implies in addition
$u_\mu = u'_\mu/(1+2\beta_A)$,
where the prime refers to the $\hat W'_\mu$ and
$\hat B'_\mu$ fields hidden in the covariant derivative within $u'_\mu$.
This is an $\mO(p)$ effect (the coupling $\beta_A$ carries the additional
chiral suppression assigned to the fermion bilineal $J_A$),
which propagates to the LO Goldstone term:
\be
\frac{v^2}{4}\;\mF_u(h/v)\;\bra u_\mu u^\mu\ket\; =\;
\frac{v^2}{4\, (1+2\beta_A)^2}\;\mF_u(h/v)\;\bra u'_\mu u'^{\,\mu}\ket\; =\;
\frac{v'^{\, 2}}{4}\;\mF'_u(h/v')\;\bra u'_\mu u'^{\, \mu}\ket\, .
\ee
One gets finally a formally identical Goldstone Lagrangian with the redefinitions
\be
v'\, =\, \frac{v}{1+2\beta_A}\, ,\qquad\qquad
\varphi'\, =\, \frac{\varphi}{1+2\beta_A}\, , \qquad\qquad
c'^{\, (u)}_n\, =\, \frac{c_n^{(u)}}{(1+2\beta_A)^n}\, ,
\ee
where $c'^{\, (u)}_n$ are the expansion coefficients of $\mF'_u(h/v')$ in powers of $h/v'$,
defined in Eq.~\eqn{eq:Fhu_V}. The Goldstone fields $\varphi$
have been rescaled to compensate the factor
that arises from relabeling $v$ in such a way that $u(\varphi/v)=u(\varphi'/v')$.

\subsection{Algebraic identities}
\label{app:algebra}

To reduce the number of EWET operators we have used the following $SU(2)$ algebraic identity ($x=x^j\sigma^j$; $x=a, b, c, d$)
\bel{eq:SU2trace}
2\,\bra a b c d \ket
\; =\; \bra a b \ket\, \bra c d \ket \, -\,
\bra a c \ket\, \bra b d \ket \, +\,\bra a d \ket\, \bra b c \ket \, .
\ee
Some single-trace operators have been simplified thanks to the Cayley-Hamilton
relation for $2\times 2$ matrices,
\bel{eq:CayleyHamilton}
a^2\, -\, a\, \bra a \ket \, +\, \frac{1}{2} \left( \bra a \ket^2 - \bra a^2 \ket \right)\,=\, 0\, ,
\ee
which implies
\bel{eq:CayleyHamilton2}
\left\{ a , b \right\}\; =\; a\,\bra b \ket\, +\, \bra a \ket\, b\, +\, \bra a b \ket\, - \,
\bra a \ket\,\bra b \ket\, .
\ee
{}From \eqn{eq:CayleyHamilton2}, one easily derives the useful equality
\bear
\bra \{a, b\}\, c \ket  &=& \bra a \ket \, \bra b c\ket\, +\,
\bra b \ket \, \bra a c\ket\, +\,\bra c \ket \, \bra a b\ket \, -\, \bra a\ket \, \bra b\ket \, \bra c\ket\, .
\eear
In particular, if $\bra b\ket=\bra c\ket =0$ one has
\bear
\bra \{a, b\}\, c\ket  &=& \bra a \ket \, \bra b c\ket \, .
\eear
Thus, the traceless condition $\bra S\ket =\bra u^\mu \ket =0$ implies
\bear
\bra S u_\mu u^\mu \ket &=& 0\, ,
\eear
being this $U(N)$ Resonance Chiral Theory
operator~\cite{Ecker:1988te}  absent in $SU(2)$.
Likewise, in the case of fermionic operators we have the Cayley-Hamilton relations:
\bear
\bra S\, \{ J_A^\mu , u_\mu\}\ket = \bra S u_\mu \ket \,\, \bra J_A^\mu\ket\, ,
\qquad\qquad
\bra S\, \{ J_V^\mu , u_\mu\}\ket = \bra S u_\mu \ket \,\, \bra J_V^\mu\ket\, .
\eear

For the odd-intrinsic parity sector with the Levi-Civita tensor,
one can make use of the Schouten identity~\cite{Schouten}:
\be
A^\rho\, \epsilon^{\mu\nu\alpha\beta}
\; =\;  A^\mu\,  \epsilon^{\rho\nu\alpha\beta}
+ A^\nu\,  \epsilon^{\mu\rho\alpha\beta}
+ A^\alpha\,  \epsilon^{\mu\nu\rho\beta}
+ A^\beta\,  \epsilon^{\mu\nu\alpha\rho}  \, .
\ee

The basic Goldstone tensors satisfy the following useful relations~\cite{Ecker:1988te,Bijnens:1999sh,odd-Op6-chpt}:
\begin{gather}
\nabla^\nu u^\mu - \nabla^\mu u^\nu \; =\; f_-^{\mu \nu} \, ,
\nn\\
\left[ \nabla_\mu ,\nabla_\nu  \right] X \; =\;  [\Gamma_{\mu\nu} , X]\, ,
\qquad\qquad
\Gamma_{\mu\nu} \; =\; \frac{1}{4} [u_\mu ,u_\nu ] - \frac{i}{2} f_{+\, \mu\nu}\, ,
\nn\\
\nabla_\rho \nabla_\mu u^\rho \; =\; \nabla_\mu (\nabla_\rho u^\rho) \,
+\, [\Gamma_{\rho \mu} , u^\rho ]\, ,
\nn\\[5pt]
\nabla^2 u_\mu \; =\; \nabla_\mu (\nabla^\rho u_\rho) \,
+\, \nabla^\rho f_{-\, \mu\rho} \, +  [\Gamma_{\rho \mu} ,u^\rho]\, .
\end{gather}
Whenever possible, we use them to express the results in terms of the tensors $f_\pm^{\mu\nu}$, proportional to the gauge fields.

\section{Chiral power counting of the low-energy EWET}
\label{app:power-counting}

A generic low-energy Lagrangian operator can be characterized as
\be
\Delta\mL_{djk}  \,\sim\,  f_\ell \,\, \, p^d \,\left(\Frac{\psi }{v}\right)^j\,
\left(\Frac{\phi}{v}\right)^k\, ,
\ee
where $\phi$ denotes any bosonic field ($\varphi, h, \vec{W}_\mu, B_\mu$) and $\psi$ any fermionic
or antifermionic field. The factor $p^d$ accounts for any explicit light scales
($\partial_\mu$, $m_{W}$, $m_Z$, $m_h$, $m_\psi$) or couplings ($g$, $g'$, $y_\xi$)
appearing in the operator, and $f_\ell$ is the corresponding LEC with the appropriate dimension.
This operator will be assigned a chiral dimension $\hat d = d + j/2$
and its impact in the low-energy amplitudes is explained below.

Let us consider a connected Feynman diagram $\Gamma$ with $L$ loops, $I_B$ internal bosonic lines, $I_F$ internal fermionic lines,
$E_B$ external bosons, $E_F$ external fermions and $N_{djk}$ vertices of type $\Delta\mL_{djk}$. The total number of internal and external lines is given by $I=I_B+I_F$ and $E=E_B+E_F$, respectively. These quantities satisfy the topological relations:
\bear
\sum_{d,j,k}  \, j\, N_{djk}  &\! =&\! 2 I_F + E_F\, ,
\nn\\
\sum_{d,j,k}  \, k\, N_{djk}  &\! =&\! 2 I_B + E_B\, ,
\nn\\
L\,=\, I+1 -V &\! =&\! I_B+I_F +1  - \sum_{d,j,k} N_{djk}\, ,
\label{eq:topological-rel}
\eear
being $V = \sum_{d,j,k} N_{djk}$ the total number of vertices in the diagram.

Replacing the external lines by the corresponding fields, the diagram $\Gamma$ corresponds to an operator of the EWET with an infrared dimension $\hat{d}_\Gamma$. Adopting a mass-independent regularization scheme such as dimensional regularization, where no cut-offs are involved, one can apply a naive power-counting to determine the scaling behaviour of the diagram \cite{Georgi:1994qn,Pich:1998xt}. A standard dimensional analysis \cite{Weinberg:1978kz} shows that
\bear
\Gamma & \sim & \Int \left( \Frac{d^4p}{(2\pi)^d}  \right)^L   \,\,\,
\Frac{1}{(p^2)^{I_B}\, (p)^{I_F}}  \,\,\,
\left(\prod_{d,j} (p^d)^{N_{dj}} \right)  \,\,\,
\left( p^\frac{1}{2} \right)^{E_F}
\nn\\[10pt]
&\sim & p^{4L - 2I_B -I_F  + \sum_{d,j} d\, N_{dj} +\frac{1}{2} E_F}
\quad = \quad
p^{2+2L  + \sum_{d,j} (d-2)\,  N_{dj} + I_F + \frac{1}{2} E_F  }
\nn\\[10pt]
& = &
p^{2+2L  +  \sum_{d,j} \left(d+\frac{1}{2} j -2\right)\, N_{dj}} \, .
\eear
with $N_{dj} = \sum_k N_{djk}$.
Therefore, $\Gamma$ scales like $p^{\hat{d}_\Gamma}$ with
\be
\hat{d}_\Gamma\, =\, 2 + 2L + \sum_{\hat{d}} (\hat{d} -2)\, N_{\hat{d}} \, ,
\ee
where $N_{\hat{d}}$ indicates the number of vertices with a given value of $\hat{d}$.

We can complete the previous formal estimate with the scales and factors that will naively accompany the $p^{\hat{d}_\Gamma}$ scaling behaviour:
\bear
\Gamma &\sim & \Frac{1}{(16\pi^2)^L}\,\,\,
\prod_{d,j,k} \left( \,\,  \Frac{f_\ell}{v^{j+k} }\,\,  \right)^{N_{djk}}
\; =\;
\Frac{1}{(16\pi^2)^L}\,\,\,
\left[ \prod_{d,j,k} \left(\Frac{f_\ell}{v^2}\right)^{N_{djk}}\, \right]
\,\,\, \left(\Frac{1}{v}\right)^{\sum_{d,j,k} (j+k-2) N_{djk}} \,
\nn\\[10pt]
& = &
\Frac{1}{(16\pi^2 v^2)^L}\,\,\,
\left[ \prod_{d,j,k} \left(\Frac{f_\ell}{v^2}\right)^{N_{djk}}\, \right]
\,\,\, \Frac{1}{v^{E-2} } \, ,
\eear
using the relations in Eq.~(\ref{eq:topological-rel}).

Therefore, the contribution from this diagram scales like
\be\label{eq:counting_amplitude}
\Gamma \,\sim\, \Frac{p^2}{v^{E-2}} \,\,\,
\left(\Frac{p^2}{16\pi^2 v^2}\right)^L  \,\,\,
 \prod_{\hat{d}} \left(
\Frac{f_\ell\,\,  p^{\, \hat{d}-2}   }{v^2}\right)^{N_{\hat{d}}}
\, .
\ee

 \section{Antisymmetry field formalism for spin-1 particles}
\label{app:antisymmetric}

A spin-1 particle can be described through an antisymmetric tensor field
$V_{\mu\nu} = -V_{\nu\mu}$, with the Lagrangian \cite{Ecker:1988te,Gasser:1983yg}
\be
\mL^{\text{Kin}}_V \; = \; -\Frac{1}{2}\, \partial^\mu V_{\mu\nu}\,\partial_\lambda
V^{\lambda\nu} \, + \, \Frac{1}{4}\, M_V^2\, V_{\mu\nu} V^{\mu\nu} \,
\, ,
\ee
which has the classical free-field equations of motion,
\bel{eq.EOMantisym}
\partial^\mu \partial_\lambda \, V^{\lambda\nu}\,
- \, \partial^\nu \partial_\lambda \, V^{\lambda\mu}\,
+ \, M_V^2 \, V^{\mu\nu}\, = \, 0 \, ,
\ee
implying
\be
\partial_\mu \, (\partial^2+M_V^2)\, V^{\mu\nu}\, = \, 0\,  .
\ee

The corresponding free propagator in momentum space takes then the form of
a four-index antisymmetric tensor:
\be
\ba{rl}
\bra V^{\mu\nu}V^{\rho\sigma}\ket_F \, = \,i\,
\Delta^{\mu\nu,\rho\sigma}(q)  &= \,
\, \Frac{2i}{M_V^2-q^2} \;
\mA^{\mu\nu,\rho\sigma}(q)
\, + \, \Frac{2i}{M_V^2} \; \Omega^{\mu\nu,\rho\sigma}(q)
\\
\\
& = \, \Frac{2i}{M_V^2-q^2}\,\left\{ \mI^{\mu\nu,\rho\sigma}
-\Frac{q^2}{M_V^2}\;\Omega(q)^{\mu\nu,\rho\sigma}\right\}
\\
\\
& = \, \Frac{2i}{M_V^2} \; \mI^{\mu\nu,\rho\sigma} \, +\,
\Frac{2i}{M_V^2-q^2}\; \Frac{q^2}{M_V^2}\;\mA(q)^{\mu\nu,\rho\sigma} \, ,
\ea
\label{eq:Apropagator}
\ee
with
\bear
\label{eq.defA}
\mA_{\mu\nu,\rho\sigma}(q) &\equiv &\Frac{1}{2q^2} \,
\left[\, g_{\mu\rho}q_\nu q_\sigma -
g_{\rho\nu}q_\mu q_\sigma - (\rho \leftrightarrow \sigma) \,\right]
\nn\\
&=&
 \Frac{1}{2}\, P_T(q)^{\mu\rho}\,  P_L(q)^{\nu\sigma}
 - \Frac{1}{2}\, P_T(q)^{\mu\sigma}\,  P_L(q)^{\nu\rho}
 - \Frac{1}{2}\, P_T(q)^{\nu\rho}\,  P_L(q)^{\mu\sigma}
 + \Frac{1}{2}\, P_T(q)^{\nu\sigma}\,  P_L(q)^{\mu\rho}
\nn\\
&=&
\, \Frac{1}{2}\, g^{\mu\rho}\,  P_L(q)^{\nu\sigma}\,
\, -\, \Frac{1}{2}\, g^{\mu\sigma}\,  P_L(q)^{\nu\rho}\,
\, -\, \Frac{1}{2}\, g^{\nu\rho}\,  P_L(q)^{\mu\sigma}\,
\, +\, \Frac{1}{2}\, g^{\nu\sigma}\,  P_L(q)^{\mu\rho}\,
\, ,
\nn\\[10pt]
\Omega_{\mu\nu,\rho\sigma}(q) &\equiv & -\,\Frac{1}{2q^2}\,
\left[ g_{\mu\rho}\, q_\nu q_\sigma -
g_{\rho\nu}\, q_\mu q_\sigma  - q^2\, g_{\mu\rho}g_{\nu\sigma} -
(\rho\leftrightarrow\sigma) \,\right]
\nn\\
&=&
\, \Frac{1}{2} \, P_T(q)^{\mu\rho}\,  P_T(q)^{\nu\sigma}\,
\, -\, \Frac{1}{2} \, P_T(q)^{\mu\sigma}\,  P_T(q)^{\nu\rho}\,
\, ,
\nn\\[10pt]
\mI_{\mu\nu,\rho\sigma} &\equiv & \Frac{1}{2}\, \left(
g_{\mu\rho}\, g_{\nu\sigma} - g_{\mu\sigma}\, g_{\nu\rho}\right) \, ,
\eear
where $P_T^{\mu\nu}(q)=g^{\mu\nu}-q^\mu q^\nu/q^2$ and
$P_L^{\mu\nu}(q)= q^\mu q^\nu /q^2$ are the usual transverse and longitudinal Lorentz projectors.

Eq.~\eqn{eq:Apropagator} can be compared with the standard gauge boson propagator,
\bear
i \Delta_W^{\mu\nu}(q) & =&
\Frac{i}{M_W^2-q^2}\; P_T^{\mu\nu}(q) \,  + \,
\Frac{i\, \xi}{\xi\, M_W^2 - q^2} \; P_L^{\mu\nu}(q)
\nn\\[10pt]
& = & \Frac{i}{M_W^2-q^2} \,
\left\{ g^{\mu\nu} \, - \, \Frac{(\xi-1)\, q^2}{\xi\, M_W^2-q^2}\; P_L^{\mu\nu}(q)
\right\} \, ,
\eear
which in the unitary gauge ($\xi\to\infty$) reduces to the familiar Proca expression,
\bear
i  \Delta_W^{\mu\nu}(q) & =&
 \Frac{i}{M_W^2-q^2}\; P_T^{\mu\nu}(q) \,  + \,
\Frac{i }{M_W^2 } \; P_L^{\mu\nu}(q)
\nn\\[10pt]
& = & \Frac{i}{M_W^2-q^2} \,
\left\{ g^{\mu\nu} \, - \, \Frac{q^2}{M_W^2}\; P_L^{\mu\nu}(q)  \right\} \, .
\eear

The former antisymmetric  tensors obey the following properties:
\be
\ba{c}
\Omega\cdot \mA\, =\, \mA\cdot \Omega\, =\, 0 \quad\; , \quad\;
\mA\cdot \mA\, = \,\mA \quad\; , \quad\;
\Omega \cdot \Omega\, = \,\Omega \quad\; , \quad\;
\mA \,+\,\Omega\, = \,\mI \, ,
\\[10pt]
q^\mu\,\Omega_{\mu\nu,\rho\sigma}(q) \; = \;
q^\nu\,\Omega_{\mu\nu,\rho\sigma}(q) \; = \;
q^\rho\,\Omega_{\mu\nu,\rho\sigma}(q) \; = \;
q^\sigma\,\Omega_{\mu\nu,\rho\sigma}(q) \; = \; 0\, .
\ea
\ee

Finally, it is interesting to consider the matrix element for an outgoing vector
of momentum  $p$ and polarization $\epsilon^\mu_{_{(i)}}(p)$:
\be
\bra 0\, |\, V^{\mu\nu}\, |\, V(p,\epsilon_{_{(i)}})\ket \,
= \, \epsilon^{\mu\nu}_{_{(i)}}(p)\; = \;
\Frac{i}{M_V} \, \left[ p^\mu\epsilon_{_{(i)}}^\nu(p)
-p^\nu\epsilon_{_{(i)}}^\mu(p)\right] \, .
\ee
The summation over the physical vector polarizations
for a massive vector ($\epsilon\cdot p=0  , \, p^2=M_V^2   $) yields:
\be
\sum_{i=1,2,3} \, \epsilon^{\mu\nu}_{_{(i)}}(p)\, \epsilon^{\rho\sigma}_{_{(i)}}(p)^* \;
= \; - \,
 2 \; \mA(p)^{\mu\nu,\rho\sigma}\, ,
\ee
where we have employed the relation \
$\sum_i\epsilon^{\alpha}_{_{(i)}}(p)\,\epsilon^{\beta}_{_{(i)}}(p)^*\, =\,\left(-g^{\alpha\beta}+\Frac{p^\alpha p^\beta}{M_V^2} \right)$.

\section{Relation between spin-1 resonance formulations}
\label{app:PA-correspondence}

Let us consider a generic (vector or axial-vector) spin-1 triplet massive state, described in terms of a four-vector Proca field $\hat{R}_\mu = R^a_\mu \,\sigma^a/\sqrt{2}$ and the Lagrangian
\be
\mL^{\rm (P)}[\hat{R},\phi_j] \; =\;
\mL^{\rm (P)}_{\hat{R}}[\hat{R},\phi_j] \; +\; \mL^{\rm (P)}_{\text{non-R}}[\phi_j] \, ,
\ee
with
\be
\mL^{\rm (P)}_{\hat{R}}[\hat{R},\phi_j]\; =\; -\Frac{1}{4}\,\bra \hat{R}_{\mu\nu}\, \hat{R}^{\mu\nu} \ket
+\Frac{1}{2}\, M_R^2\, \bra \hat{R}_\mu \hat{R}^\mu\ket
+  \bra \hat{R}_{\mu}\, \hat{\chi}_{\hat{R}}^\mu
+ \hat{R}_{\mu\nu}\, \hat{\chi}_{\hat{R}}^{\mu\nu} \ket
\, ,
\label{eq:LagP-R}
\ee
and $\hat{R}_{\mu\nu}=\nabla_\mu \hat{R}_\nu -\nabla_\nu \hat{R}_\mu$.
The term
$\mL^{\rm (P)}_{\text{non-R}}[\phi_j]$ and the chiral structures $\hat{\chi}_{\hat{R}}^\mu$ and $\hat{\chi}_{\hat{R}}^{\mu\nu}$ only contain light SM fields $\phi_j$.

Quantum fields are integration variables in the path-integral formulation of the generating functional. Focusing only on the integration over the four-vector $\hat{R}_\mu$ configurations,
\bear
Z[\phi_j] &=& \mN \; \Int [d\hat{R}]\;
\exp\left\{
i\Int{\rm d^d x}\,\,  \mL^{\rm (P)}[\hat{R},\phi_j]
\right\}
\nn\\[5pt]
&=&  \mN' \; \Int [dR] \,[d\hat{R}]\;
\exp\left\{
i\Int{\rm d^d x}\,
\left( \mL^{\rm (P)}[\hat{R},\phi_j]  \,+\, \Frac{1}{4} \bra R_{\mu\nu} R^{\mu\nu} \ket
\right)\right\} \, ,
\eear
where in the second line we have introduced the term $\bra R_{\mu\nu} R^{\mu\nu} \ket$ which, after integrating over the auxiliary antisymmetric tensor field $R_{\mu\nu}$, produces just a global normalization factor.

Making the change of variables~\cite{Bijnens:1995,Kampf:2006}
$R^{\mu\nu} \to M_R R^{\mu\nu} -\hat{R}^{\mu\nu}
+  (2 \,\hat{\chi}_{\hat{R}}^{\mu\nu}-\bra \hat{\chi}_{\hat{R}}^{\mu\nu}\ket )$,
in the auxiliary field, $Z[\phi_j]$ adopts the form
\bear
Z[\phi_j] &=&  \mN'' \; \Int [dR] \,[d\hat{R}]\;
\exp\left\{
i\Int{\rm d^d x}\,
\left( \Delta \mL^{\rm (A)}[R,\phi_j]  + \Frac{M_R^2}{2}\, \bra \hat{R}_{\mu} \hat{R}^{\mu}\ket
+ \bra \hat{R}_\mu \mJ^\mu \ket\right)\right\}
\nn\\[8pt]
&=&
 \tilde{\mN} \; \Int [dR]\;
\exp\left\{
i\Int{\rm d^d x}\,
\left( \Delta  \mL^{\rm (A)}[R,\phi_j] - \Frac{1}{2 M_R^2}\,
\left[\bra \mJ_\mu \mJ^\mu \ket  - \Frac{1}{2}\bra \mJ_\mu \ket^2\right]
\right)\right\} \, ,\qquad
\label{eq.gen-functional1}
\eear
with the convenient definitions~\cite{Kampf:2006}
\bear
\Delta \mL^{\rm (A)}[R,\phi_j] &=&
\Frac{1}{4}\, M_R^2\, \bra R_{\mu\nu} R^{\mu\nu}\ket \, +\, M_R\, \bra R_{\mu\nu} \hat{\chi}^{\mu\nu}_{\hat{R}}\ket
\, +\, \left( \bra \hat{\chi}_{\hat{R}\,\mu\nu}\, \hat{\chi}_{\hat{R}}^{\mu\nu}\ket -\Frac{1}{2}\,\bra  \hat{\chi}_{\hat{R}\,\mu\nu}\ket
\bra  \hat{\chi}^{\mu\nu}_{\hat{R}}\ket\right) \, ,
\nn\\[5pt]
\mJ^\mu &=& \hat{\chi}_{\hat{R}}^\mu\,  +\, M_R\, \nabla_{\nu} R^{\nu\mu} \, .
\eear
In the last line of Eq.~(\ref{eq.gen-functional1}),
we have performed the Gaussian integration over $\hat{R}$.

The generating functional can be now rewritten as
\bear
Z[\phi_j] &=& \widetilde{\mN} \; \Int [dR]\;
\exp\left\{
i\Int{\rm d^d x} \,\, \mL^{\rm (A)}[R,\phi_j]
\right\} \, ,
\eear
in terms of the antisymmetric tensor field $R_{\mu\nu}$ Lagrangian
\bear
\mL^{\rm (A)}[R,\phi_j] &=& \mL^{\rm (A)}_{R}[R,\phi_j] \; +\;
\mL^{\rm (A)}_{\text{non-R}}[\phi_j] \, ,
\eear
where
\bear
\mL^{\rm (A)}_{R}[R,\phi_j] &=& -\Frac{1}{2}\,\bra  \nabla^\mu R_{\mu\lambda}\, \nabla_\nu R^{\nu\lambda}  \ket
\, +\,\Frac{1}{4}\, M_R^2\, \bra R_{\mu\nu} R^{\mu\nu}\ket\, +\, \bra R_{\mu\nu} \chi_R^{\mu\nu}\ket \, ,
\nonumber \\[10pt]
\mL^{\rm (A)}_{\text{non-R}}[\phi_j] &=& \mL^{\rm (P)}_{\text{non-R}}[\phi_j]\, +
\,\bra \hat{\chi}_{\hat{R}\,\mu\nu}\,  \hat{\chi}_{\hat{R}}^{\mu\nu}\ket
\, -\,\Frac{1}{2}\, \bra  \hat{\chi}_{\hat{R}}^{\mu\nu}\ket\bra \hat{\chi}_{\hat{R}\,\mu\nu}\ket
\nn\\[5pt] &-&
\Frac{1}{2M_R^2}\, \left(  \bra \hat{\chi}_{\hat{R}\,\mu} \, \hat{\chi}_{\hat{R}}^{\mu}\ket
-\Frac{1}{2}\, \bra \hat{\chi}_{\hat{R}}^{\mu}\ket \bra \hat{\chi}_{\hat{R}\,\mu}\ket\right) \, ,
\label{eq:LagA-R}
\eear
with
\be
\chi_R^{\mu\nu} \; =\; \Frac{1}{2M_R}\, \left(\nabla^\mu \hat{\chi}_{\hat{R}}^\nu - \nabla^\nu \hat{\chi}_{\hat{R}}^\mu \right)
\, +\, M_R\, \hat{\chi}_{\hat{R}}^{\mu\nu}\, .
\ee

We can easily generalize this result to an arbitrary number of triplet $\hat{R}$ and singlet $\hat{R}_1$ Proca fields, described by the Lagrangian
\bear
\mL^{\rm (P)} &=& \sum_{\hat{R}} \mL^{\rm (P)}_{\hat{R}}[\hat R,\phi_j]
\, +\, \sum_{\hat{R}_1} \mL^{\rm (P)}_{\hat{R}_1}[\hat R_1,\phi_j]
\, +\,\mL^{\rm (P)}_{\text{non-R}}[\phi_j] \, ,
\eear
 with the triplet resonance contributions in Eq.~\eqn{eq:LagP-R} and the singlet resonance terms
\bear
\mL^{\rm (P)}_{\hat{R}_1}[\hat R_1,\phi_j] &=& -\Frac{1}{4}\,\hat{R}_{1\, \mu\nu} \,\hat{R}_1^{\mu\nu}
\, +\, \Frac{1}{2}\, M_R^2\, \hat{R}_{1\, \mu} \hat{R}_1^\mu \, +\,
\hat{R}_{1\, \mu}\, \hat{\chi}_{\hat{R}_1}^\mu
\, +\, \hat{R}_{1\, \mu\nu} \,\hat{\chi}_{\hat{R}_1}^{\mu\nu}\, .
\eear

Performing for each separate spin-1 field the previous formal manipulations, the generating functional can be written in terms of
an equivalent Lagrangian in the antisymmetric tensor formalism,
\bear
\mL^{\rm (A)} &=& \sum_R \mL^{\rm (A)}_{R}[R,\phi_j]
 \, +\,  \sum_{R_1} \mL^{\rm (A)}_{R_1}[R_1,\phi_j]
 \, +\, \mL^{\rm (A)}_{\text{non-R}}[\phi_j] \, ,
\eear
with the triplet resonance contributions in Eq.~\eqn{eq:LagA-R} and the singlet resonance terms
\bear
\mL^{\rm (A)}_{R_1}[R_1,\phi_j] &=& -\Frac{1}{2}\,\partial^\mu R_{1\,\mu\lambda} \,\partial_\nu R_1^{\nu\lambda}
\, +\,\Frac{1}{4}\, M_{R_1}^2\, R_{1\,\mu\nu}\, R_1^{\mu\nu}  \, +\,
R_{1\,\mu \nu }\, \chi_{R_1}^{\mu\nu}  \, ,
\eear
with
\be
\chi_{R_1}^{\mu\nu} \; =\; \Frac{1}{2M_{R_1}}\, (\partial^\mu \hat{\chi}_{\hat{R}_1}^\nu
- \partial^\nu \hat{\chi}_{\hat{R}_1}^\mu )
\, +\, M_{R_1}\, \hat{\chi}_{\hat{R}_1}^{\mu\nu}\, .
\ee
The Lagrangian piece without resonance fields is given by
\bear
\mL^{\rm (A)}_{\text{non-R}}[\phi_j] &=& \sum_{   \hat{R}   }
\left[ \bra \hat{\chi}_{\hat{R}\,\mu\nu}\,  \hat{\chi}_{\hat{R}}^{\mu\nu}\ket
-\Frac{1}{2}\,\bra \hat{\chi}_{\hat{R}\,\mu\nu}\ket \bra \hat{\chi}_{\hat{R}}^{\mu\nu}\ket
\, -\, \Frac{1}{2M_R^2}\, \left(  \bra \hat{\chi}_{\hat{R}\,\mu}\,  \hat{\chi}_{\hat{R}}^{\mu}\ket
-\Frac{1}{2}\, \bra \hat{\chi}_{\hat{R}\,\mu}\ket \bra \hat{\chi}_{\hat{R}}^{\mu}\ket\right)
\right]
\nn\\[5pt]
&+&
\sum_{    \hat{R}_1     }
\left[ \hat{\chi}_{\hat{R}_1\,\mu\nu}\,\hat{\chi}_{\hat{R}_1}^{\mu\nu}  \, -\, \Frac{1}{2M_{R_1}^2}\, \hat{\chi}_{\hat{R}_1\,\mu}\,\hat{\chi}_{\hat{R}_1}^{\mu}\right]
\, +\, \mL^{\rm (P)}_{\text{non-R}}[\phi_j] \, .
\eear

\section{Higgsless bosonic operators at short distances}
\label{app:more-high-energy}

We analyze next the high-energy behaviour of some selected Green functions, which are sensitive to specific LECs, and compare the results obtained with the antisymmetric and Proca formalisms for spin-1 fields.

\subsection{Two-Goldstone scattering amplitudes}

The LECs $\mF_4$ and $\mF_5$ contribute to the two-Goldstone scattering amplitudes
\be
T[\varphi^a (p_1)\, \varphi^b(p_2) \to \varphi^c(p_3)\, \varphi^d(p_4)]
\; =\;
A(s,t,u)\,\delta_{ab}\delta_{cd} \,+\, A(t,s,u)\,\delta_{ac}\delta_{bd}
\,+\, A(u,s,t)\,\delta_{ad}\delta_{bc} \, .
\ee
This generic structure is a consequence of the $SU(2)_{L+R}$ and crossing symmetries,
with $s$, $t$ and $u$ the standard Mandelstam variables.

At LO, the two different spin-1 effective theories, SDET-A and SDET-P, give the results:
\bear
A (s,t,u)^{\rm SDA} &=&
 \Frac{G_V^2}{v^4}\,\left[\, \Frac{s^2-u^2}{t-M_V^2} \,
+\, \Frac{s^2-t^2}{u-M_V^2} \, \right]\; +\;
\Frac{\widetilde{G}_A^2}{v^4}\,\left[\, \Frac{s^2-u^2}{t-M_A^2} \,
+\, \Frac{s^2-t^2}{u-M_A^2}\,\right]
\nn\\ &+&
\Frac{2c_d^2}{v^4}\,\Frac{s^2}{M_{S_1}^2-s} \; +\;
\Frac{s}{v^2}  \; +\;   \Frac{4}{v^4}\,\left[ 2\,\mF_5^{\rm SDA}\, s^2 \, +\, \mF_4^{\rm SDA}\,  (t^2+u^2)\,\right]\, ,
\nn\\[8pt]
A (s,t,u)^{\rm SDP} &=&
\Frac{g_{\hat V}^2}{ v^4 } \, \left[\, \Frac{ t\, (s^2-u^2) }{t-M_{V}^2}
\, + \, \Frac{u\, (s^2-t^2)}{u-M_{V}^2}\,\right]
\; +\;
\Frac{\widetilde{g}_{\hat A}^2}{ v^4 } \, \left[\, \Frac{ t\, (s^2-u^2) }{t-M_{A}^2}
\, + \, \Frac{u\, (s^2-t^2)}{u-M_{A}^2}\,\right]
\nn\\  &+&
\Frac{2c_d^2}{v^4}\,\Frac{s^2}{M_{S_1}^2-s}  \; +\; \Frac{s}{v^2} \; +\;
\Frac{4}{v^4}\,\left[ 2\,\mF_5^{\rm SDP}\, s^2 \, +\, \mF_4^{\rm SDP}\,  (t^2+u^2)\,\right] \, .
\eear
The scalar-exchange contribution is obviously identical in both cases and grows linearly with $s\sim E^2$, which a priori does not violate the Froissart bound on the cross section. A similar growing with energy appears in the antisymmetric spin-1 contribution. However, the Proca realization gives a much worse behaviour $A\sim E^4$. The local $\mF_{4,5}$ terms generate in both cases a quadratic dependence with the Mandelstam variables.

To satisfy unitarity, the forward scattering amplitudes
must obey a once-subtracted dispersion relation. The pieces growing as $E^4$ must then cancel in both EFTs, which sets a relation between the local terms and the spin-1 contribution.
In the antisymmetric realization one finds that  $\mF_{4,5}^{\rm SDA}$ must vanish, whereas in the Proca formalism one needs non-zero $\mF_{4,5}^{\rm SDP}$ couplings:
\bear\label{eq:F45SDA}
\mF_{4}^{\rm SDA} &=& \mF_{5}^{\rm SDA}\,=\,  0 \, ,
\nn\\
\mF_4^{\rm SDP}&=&
\Frac{g_{\hat V}^2}{4} \, + \, \Frac{\widetilde g_{\hat A}^2}{4}\, ,
\qquad\qquad\qquad
\mF_5^{\rm SDP} \,=\, -\, \Frac{g_{\hat V}^2}{4} \, - \, \Frac{\widetilde g_{\hat A}^2}{4}\, .
\label{eq:F45SDP}
\eear

The two spin-1 descriptions give then the same scattering amplitudes with $g_{\hat R} = G_R/M_R$ and $\widetilde g_{\hat R} = \widetilde G_R/M_R$
 ($R=V,A$), in agreement with the relations~(\ref{eq.P-A-relations})
between the Proca and antisymmetric Lagrangians.
The infrared behaviour determines the final predictions for the EWET LECs
 in Table~\ref{tab:bLECs-final}:
$\mF_{4}$ takes the value quoted in Table~\ref{tab:bLECs-A},
while $\mF_{5}$ receives in addition the spin-0 contribution given in Table~\ref{tab:spin-0-Op4-LEC}.

\subsection{Two-point current correlators}

Let us consider the two-point correlation functions of the vector and axial-vector currents in Eqs.~\eqn{eq:Vcurrent} and \eqn{eq:Acurrent},
\be
i \int {\rm d}^4x\; {\rm e}^{iq(x-y)}\, \bra 0\,|\, T\left[ \mJ^{\mu}_{a} (x) \, \mJ'^{\,\nu}_{b}(y)^\dagger \right]  |\,0\ket =
\delta_{ab}\, ( -g^{\mu\nu} q^2 + q^\mu q^\nu)\,\Pi_{\mJ \mJ'} (q^2)  \, \quad (\mJ, \mJ' = \mV, \mA)\, .
\ee
At LO, one obtains
\bear
\Pi_{\mV \mV}(q^2) &=& \left\{ \bat
\Frac{F_V^2}{M_V^2-q^2} \, + \, \Frac{\widetilde F_A^2}{M_A^2-q^2}
\, - \, 2\,\left(\mF_1^{\rm SDA} + 2\, \mF_2^{\rm SDA}\right)
& \qquad\qquad\mbox{\small  (SDET-A)}\, ,
\\[12pt]
\Frac{f_{\hat V}^2\, q^2}{M_{V}^2-q^2}
\, + \, \Frac{\widetilde f_{\hat A}^2\, q^2}{M_{A}^2-q^2}
\, - \, 2\,\left(\mF_1^{\rm SDP} + 2\, \mF_2^{\rm SDP}\right)
& \qquad\qquad\mbox{\small  (SDET-P)}\, ,
\ea\right.
\label{eq:PiVV}
\\[15pt]
\Pi_{\mA \mA}(q^2) &=& \left\{ \bat
\Frac{F_A^2}{M_A^2-q^2} \, + \, \Frac{\widetilde F_V^2}{M_V^2-q^2} \,
- \, \Frac{v^2}{q^2}
\,+ \, 2\,\left(\mF_1^{\rm SDA} - 2\, \mF_2^{\rm SDA}\right)
& \qquad\quad\mbox{\small  (SDET-A)}\, ,
\\[12pt]
\Frac{f_{\hat A}^2\, q^2}{M_{A}^2-q^2}
\, + \, \Frac{\widetilde f_{\hat V}^2\, q^2}{M_{V}^2-q^2}
\,  - \, \Frac{v^2}{q^2}
\, + \, 2\,\left(\mF_1^{\rm SDP} - 2\, \mF_2^{\rm SDP}\right)
& \qquad\quad\mbox{\small  (SDET-P)}\, ,
\ea\right.\quad\;
\label{eq:PiAA}
\\[15pt]
\Pi_{\mV \mA}(q^2) &=& \left\{ \bat
-\,\Frac{F_V\, \widetilde F_V}{M_V^2-q^2} \, -\, \Frac{F_A\, \widetilde F_A}{M_A^2 - q^2}
\, + \, 4\,\widetilde \mF_2^{\rm SDA}
& \qquad\qquad\mbox{\small  (SDET-A)}\, ,
\\[12pt]
-\,\Frac{f_{\hat V}\,\widetilde f_{\hat V}\, q^2}{M_{V}^2-q^2}
\, - \, \Frac{f_{\hat A}\, \widetilde f_{\hat A}\, q^2}{M_{A}^2 - q^2}
\, + \, 4\,\widetilde \mF_2^{\rm SDP}
& \qquad\qquad\mbox{\small  (SDET-P)}\, .
\ea\right.
\label{eq:PiVA}
\eear
The couplings of the two formalisms being related by Eqs.~\eqn{eq.P-A-relations}.

The difference $\Pi_{\mV \mV}(q^2)-\Pi_{\mA \mA}(q^2)$ is an order parameter of EWSB. Its short-distance OPE can only receive non-zero contributions from operators which break chiral symmetry and, therefore, vanishes very fast at large values of $t=q^2$ (as $1/t^3$ in asymptotically-free theories \cite{Bernard:1975cd}).
Requiring only the softer condition that it satisfies an unsubtracted dispersion relation implies
\bear\label{eq:F1SDA}
\mF_1^{\rm SDA} &=& 0 \, ,
\nn\\
\mF_1^{\rm SDP}  &=& -\frac{1}{4}\,\left(
f_{\hat V}^2 -\widetilde f_{\hat V}^2 - f_{\hat A}^2 + \widetilde f_{\hat A}^{  2   } \right) \, .
\label{eq:F1SDP}
\eear

A similar argument applies to $\Pi_{\mV \mA}(q^2)$. Imposing that it vanishes at large $q^2$ leads to
\bear
\widetilde\mF_2^{\rm SDA} &=& 0 \, ,
\nn\\
\widetilde\mF_2^{\rm SDP}  &=&
-\frac{1}{4}\,\left( f_{\hat V}\,\widetilde f_{\hat V} + f_{\hat A}\,\widetilde f_{\hat A}\right)\, .
\eear

If one further requires that the separate $\Pi_{\mV \mV}(q^2)$ and $\Pi_{\mA \mA}(q^2)$ correlators vanish at large energies, one gets in addition
\bear\label{eq:F2SDA}
\mF_2^{\rm SDA} &=& 0 \, ,
\nn\\
\mF_2^{\rm SDP}  &=&
-\frac{1}{8}\,\left( f_{\hat V}^2 +\widetilde f_{\hat V}^2 + f_{\hat A}^2 +\widetilde f_{\hat A}^2\right)\, .
\label{eq:F2SDP}
\eear
Therefore, the three LECs, $\mF_1$, $\mF_2$ and $\widetilde\mF_2$ are saturated
by spin-1 resonance exchange in the antisymmetric formalism,
and take the values given
 in Table~\ref{tab:bLECs-final}.

Note however, that there are no strong reasons why this last condition should be fulfilled (in fact, it does not in QCD). Thus, there could exist an additional
non-zero contribution to $\Delta\mF_2^{\rm SDP} = \mF_2^{\rm SDA}$ which is not fixed by the single-resonance dynamics. Its determination would require more direct information on the underlying short-distance theory.


\end{document}